%% file: DESY_97_055_v2.tex
\documentclass[12pt]{article}
 \usepackage{epsfig}
 \usepackage{pstricks}

\newcommand{\dd}{\mbox{d}}
\newcommand{\scr}{\scriptscriptstyle}

\newcommand{\lljj}{\left(\l_{\n}^{\dg}\l_{\n}\right)_{jj}}
\newcommand{\llii}{\left(\l_{\n}^{\dg}\l_{\n}\right)_{ii}}
\newcommand{\llji}{\left(\l_{\n}^{\dg}\l_{\n}\right)_{ji}}
\newcommand{\llnj}{\left(\l_{\n}^{\dg}\l_{\n}\right)_{nj}}
\newcommand{\mmjj}{{(m_{\scr D}^{\dg}m_{\scr D})_{jj}}}

\newcommand{\lnaj}{\ln\left({x+a_j\over a_j}\right)}
\newcommand{\lnan}{\ln\left({x+a_n\over a_n}\right)}
\newcommand{\lnpropj}{\ln\left({(x-a_j)^2+a_j\widetilde{c_j}\over
                             a_j^2+a_j\widetilde{c_j}}\right)}
\newcommand{\lnpropn}{\ln\left({(x-a_n)^2+a_n\widetilde{c_n}\over
                             a_n^2+a_n\widetilde{c_n}}\right)}
\newcommand{\atnj}{\left[
            \arctan\left({x-a_j\over\sqrt{a_j\widetilde{c_j}}}\right)
           +\arctan\left(\sqrt{a_j\over\widetilde{c_j}}\right)\right]}
\newcommand{\atnn}{\left[
            \arctan\left({x-a_n\over\sqrt{a_n\widetilde{c_n}}}\right)
           +\arctan\left(\sqrt{a_n\over\widetilde{c_n}}\right)\right]}
\newcommand{\lkin}{\l_{ij}}
\newcommand{\lnNN}{\mbox{L}_{ij}}
\newcommand{\sn}{\widetilde{N_j^c}}
\newcommand{\snj}{\widetilde{N_j^c}}
\newcommand{\sni}{\widetilde{N_i^c}}
\newcommand{\snone}{\widetilde{N_1^c}}
\newcommand{\sur}{\widetilde{U^c}}
\newcommand{\Gnj}{\G_{\scr N_j}}
\newcommand{\Gsnj}{\G_{\scr\snj}^{(2)}}
\newcommand{\Gtr}{\G_{\scr\snj}^{(3)}}
\newcommand{\gnj}{\g_{\scr N_j}}
\newcommand{\gsnj}{\g_{\scr\snj}^{(2)}}
\newcommand{\gtr}{\g_{\scr\snj}^{(3)}}
\newcommand{\gnone}{\g_{\scr N_1}}
\newcommand{\gsnone}{\g_{\scr\snone}^{(2)}}
\newcommand{\gtrone}{\g_{\scr\snone}^{(3)}}
\newcommand{\Ynj}{Y_{\scr N_j}}
\newcommand{\Yni}{Y_{\scr N_i}}
\newcommand{\Ynjeq}{Y_{\scr N_j}^{\rm eq}}
\newcommand{\Ynieq}{Y_{\scr N_i}^{\rm eq}}
\newcommand{\Ynjratio}{{\Ynj\over\Ynjeq}}
\newcommand{\Yniratio}{{\Yni\over\Ynieq}}
\newcommand{\Yp}{Y_{j_+}}
\newcommand{\Ym}{Y_{j_-}}
\newcommand{\Ysnjeq}{Y_{\scr \snj}^{\rm eq}}
\newcommand{\Ypratio}{{\Yp\over\Ysnjeq}}
\newcommand{\Ypiratio}{{Y_{i_+}\over Y_{\scr\sni}^{\rm eq}}}
\newcommand{\Ymratio}{{\Ym\over\Ysnjeq}}
\newcommand{\Ymiratio}{{Y_{i_-}\over Y_{\scr\sni}^{\rm eq}}}
\newcommand{\YL}{Y_{\scr L_f}}
\newcommand{\YLt}{Y_{\scr L_s}}
\newcommand{\Yleq}{Y_l^{\rm eq}}
\newcommand{\Ylteq}{Y_{\wt{l}}^{\rm eq}}
\newcommand{\YLratio}{{\YL\over\Yleq}}
\newcommand{\YLtratio}{{\YLt\over\Ylteq}}
\def\a{\alpha}
\def\b{\beta}

\def\d{\delta}
\def\e{\epsilon}

\def\g{\gamma}
\def\h{\eta}

\def\j{\psi}

\def\l{\lambda}
\def\m{\mu}
\def\n{\nu}

\def\p{\pi}
\def\q{\theta}
\def\r{\rho}
\def\s{\sigma}
\def\t{\tau}

\def\D{\Delta}

\def\G{\Gamma}

\def\L{\Lambda}

\def\ve{\varepsilon}

\def\co{{\cal O}}

\def\cw{{\cal W}}

\def\bo{{\raise.15ex\hbox{\large$\Box$}}}               % D'Alembertian
\def\pr{\prod}                                          % product
\def\ltap{\raisebox{-.4ex}{\rlap{$\sim$}} \raisebox{.4ex}{$<$}}   % < or ~
\def\gtap{\raisebox{-.4ex}{\rlap{$\sim$}} \raisebox{.4ex}{$>$}}   % > or ~
\def\face{{\raise.2ex\hbox{$\displaystyle \bigodot$}\mskip-2.2mu \llap {$\ddot
        \smile$}}}                                      % happy face
\def\dg{\dagger}                                     % hermitian conjugate
\def\wt#1{\widetilde{#1}}                    % big tilde
\def\Bar#1{\overline{#1}}                       % big bar
\def\leftrightarrowfill{$\mathsurround=0pt \mathord\leftarrow \mkern-6mu
        \cleaders\hbox{$\mkern-2mu \mathord- \mkern-2mu$}\hfill
        \mkern-6mu \mathord\rightarrow$}       % <--> double differential
\def\dvec#1{\vbox{\ialign{##\crcr
        \leftrightarrowfill\crcr\noalign{\kern-1pt\nointerlineskip}
        $\hfil\displaystyle{#1}\hfil$\crcr}}}           % <--> accent
\def\dt#1{{\buildrel {\hbox{\LARGE .}} \over {#1}}}     % dot-over for sp/sb
\def\beq{\begin{equation}}
\def\eeq{\end{equation}}

\def\beqx{\begin{displaymath}}
\def\eeqx{\end{displaymath}}

\def\beqa{\begin{eqnarray}}
\def\eeqa{\end{eqnarray}}
\def\NO{\nonumber}

\def\pl#1#2#3{Phys.~Lett.~{\bf B {#1}} (19{#2}) #3}
\def\np#1#2#3{Nucl.~Phys.~{\bf B {#1}} (19{#2}) #3}
\def\prl#1#2#3{Phys.~Rev.~Lett.~{\bf #1} (19{#2}) #3}
\def\pr#1#2#3{Phys.~Rev.~{\bf D {#1}} (19{#2}) #3}
\def\zp#1#2#3{Z.~Phys.~{\bf C {#1}} (19{#2}) #3}

\def\mpl#1#2#3{Mod.~Phys.~Lett.~{\bf A {#1}} (19{#2}) #3}
\def\nc#1#2#3{Nuovo Cim.~{\bf {#1}} (19{#2}) #3}

\catcode`@=11
\def\@citex[#1]#2{\if@filesw\immediate\write\@auxout{\string\citation{#2}}\fi
  \def\@citea{}\@cite{\@for\@citeb:=#2\do
    {\@citea\def\@citea{,\penalty\@m}\@ifundefined
      {b@\@citeb}{{\bf ?}\@warning
       {Citation `\@citeb' on page \thepage \space undefined}}%
\hbox{\csname b@\@citeb\endcsname}}}{#1}}

\def\citer{\@ifnextchar [{\@tempswatrue\@citexr}{\@tempswafalse\@citexr[]}}
\def\@citexr[#1]#2{\if@filesw\immediate\write\@auxout{\string\citation{#2}}\fi
  \def\@citea{}\@cite{\@for\@citeb:=#2\do
    {\@citea\def\@citea{--\penalty\@m}\@ifundefined
       {b@\@citeb}{{\bf ?}\@warning
       {Citation `\@citeb' on page \thepage \space undefined}}%
\hbox{\csname b@\@citeb\endcsname}}}{#1}}
\catcode`@=12
\begin{document}
\date{\mbox{ }}

\title{ 
{\normalsize     
\hfill \parbox{32mm}{DESY 97-055\\\tt hep-ph/9704231}}\\[25mm]
Baryon Asymmetry, Neutrino Mixing\\
and Supersymmetric SO(10) Unification\\[8mm]}
\author{
Michael Pl\"umacher\thanks{e-mail: {\tt pluemi}@{\tt mail.desy.de}}\\
Deutsches Elektronen-Synchrotron DESY,      \\
Notkestr.\,85, D-22603 Hamburg, Germany}
\maketitle

\thispagestyle{empty}

\begin{abstract}
  \noindent
  The baryon asymmetry of the universe can be explained by the
  out-of-equilibrium decays of heavy right-handed neutrinos. We
  analyse this mechanism in the framework of a supersymmetric
  extension of the Standard Model and show that lepton number
  violating scatterings are indispensable for baryogenesis, even
  though they may wash-out a generated asymmetry.  By assuming a
  similar pattern of mixings and masses for neutrinos and up-type
  quarks, as suggested by SO(10) unification, we can generate the
  observed baryon asymmetry without any fine tuning, if $(B-L)$ is
  broken at the unification scale $\Lambda_{\mbox{\tiny GUT}}\sim
  10^{16}\;$GeV and, if $m_{\n_\m} \sim 3\cdot 10^{-3}\;$eV as
  preferred by the MSW solution to the solar neutrino deficit.
\end{abstract}

\newpage  
\section{Introduction}
    The observed baryon asymmetry of the universe
    \beq
      Y_B={n_B\over s}=(0.6-1)\cdot 10^{-10}\;,
      \label{number}
    \eeq
    cannot be explained within the Standard Model, i.e.\ one has to
    envisage extensions of the Standard Model. Grand unified theories
    (GUTs) are attractive for various reasons and there have been many
    attempts to generate $Y_B$ at the GUT scale \cite{kt1}. However,
    these mechanisms seem to be incompatible with inflationary
    scenarios which require reheating temperatures well below the GUT
    scale, the influence of preheating \cite{preheating} on the baryon
    asymmetry requiring further studies.

    During the evolution of the early universe, the electroweak phase
    transition is the last opportunity to generate a baryon asymmetry
    without being in conflict with the strong experimental bounds on
    baryon number violation at low energies \cite{phase}. However, the
    thermodynamics of this transition indicates that such scenarios
    are rather unlikely \cite{jansen}.

    Therefore, the baryon asymmetry has to be generated between the
    reheating scale and the electroweak scale, where baryon plus
    lepton number $(B+L)$ violating anomalous processes are in thermal
    equilibrium \cite{sphal2}, thereby making a $(B-L)$ violation
    necessary for baryogenesis. Hence, no asymmetry can be generated
    within GUT scenarios based on the gauge group SU(5), where $(B-L)$
    is a conserved quantity.

    Gauge groups containing SO(10) predict the existence of
    right-handed neutrinos. In such theories $(B-L)$ is spontaneously
    broken, one consequence being that the right-handed neutrinos can
    acquire a large Majorana mass, thereby explaining the smallness of
    the light neutrino masses via the see-saw mechanism \cite{seesaw}.
    Heavy right-handed Majorana neutrinos violate lepton number in
    their decays, thus implementing the required $(B-L)$ breaking as
    lepton number violation. This leptogenesis mechanism was first
    suggested by Fukugita and Yanagida \cite{fy} and has subsequently
    been studied by several authors \citer{etc,covi}.

    If one assumes a similar pattern of mass ratios and mixings for
    leptons and quarks and, if $m_{\n_{\m}}\sim3\cdot10^{-3}\;$eV
    as preferred by the MSW solution to the solar neutrino problem,
    leptogenesis implies that $(B-L)$ is broken at the unification scale
    \cite{bp}.  This suggests a grand unified theory based on the
    group SO(10), or one of its extensions, which is directly broken
    into the Standard Model gauge group at the unification scale
    $\sim10^{16}\;$GeV. However, for a successful gauge coupling
    unification, such a GUT scenario requires low-energy
    supersymmetry.

    Supersymmetric leptogenesis has already been considered in
    refs.~\cite{camp,covi} in the approxi\-mation that there are no
    lepton number violating scatterings which can inhibit the
    generation of a lepton number. Another usually neglected problem
    of leptogenesis scenarios is the necessary production of
    the right-handed neutrinos after reheating. In the
    non-supersymmetric scenarios one has to assume additional
    interactions of the right-handed neutrinos for successful
    leptogenesis \cite{pluemi}.

    In this paper, we investigate supersymmetric leptogenesis within
    the framework of the mini\-mal supersymmetric standard model (MSSM)
    to which we add right-handed Majorana neutrinos, as suggested by
    SO(10) unification. In the next section we will discuss the
    neutrino decays and scattering processes that one has to take into
    account to be consistent. In section \ref{results} we will develop
    the full network of Boltzmann equations necessary to get a
    reliable relation between the input parameters and the final
    baryon asymmetry. We will show that by neglecting the lepton
    number violating scatterings one largely overestimates the
    generated asymmetry and that in our scenario the Yukawa
    interactions are strong enough to produce a thermal population of
    right-handed neutrinos at high temperatures. Finally we will see
    in section \ref{Yuk} that by assuming a similar pattern of masses
    and mixings for leptons and quarks one gets the required value for
    the baryon asymmetry without any fine tuning, provided $(B-L)$ is
    broken at the GUT scale and the Dirac mass scale for the neutrinos
    is of order of the top-quark mass, as suggested by SO(10)
    unification.

    In the appendices \ref{appA} and \ref{appB} we introduce our
    notations concerning superfields and the Boltzmann equations,
    respectively. The reduced cross sections for the scattering
    processes discussed in section \ref{theory} can be found in
    appendix \ref{appC}, while appendix \ref{appD} summarizes some
    limiting cases in which the corresponding reaction densities can
    be calculated analytically.
\section{The model \label{theory}}
\subsection{The superpotential \label{superpot}}
    In supersymmetric unification scenarios based on SO(10), the
    effective theory below the $(B-L)$ breaking scale is the MSSM
    supplemented by right-handed Majorana neutrinos.  Neglecting soft
    breaking terms, the masses and Yukawa couplings relevant for
    leptogenesis are given by the superpotential
    \beq
      \cw = {1\over2}N^cMN^c + \m H_1\e H_2 
      + H_1 \e Q \l_d D^c + H_1 \e L \l_l E^c 
      + H_2 \e Q \l_u U^c + H_2 \e L \l_{\n} N^c\;,
    \eeq
    where we have chosen a basis in which the Majorana mass matrix
    $M$ and the Yukawa coupling matrices $\l_d$ and $\l_l$ for the
    down-type quarks and the charged leptons are diagonal with real
    and positive eigenvalues.

    The vacuum expectation values of the neutral Higgs fields generate
    Dirac masses for the down-type quarks and the charged leptons
    \beq
      v_1=\left\langle H_1\right\rangle\ne0\qquad\Rightarrow\qquad
      m_d=\l_d\;v_1\quad\mbox{and}\quad m_l=\l_l\;v_1\;,
    \eeq
    and for the up-type quarks and the neutrinos
    \beq
      v_2=\left\langle H_2\right\rangle\ne0\qquad\Rightarrow\qquad
      m_u=\l_u\;v_2\quad\mbox{and}\quad m_{\scr D}=\l_{\n}\;v_2\;.
    \eeq
    The Majorana masses $M$ for the right-handed neutrinos, which have
    to be much larger than the Dirac masses $m_{\scr D}$, offer a natural 
    explanation for the smallness of the light neutrino masses via the
    see-saw mechanism \cite{seesaw}. 
    
    To get a non-vanishing lepton asymmetry, one needs non-degenerate
    Majorana masses $M_i$.  Then the scale at which the asymmetry is
    generated is given by the mass $M_1$ of the lightest right-handed
    neutrino. Hence, it is convenient to write all the masses and
    energies in units of $M_1$,
    \beq
      a_j := \left({M_j\over M_1}\right)^2\;,\qquad
      x={s\over M_1^2}\quad\mbox{and}\quad
      z={M_1\over T}\;,
    \eeq
    where $M_j$ are the masses of the heavier right-handed neutrinos,
    $s$ is the squared centre of mass energy of a scattering process
    and $T$ is the temperature.
\subsection{The decay channels of the heavy neutrinos}
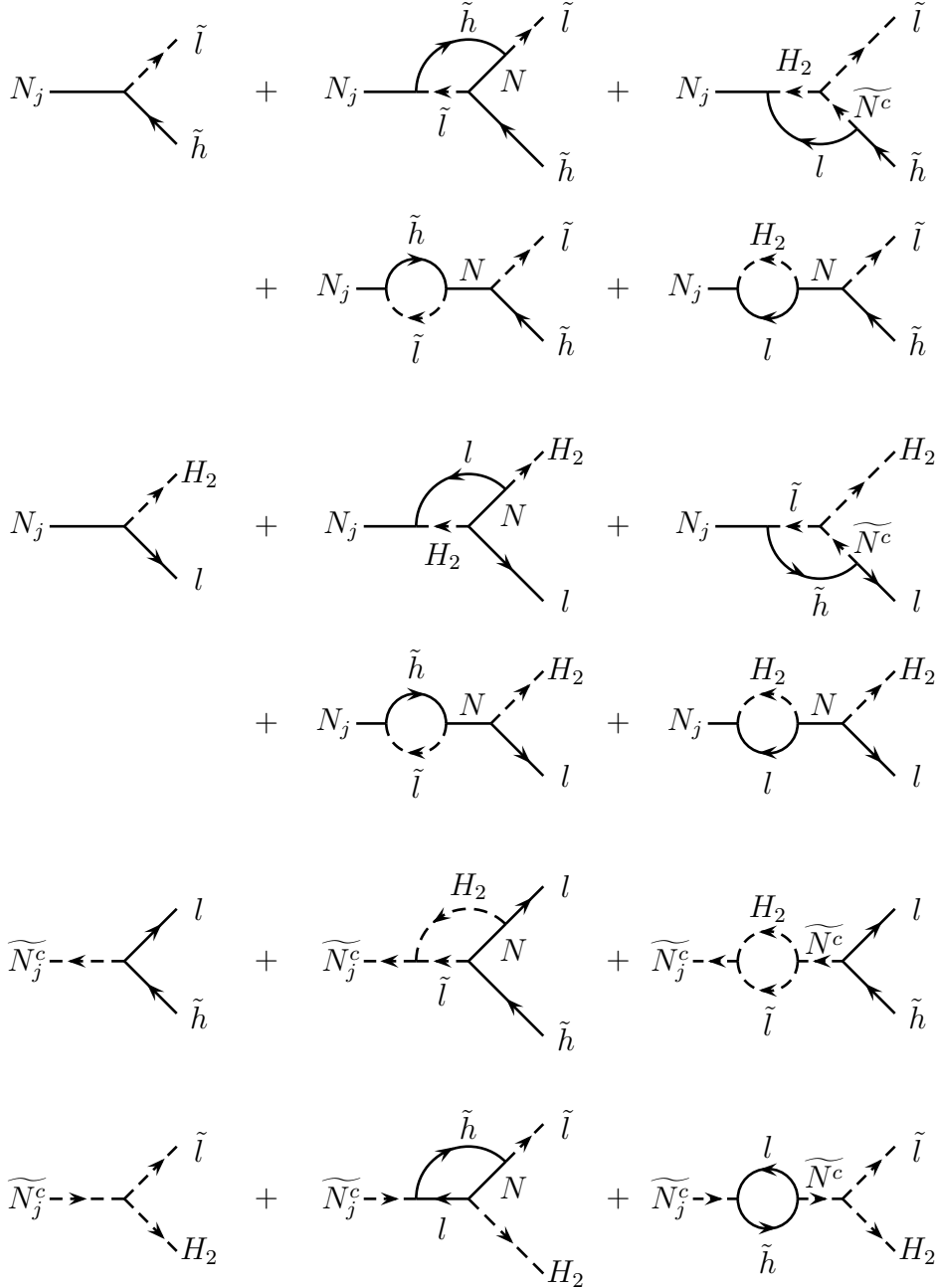
\begin{figure}[h]
      \input{Fig01.tex}
      \caption{\it Decay modes of the right-handed Majorana neutrinos
        and their scalar partners. \label{fig01}}
    \end{figure}

    The right-handed Majorana neutrinos $N_j$ can decay into a lepton
    and a Higgs boson or into a slepton and a higgsino, while their
    scalar partners $\snj$ can decay into a lepton and a higgsino or
    into a slepton and a Higgs boson (cf.~fig.~\ref{fig01}). The 
    decay widths at tree level read \cite{covi}
    \beqa
      {1\over4}\Gnj&:=&\G\Big(N_j\to\wt{l}+\Bar{\wt{h}}\;\Big)
        =\G\Big(N_j\to{\wt{l}{ }}^{\,\dg}+\wt{h}\;\Big)\NO\\
      &=&\G\Big(N_j\to l+H_2\Big)=\G\Big(N_j\to 
        \Bar{l}+H_2^{\dg}\Big)
        ={M_j\over16\p}\;{\mmjj\over v_2^2}\;,
        \label{decay1}\\[1ex]
      {1\over2}\Gsnj&:=&\G\Big(\snj\to\wt{l}+H_2\Big)
        =\G\Big(\snj\to\Bar{l}+\wt{h}\;\Big)\NO\\
      &=&\G\Big(\snj^{\,\dg}\to{\wt{l}{ }}^{\,\dg}+H_2^{\dg}\Big)
        =\G\Big(\snj^{\,\dg}\to l+\Bar{\wt{h}}\;\Big)=
        {M_j\over8\p}\;{\mmjj\over v_2^2}\;.
        \label{decay2}
    \eeqa
    According to eq.~(\ref{decay}), the reaction densities for these
    decays are then given by
    \beq
      \gnj=2\,\gsnj = {M_1^4\over4\p^3}\;{\mmjj\over v_2^2}\;
        {a_j\sqrt{a_j}\over z}\mbox{K}_1(z\sqrt{a_j})\;.
    \eeq
    All these decay modes are $CP$ violating, the dominant
    contribution to $CP$ violation coming about through interference
    between the tree level and the one-loop diagrams shown in
    fig.~\ref{fig01}.  The $CP$ asymmetries in the different decay
    channels of $N_j$ and $\snj$ can all be expressed by the same $CP$
    violation parameter $\ve_j$,
    \beqa
      \lefteqn{\ve_j:={\G\Big(N_j\to\wt{l}+\Bar{\wt{h}}\,\Big)
         -\G\Big(N_j\to{\wt{l}{ }}^{\;\dg}+\wt{h}\,\Big)\over
         \G\Big(N_j\to\wt{l}+\Bar{\wt{h}}\,\Big)
         +\G\Big(N_j\to{\wt{l}{ }}^{\;\dg}+\wt{h}\,\Big)}
      ={\G\Big(N_j\to l+H_2\Big)-\G\Big(N_j\to\Bar{l}
         +H_2^{\dg}\Big)\over\G\Big(N_j\to l+H_2\Big)
         +\G\Big(N_j\to\Bar{l}+H_2^{\dg}\Big)}}\NO\\[1ex]
      &=&{\G\Big(\snj^{\dg}\to l+\Bar{\wt{h}}\,\Big)-
         \G\Big(\snj\to \Bar{l}+\wt{h}\,\Big)\over
         \G\Big(\snj^{\dg}\to l+\Bar{\wt{h}}\,\Big)+
         \G\Big(\snj\to \Bar{l}+\wt{h}\,\Big)}
      ={\G\Big(\snj\to\wt{l}+H_2\Big)-\G\Big(\snj^{\dg}
         \to{\wt{l}{ }}^{\;\dg}+H_2^{\dg}\Big)\over\G\Big(\snj
         \to\wt{l}+H_2\Big)+\G\Big(\snj^{\dg}\to
         {\wt{l}{ }}^{\;\dg}+H_2^{\dg}\Big)}\NO\\[1ex]
      &=& -{1\over8\p v_2^2}\;{1\over\mmjj}\sum\limits_{n\ne j}
         \mbox{Im}\left[(m_{\scr D}^{\dg}m_{\scr D})^2_{nj}\right]
         \;g\Big({a_n\over a_j}\Big)\;,\label{CP}\\[1ex]
      &&\mbox{with}\quad g(x)=\sqrt{x}\left[\mbox{ln}\left({1+x\over x}
         \right)+{2\over x-1}\right]\;\approx {3\over\sqrt{x}}\quad
         \mbox{for}\quad x\gg1\;.\NO
    \eeqa
    Here $n$ is the flavour index of the intermediate heavy (s)neutrino.
    This result agrees with the one in ref.~\cite{covi} and is
    of the same order as the $CP$ asymmetry in ref.~\cite{camp}. 
    \pagebreak

    With $\ve_j$ we can parametrize the reaction densities for the
    decays and inverse decays in the following way
    \beqx
      \begin{array}[b]{r@{=}l}
      \g\Big(N_j\to\wt{l}+\Bar{\wt{h}}\,\Big)
        =\g\Big(N_j\to l+H_2\Big)
        =\g\Big({\wt{l}{ }}^{\;\dg}+\wt{h}\to N_j\Big)
        =\g\Big(\Bar{l}+H_2^{\dg}\to N_j\Big)
        &{1\over4}(1+\ve_j)\gnj\\[2ex]
      \g\Big(N_j\to{\wt{l}{ }}^{\;\dg}+\wt{h}\,\Big)
        =\g\Big(N_j\to\Bar{l}+H_2^{\dg}\Big)
        =\g\Big(\wt{l}+\Bar{\wt{h}}\to N_j\,\Big)
        =\g\Big(l+H_2\to N_j\Big)&{1\over4}(1-\ve_j)\gnj\\[2ex]
      \g\Big(\snj\to\wt{l}+H_2\Big)
        =\g\Big(\snj^{\,\dg}\to l+\Bar{\wt{h}}\,\Big)
        =\g\Big({\wt{l}{ }}^{\;\dg}+H_2^{\dg}\to\snj^{\,\dg}\Big)
        =\g\Big(\Bar{l}+\wt{h}\to\snj\Big)
        &{1\over2}(1+\ve_j)\gsnj\\[2ex]
      \g\Big(\snj^{\dg}\to{\wt{l}{ }}^{\;\dg}+H_2^{\dg}\Big)
        =\g\Big(\snj\to \Bar{l}+\wt{h}\,\Big)
        =\g\Big(\wt{l}+H_2\to\snj\Big)
        =\g\Big(l+\Bar{\wt{h}}\to\snj^{\,\dg}\Big)
        &{1\over2}(1-\ve_j)\gsnj
      \end{array}
    \eeqx
        
    \begin{figure}[t]
      \input{Fig02.tex}
      \caption{\it Contributions of the scalar potential to the decay
        width and the  interactions of a scalar neutrino.\label{fig02}}
    \end{figure}
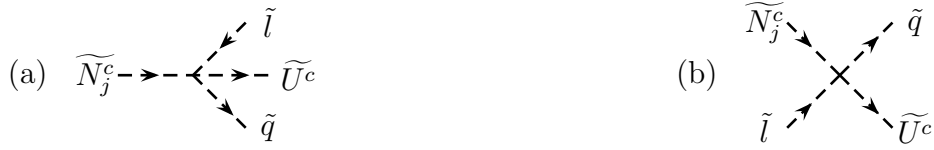    
    \begin{figure}[t]
      \input{Fig03.tex}
      \caption{\it $L$ violating processes mediated by a virtual
        Majorana neutrino or its scalar partner. \label{fig03}}
    \end{figure}
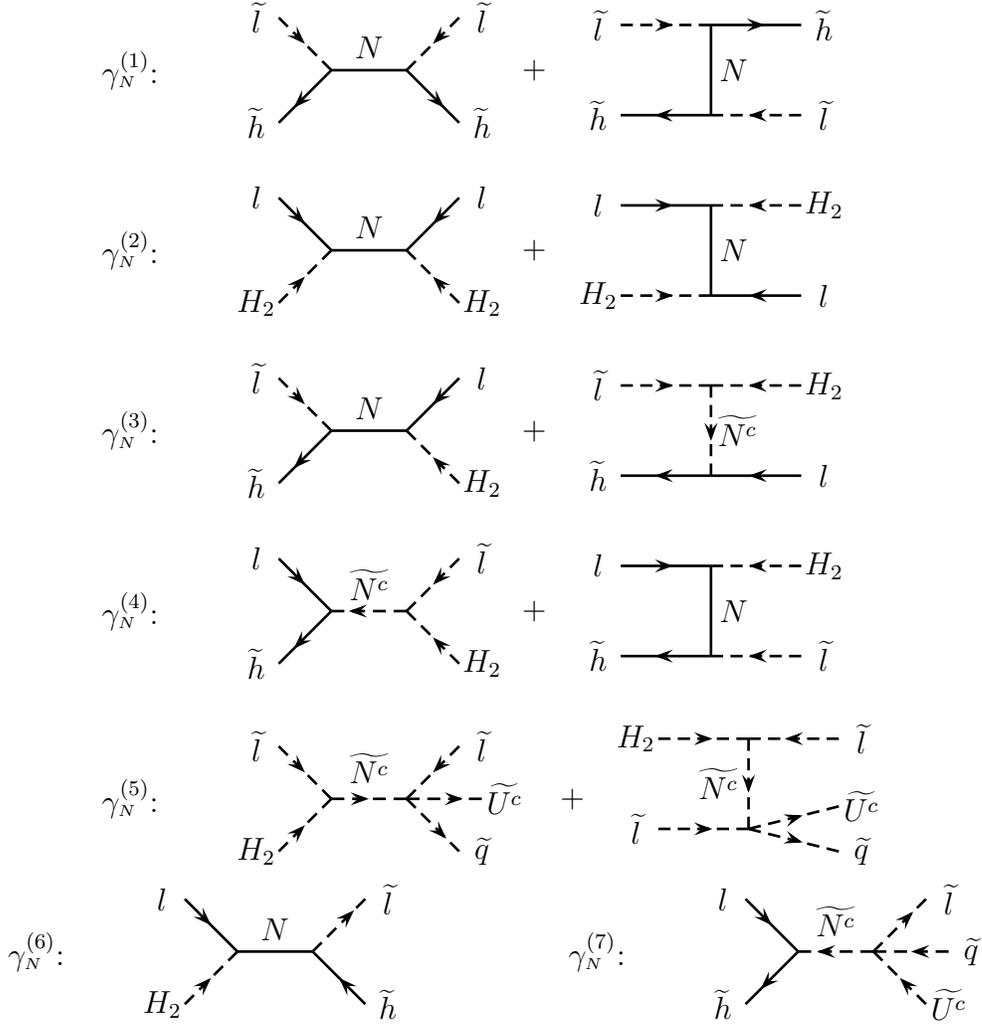
    Additionally, the scalar potential contains quartic scalar
    couplings, which enable the decay of $\snj$ into three particles
    via the diagram shown in fig.~\ref{fig02}a. The partial width
    for this decay is given by
    \beq
      \Gtr:=\G\Big(\snj^{\;\dg}\to \wt{l}+\sur^{\dg}
        +\wt{q}^{\dg}\;\Big)={3\,\a_uM_j\over64\p^2}\;
        {\mmjj\over v_2^2}\quad\mbox{with}\quad
        \a_u={\mbox{Tr}\Big(\l_u^{\dg}\l_u\Big)\over4\p}\;,
    \eeq
    and the corresponding reaction density reads
    \beq
      \gtr= {3\,\a_uM_1^4\over128\p^4}\;{\mmjj\over v_2^2}\;
        {a_j\sqrt{a_j}\over z}\mbox{K}_1(z\sqrt{a_j})
        ={3\,\a_u\over16\p}\,\gsnj\;.
    \eeq
    Since the Yukawa coupling of the top quark and its scalar partner
    is large, $\a_u$ can be of order one. But even then
    $\gtr$ is much smaller than $\gsnj$. Hence, the
    three particle decays give only a small correction, which we have
    taken into account for completeness. However, we have neglected
    the $CP$ asymmetry in this decay which comes about through similar
    one-loop diagrams as in fig.~\ref{fig01}.

    The dimensionless squared total decay widths of $N_j$ and $\snj$
    are then finally given by
    \beqa
      c_j&:=&\left({\Gnj\over M_1}\right)^2
        ={a_j\over16\p^2}\;{\mmjj^2\over v_2^4}\;,\\[1ex]
      \wt{c_j}&:=&\left({\Gsnj+\Gtr\over M_1}\right)^2
        ={a_j\over16\p^2}\;{\mmjj^2\over v_2^4}\;
        \left[1+{3\,\a_u\over16\p}\right]^2\;.
    \eeqa

    The vertex in fig.~\ref{fig02}a also gives $2\to2$ scattering
    processes involving one scalar neutrino, like
    $\snj+\wt{l}\to\wt{q}+\sur$ (cf.~fig.~\ref{fig02}b). The
    reduced cross section for this process reads
    \beq
       \hat{\s}_{22_j}(x)=3\a_u\;{\mmjj\over v_2^2}\;{x-a_j\over x}\;.
    \eeq
    For the processes $\snj+\wt{q}^{\dg}\to{\wt{l}{ }}^{\,\dg}+\sur$
    and $\snj+\sur^{\dg}\to{\wt{l}{ }}^{\,\dg}+\wt{q}$, the
    corresponding back reactions and the $CP$ conjugated processes we
    find the same result. The corresponding reaction density can then
    be calculated according to eq.~(\ref{22scatt}). One finds
    \beq
      \g_{22_j}(z)={3\,\a_uM_1^4\over16\p^4}\;{\mmjj\over v_2^2}\;
      {\sqrt{a_j}\over z^3}
      \,\mbox{K}_1(z\sqrt{a_j})={3\,\a_u\over4\p\,a_jz^2}\;\gnj(z)\;.
    \eeq
    Hence, $\g_{22_j}$ will be much larger than $\gnj$ for small
    $a_jz^2$, i.e.\ for high temperatures $T\gg M_j$. Together with
    similar scatterings which we are going to discuss in section
    \ref{Ntopsection}, these processes will therefore be very effective
    in bringing the heavy (s)neutrinos into thermal equilibrium at
    high temperatures where decays and inverse decays are suppressed
    by a time dilatation factor.

\subsection{Lepton number violating scatterings mediated by the
    right-handed neutrinos}
    Using the tree level vertices from figs.~\ref{fig01} and
    \ref{fig02} as building blocks we can construct lepton number
    violating scatterings mediated by a heavy (s)neutrino. Although of
    higher order than the tree level decays, these diagrams have to be
    taken into consideration to avoid the generation of an asymmetry in
    thermal equilibrium (cf.~ref.~\cite{kw}). In this section we will
    only mention the different processes which have to be considered.
    The corresponding reduced cross sections can be found in
    appendix~\ref{appC} and the reaction densities are discussed in 
    appendix~\ref{appD}.

    By combining two of the decay vertices (cf.~fig.~\ref{fig01} and
    fig.~\ref{fig02}a) one gets the processes that we have shown in
    fig.~\ref{fig03} and the corresponding $CP$ conjugated processes.
    We will use the following abbreviations for the reaction densities
    \beqx
      \begin{array}{l@{\qquad\qquad}l}
        \g_{\scr N}^{(1)}=\g\Big(\wt{l}+\Bar{\wt{h}}\leftrightarrow 
          {\wt{l}{}}^{\;\dg}+\wt{h}\Big)\;, &
        \g_{\scr N}^{(2)}=\g\Big(l+H_2\leftrightarrow
          \Bar{l}+H_2^{\dg}\Big)\;,\\[1ex]
        \g_{\scr N}^{(3)}=\g\Big(\wt{l}+\Bar{\wt{h}}\leftrightarrow
          \Bar{l}+H_2^{\dg}\Big)\;, &
        \g_{\scr N}^{(4)}=\g\Big(l+\Bar{\wt{h}}\leftrightarrow
          {\wt{l}{}}^{\;\dg}+H_2^{\dg}\Big)\;,\\[1ex]
        \g_{\scr N}^{(5)}=\g\Big(\wt{l}+H_2\leftrightarrow{\wt{l}{}}^{\;\dg}
          +\sur+\wt{q}\Big)\;, &
        \g_{\scr N}^{(6)}=\g\Big(l+H_2\leftrightarrow\wt{l}+\Bar{\wt{h}}\;
          \Big)\;,\\[1ex]
        \g_{\scr N}^{(7)}=\g\Big(l+\Bar{\wt{h}}\leftrightarrow
          \wt{l}+\wt{q}^{\,\dg}+\sur^{\dg}\Big)\;. &
      \end{array}
    \eeqx
    The contributions from on-shell (s)neutrinos contained in these
    reactions have already been taken into account as inverse decay
    followed by a decay. Hence, one has to subtract the contributions
    from real intermediate states to avoid a double counting of
    reactions \cite{kw}.  

    {}From the scattering vertex in fig.~\ref{fig02}b and the decay
    vertices we can construct the following processes
    \begin{figure}[t]
      \input{Fig04.tex}
      \caption{\it Diagrams contributing to the lepton number
        violating scatterings via heavy sneutrino exchange.
        \label{fig04}}
    \end{figure}
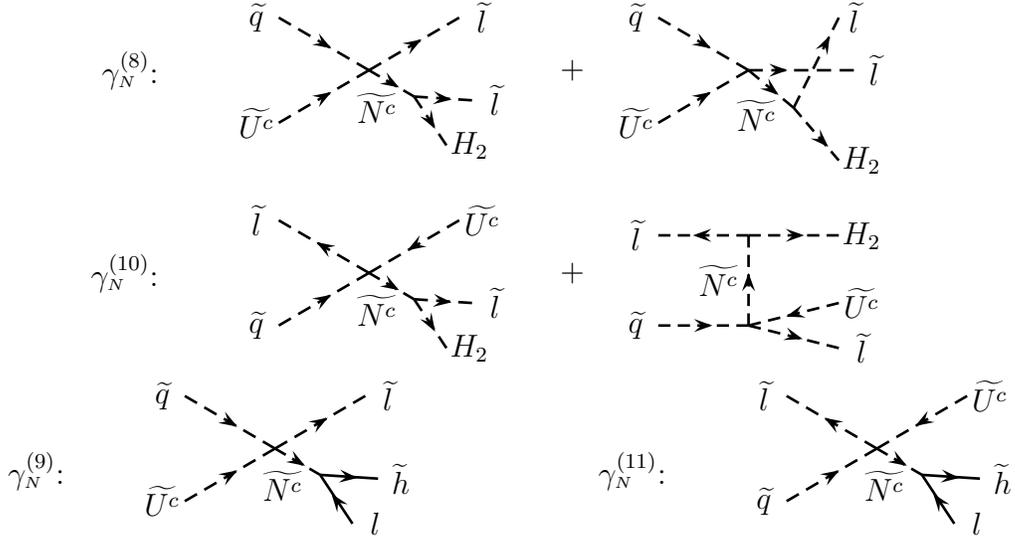
    \beqa
       \g_{\scr N}^{(8)}&=&\g\Big(\sur+\wt{q}\leftrightarrow\wt{l}+\wt{l}
         +H_2\Big)\;,\qquad\qquad
         \g_{\scr N}^{(9)}=\g\Big(\wt{q}+\sur\leftrightarrow\wt{l}+
         \Bar{l}+\wt{h}\Big)\;,\NO\\[1ex]
       \g_{\scr N}^{(10)}&=&\g\Big({\wt{l}{}}^{\;\dg}+\wt{q}\leftrightarrow
         \wt{l}+\sur^{\dg}+H_2\Big)
         =\g\Big({\wt{l}{}}^{\;\dg}+\sur\leftrightarrow
         \wt{l}+\wt{q}^{\,\dg}+H_2\Big)\;,\NO\\[1ex]
       \g_{\scr N}^{(11)}&=&\g\Big({\wt{l}{}}^{\;\dg}+\wt{q}\leftrightarrow
         \Bar{l}+\wt{h}+\sur^{\dg}\Big)
         =\g\Big({\wt{l}{}}^{\;\dg}+\sur\leftrightarrow
         \Bar{l}+\wt{h}+\wt{q}^{\,\dg}\Big)\;.\NO
     \eeqa
     In fig.~\ref{fig04} we have shown one typical diagram for each of
     these reaction densities. Again, these diagrams have on-shell
     contributions which have to be subtracted, since they can be
     described as decay of a sneutrino which has been produced in a
     scattering process.

     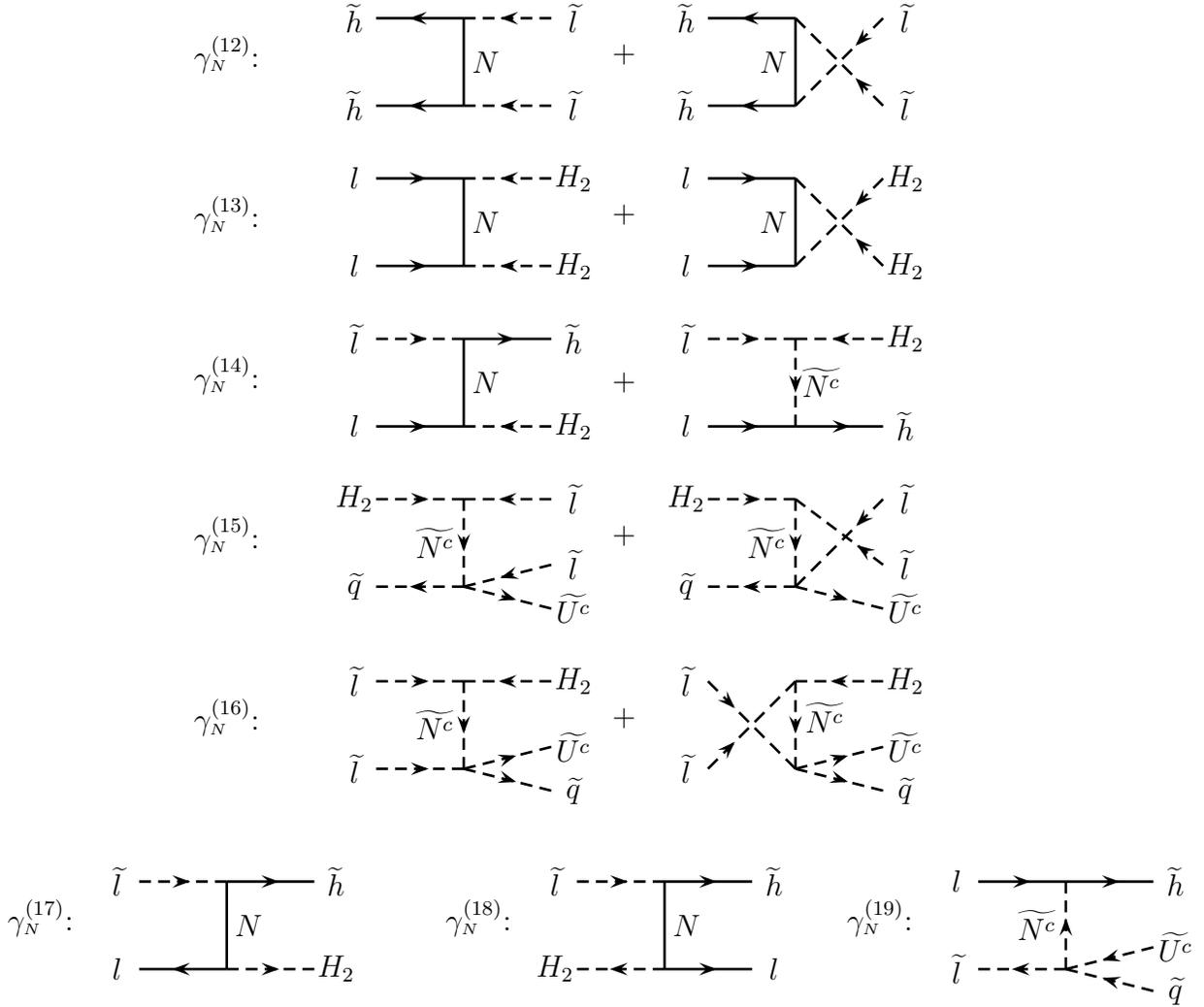
\begin{figure}[ht]
      \input{Fig05.tex}
      \caption{\it $L$ violating processes mediated by a virtual
        Majorana neutrino in the $t$-channel. \label{fig05}}
    \end{figure}
     Up to now we have only considered processes with a neutrino
     or its scalar partner in the $s$-channel. In fig.~\ref{fig05} we
     have shown a selection of diagrams without on-shell
     contributions. The corresponding reaction densities will be
     denoted by
    \beqa
      \g_{\scr N}^{(12)}&=&\g\Big(\Bar{\wt{h}}+\Bar{\wt{h}}\leftrightarrow
         {\wt{l}{}}^{\;\dg}+{\wt{l}{}}^{\;\dg}\;\Big)\;,\qquad\qquad
        \g_{\scr N}^{(13)}=\g\Big(l+l\leftrightarrow H_2^{\dg}+H_2^{\dg}
        \Big)\;,\NO\\[1ex]
      \g_{\scr N}^{(14)}&=&\g\Big(\wt{l}+l\leftrightarrow\wt{h}
        +H_2^{\dg}\Big)\;,\qquad\qquad
        \g_{\scr N}^{(16)}=\g\Big(\wt{l}+\wt{l}\leftrightarrow\sur+\wt{q}
        +H_2^{\dg}\Big)\;,\NO\\[1ex]
      \g_{\scr N}^{(15)}&=&\g\Big(H_2+\wt{q}^{\,\dg}\leftrightarrow
        {\wt{l}{}}^{\;\dg}+{\wt{l}{}}^{\;\dg}+\sur\Big)
        =\g\Big(H_2+\sur^{\dg}\leftrightarrow{\wt{l}{}}^{\;\dg}
        +{\wt{l}{}}^{\;\dg}+\wt{q}\Big)\;,\NO\\[1ex]
      \g_{\scr N}^{(17)}&=&\g\Big(\wt{l}+\Bar{l}\leftrightarrow
        \wt{h}+H_2\Big)\;,\qquad\qquad
      \g_{\scr N}^{(18)}=\g\Big(\wt{l}+H_2^{\dg}\leftrightarrow
        \wt{h}+l\Big)\;,\NO\\[1ex]
      \g_{\scr N}^{(19)}&=&\g\Big(l+{\wt{l}{}}^{\;\dg}\leftrightarrow
        \wt{h}+\wt{q}^{\,\dg}+\sur^{\dg}\Big)
      =\g\Big(l+\wt{q}\leftrightarrow\wt{l}+\sur^{\dg}+\wt{h}
        \Big)\NO\\[1ex]
      &=&\g\Big(l+\sur\leftrightarrow\wt{l}+\wt{q}^{\,\dg}+\wt{h}
        \Big)
      =\g\Big({\wt{l}{}}^{\;\dg}+\Bar{\wt{h}}\leftrightarrow
        \Bar{l}+\wt{q}^{\,\dg}+\sur^{\dg}\Big)\NO\\[1ex]
      &=&\g\Big(\wt{q}+\Bar{\wt{h}}\leftrightarrow\Bar{l}+\wt{l}+
        \sur^{\dg}\Big)
      =\g\Big(\sur+\Bar{\wt{h}}\leftrightarrow\Bar{l}+\wt{l}+
        \wt{q}^{\,\dg}\Big)\;.\NO
    \eeqa
    At first sight one may think that these diagrams could be neglected,
    since they are suppressed at intermediate temperatures, i.e.\
    intermediate energies $x\approx a_j$. However, they give an
    important contribution to the effective lepton number violating
    interactions at low energies and therefore have to be taken into
    consideration.
\subsection{Interactions with a top or a stop \label{Ntopsection}}
    The Yukawa coupling of the top quark is large. Thus we have to
    consider the lepton number violating interactions of a
    right-handed neutrino with a top quark or its scalar partner. 
    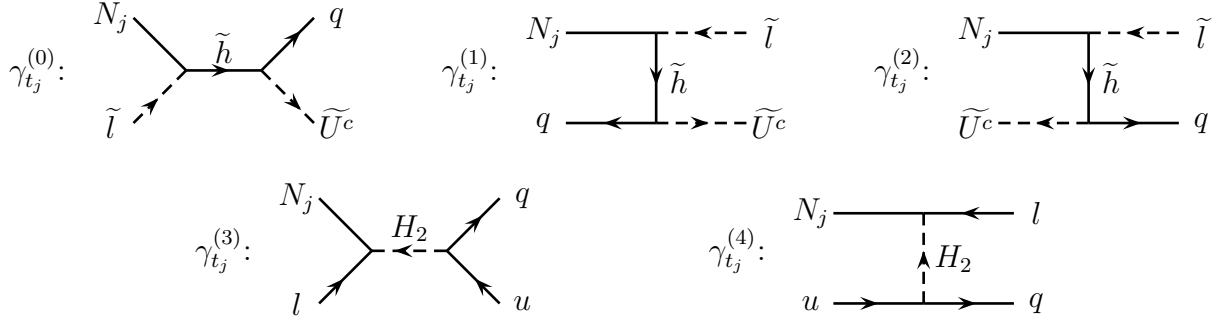
\begin{figure}[ht]
      \input{Fig06.tex}
      \caption{\it Neutrino-(s)top scattering. \label{Ntop}}
    \end{figure}

    \noindent 
    For the neutrino we have to take into account the following
    processes (cf.~fig.~\ref{Ntop})
    \beqa
      \g_{t_j}^{(0)}&=&\g\Big(N_j+\wt{l}\leftrightarrow q+\sur\Big)
        =\g\Big(N_j+\wt{l}\leftrightarrow\wt{q}+\Bar{u}\Big)
        \;,\NO\\[1ex]
      \g_{t_j}^{(1)}&=&\g\Big(N_j+\Bar{q}\leftrightarrow{\wt{l}{}}^{\;\dg}
        +\sur\Big)=\g\Big(N_j+u\leftrightarrow{\wt{l}{}}^{\;\dg}
        +\wt{q}\Big)\;,\NO\\[1ex]
      \g_{t_j}^{(2)}&=&\g\Big(N_j+\sur^{\dg}\leftrightarrow
        {\wt{l}{}}^{\;\dg}+q\Big)=\g\Big(N_j+\wt{q}^{\,\dg}
        \leftrightarrow{\wt{l}{}}^{\;\dg}+\Bar{u}\Big)\;,\NO \\[1ex]
      \g_{t_j}^{(3)}&=&\g\Big(N_j+l\leftrightarrow q+\Bar{u}\Big)
        \;,\NO\\[1ex]
      \g_{t_j}^{(4)}&=&\g\Big(N_j+u\leftrightarrow\Bar{l}+q\Big)=
        \g\Big(N_j+\Bar{q}\leftrightarrow\Bar{l}+\Bar{u}\Big)\;.\NO
    \eeqa
    At this order of perturbation theory these processes are $CP$
    invariant. Hence, we have the same reaction densities for the $CP$
    conjugated processes.

    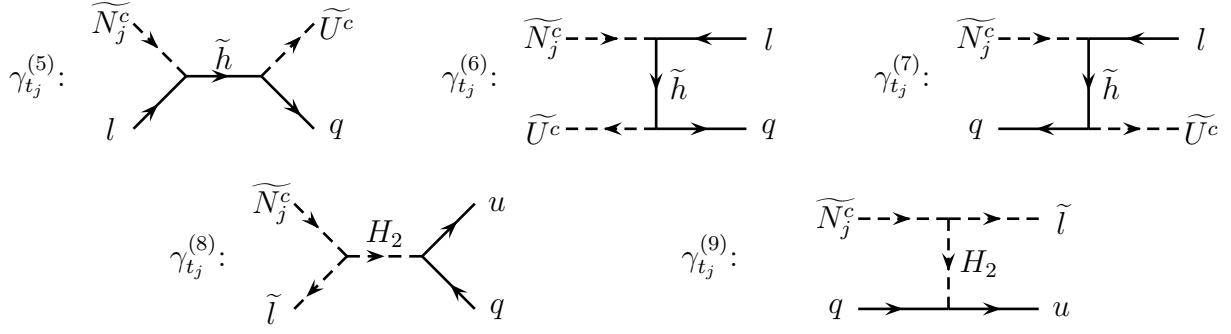
\begin{figure}[ht]
      \input{Fig07.tex}
      \caption{\it Sneutrino-(s)top scattering. \label{Nttop}}
    \end{figure}
    For the scalar neutrinos we have similarly (cf.~fig.~\ref{Nttop})
    \beqa
      \g_{t_j}^{(5)}&=&\g\Big(\snj+l\leftrightarrow q+\sur\Big)
        =\g\Big(\snj+l\leftrightarrow\wt{q}+\Bar{u}\Big)
        \;,\NO\\[1ex]
      \g_{t_j}^{(6)}&=&\g\Big(\snj+\sur^{\dg}\leftrightarrow\Bar{l}+q
        \Big)=\g\Big(\snj+\wt{q}^{\,\dg}\leftrightarrow\Bar{l}
        +\Bar{u}\Big)\;,\NO\\[1ex]
      \g_{t_j}^{(7)}&=&\g\Big(\snj+\Bar{q}\leftrightarrow\Bar{l}+\sur\Big)
        =\g\Big(\snj+u\leftrightarrow\Bar{l}+\wt{q}\Big)\;,\NO \\[1ex]
      \g_{t_j}^{(8)}&=&\g\Big(\snj+{\wt{l}{}}^{\;\dg}\leftrightarrow
        \Bar{q}+u\Big)\;,\NO\\[1ex]
      \g_{t_j}^{(9)}&=&\g\Big(\snj+q\leftrightarrow\wt{l}+u\Big)=
        \g\Big(\snj+\Bar{u}\leftrightarrow\wt{l}+\Bar{q}\Big)\;.\NO
    \eeqa
    The quartic scalar couplings of the sneutrinos give additional 
    $2\to3$, $3\to3$ and $2\to4$ processes, which can be neglected
    since they are phase space suppressed. 
    \begin{figure}[ht]
      \input{Fig08.tex}
      \caption{\it Neutrino pair annihilation. \label{NN}}
    \end{figure}
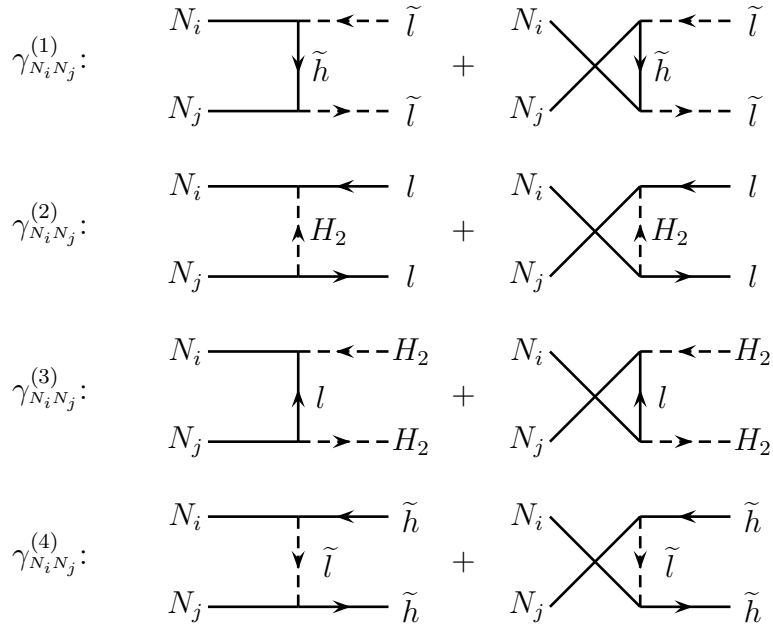
\subsection{Neutrino pair creation and annihilation}
    The Yukawa couplings of the right-handed neutrinos also allow
    lepton number conserving processes like the neutrino pair creation
    and annihilation. 

    For the neutrinos we have the processes depicted in fig.~\ref{NN},
    \beqx
      \begin{array}{l@{\qquad\qquad}l}
        \g_{\scr N_iN_j} ^{(1)}=\g\Big(N_i+N_j\leftrightarrow\wt{l}+
          {\wt{l}{}}^{\;\dg}\;\Big)\;,&
          \g_{\scr N_iN_j} ^{(2)}=\g\Big(N_i+N_j\leftrightarrow l+\Bar{l}
          \;\Big)\;,\\[1ex]
        \g_{\scr N_iN_j} ^{(3)}=\g\Big(N_i+N_j\leftrightarrow H_2+H_2^{\dg}
          \Big)\;,&
          \g_{\scr N_iN_j} ^{(4)}=\g\Big(N_i+N_j\leftrightarrow\wt{h}
          +\Bar{\wt{h}}\;\Big)\;.
      \end{array}
    \eeqx
    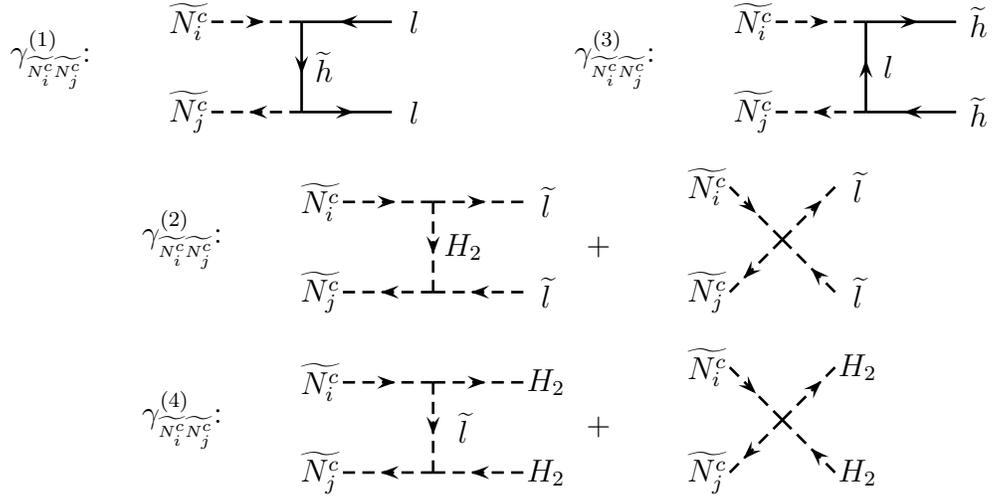
\begin{figure}[ht]
      \input{Fig09.tex}
      \caption{\it Sneutrino pair annihilation. \label{NtNt}}
    \end{figure}

    For the scalar neutrinos we have similar diagrams and additional
    contributions from the scalar potential (cf.~fig.~\ref{NtNt}). We
    have the following reaction densities
    \beqx
      \begin{array}{l@{\qquad\qquad}l}
        \g_{\scr\sni\snj} ^{(1)}=\g\Big(\sni+\snj^{\,\dg}
          \leftrightarrow l+\Bar{l}\;\Big)\;,&
          \g_{\scr\sni\snj} ^{(2)}=\g\Big(\sni+\snj^{\,\dg}
          \leftrightarrow\wt{l}+{\wt{l}{}}^{\;\dg}\;\Big)\;,\\[1ex]
        \g_{\scr\sni\snj} ^{(3)}=\g\Big(\sni+\snj^{\,\dg}
          \leftrightarrow\wt{h}+\Bar{\wt{h}}\;\Big)\;,&
          \g_{\scr\sni\snj} ^{(4)}=\g\Big(\sni+\snj^{\,\dg}
          \leftrightarrow H_2+H_2^{\dg}\Big)\;.
      \end{array}
    \eeqx
    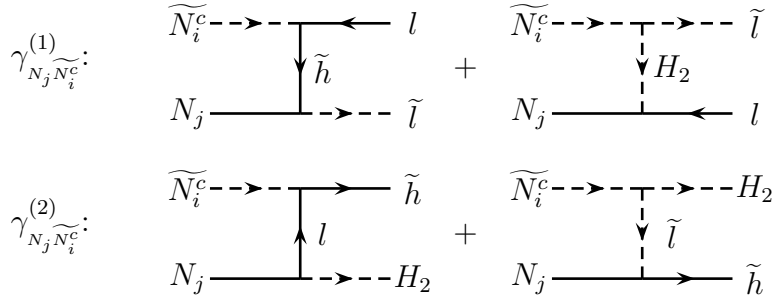
\begin{figure}[ht]
      \input{Fig10.tex}
      \caption{\it Neutrino-sneutrino scattering. \label{NNt}}
    \end{figure}

    Finally, there are neutrino-sneutrino scattering processes 
    (cf.~fig.~\ref{NNt}),
    \beqx
      \g_{\scr N_j\sni}^{(1)}=\g\Big(\sni+N_j\leftrightarrow 
        \Bar{l}+\wt{l}\;\Big)\;,\qquad\qquad
      \g_{\scr N_j\sni}^{(2)}=\g\Big(\sni+N_j\leftrightarrow
        \wt{h}+H_2\Big)\;.
    \eeqx
    Such diagrams also give neutrino-sneutrino transitions like
    $\sni+l\leftrightarrow N_j+\wt{l}$. These processes transform
    neutrinos into sneutrinos and leptons into sleptons, i.e.\ they
    tend to balance out the number densities of the fermions and their
    supersymmetric partners, but they cannot wash out any generated
    asymmetry. As we will see in the next chapter, the number
    densities of the neutrinos and the scalar neutrinos are already
    equal without taking into account these interactions, while the
    equality of the number densities of leptons and sleptons
    is ensured by MSSM-processes (cf.~section~\ref{MSSMsect}).
    Finally, the dominant contributions to these neutrino-sneutrino
    transitions come from inverse decays, decays and scatterings off a
    (s)top which we have already considered. Hence, we can neglect
    these additional processes.
\subsection{MSSM processes \label{MSSMsect}}
    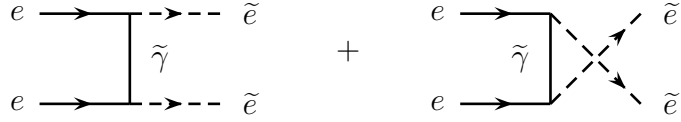
\begin{figure}[t]
      \input{Fig11.tex}
      \caption{\it Example of a $L_f$ and $L_s$ violating MSSM 
        process. \label{MSSMproc}}
    \end{figure}
    In the MSSM the fermionic lepton number $L_f$ and the lepton
    number stored in the scalar leptons $L_s$ are not separately
    conserved. There are processes transforming leptons into scalar
    leptons and vice versa. As an example we have considered the
    process $e+e\leftrightarrow\wt{e}+\wt{e}$
    (cf.~fig.~\ref{MSSMproc}). For large temperatures, i.e.\ $s\gg
    m_{\wt{\g}}^2$, the reduced cross section for this process is
    given by \cite{keung}
    \beq
      \hat{\s}_{\mbox{\tiny MSSM}}\approx 4\p\a^2\left[
      \ln\left(s\over m_{\wt{\g}}^2\right)-4\right]\;.
    \eeq
    This translates into the following reaction density
    \beq
       \g_{\mbox{\tiny MSSM}}\approx {M_1^4\,\a^2\over4\p^3}\,{1\over z^4}
       \left[\ln\left({4\over z^2a_{\wt{\g}}}\right)
       -2\g_{\mbox{\tiny E}}-3\right]\;,
    \eeq
    where we have introduced the dimensionless squared photino mass
    \beq
      a_{\wt{\g}} := \left({m_{\wt{\g}}\over M_1}\right)^2\;.
    \eeq
    These processes are in thermal equilibrium if the reaction
    rates are larger than the Hubble parameter $H$. This condition
    gives a very weak upper bound on the photino mass,
    \beq
      m_{\wt{\g}}\;\ltap\;2.5\times10^{9}\;\mbox{GeV}\;
        \left({T\over10^{10}\;\mbox{GeV}}\right)\;
        \exp\left[{-{1\over412}\left({T\over10^{10}\;\mbox{GeV}}
        \right)}\right]\;.
    \eeq
    In the calculations we assume $m_{\wt{\g}}=100\;$GeV.

\section{Results \label{results}}
\subsection{The Boltzmann equations \label{boltzeq}}
    Now that we have identified all the relevant processes we can
    write down the network of Boltzmann equations which governs the
    time evolution of the neutrino and sneutrino number densities and
    of the lepton asymmetry. For the scalar neutrinos and their
    antiparticles it is convenient to use the sum and the difference
    of the particle numbers per comoving volume element as independent
    variables,
    \beq
      Y_{j\pm} := Y_{\scr\snj}\pm Y_{\scr\snj^{\dg}}\;.
    \eeq
    Furthermore, we have to discern the lepton asymmetry stored in the
    Standard Model particles $\YL$ and the asymmetry $\YLt$ in
    the scalar leptons.

    For the neutrinos $N_j$ one has 
    \beqa
      \lefteqn{{\mbox{d}\Ynj\over\mbox{d}z}={-z\over sH(M_1)}\left\{
        \left(\Ynjratio-1\right)\left[\gnj+4\g_{t_j}^{(0)}+4\g_{t_j}^{(1)}
        +4\g_{t_j}^{(2)}+2\g_{t_j}^{(3)}+4\g_{t_j}^{(4)}\right]\right.}
        \\[1ex]
      &&\left.\qquad\qquad+\sum\limits_i\left[\left(\Ynjratio\Yniratio
        -1\right)\sum\limits_{k=1}^4\g_{\scr N_iN_j}^{(k)}
        +\left(\Ynjratio\Ypiratio-2\right)
        \sum\limits_{k=1}^2\g_{\scr N_j\sni}^{(k)}
        \right]\right\}\NO\;.
    \eeqa
    Correspondingly the Boltzmann equations for the scalar neutrinos
    read 
    \beqa
      \lefteqn{{\mbox{d}\Yp\over\mbox{d}z}={-z\over sH(M_1)}\left\{
        \left(\Ypratio-2\right)\left(\gsnj+\gtr+3\g_{22_j}+2\g_{t_j}^{(5)}
        +2\g_{t_j}^{(6)}+2\g_{t_j}^{(7)}+\g_{t_j}^{(8)}+2\g_{t_j}^{(9)}
        \right)\right.}\NO\\[1ex]
      &&\qquad\qquad{}+{1\over2}\Ymratio\YLtratio\left(\g_{22_j}
        -\g_{t_j}^{(8)}\right)+\Ymratio\YLratio\g_{t_j}^{(5)}\\[1ex]
      &&\left.\qquad\qquad+\sum\limits_i
        \left[{1\over2}\left(\Ypratio\Ypiratio-\Ymratio\Ymiratio-4\right)
        \sum\limits_{k=1}^4\g_{\scr\sni\snj}^{(k)}
        +\left(\Ypratio\Yniratio-2\right)\sum\limits_{k=1}^2
        \g_{\scr N_i\snj}^{(k)}\right]
        \right\}\;,\NO\\[3ex]
      \lefteqn{{\mbox{d}\Ym\over\mbox{d}z}={-z\over sH(M_1)}\left\{
        \Ymratio\left(\gsnj+\gtr+3\g_{22_j}+2\g_{t_j}^{(5)}+2\g_{t_j}^{(6)}
        +2\g_{t_j}^{(7)}+\g_{t_j}^{(8)}+2\g_{t_j}^{(9)}\right)
        \right.}\NO\\[1ex]
      &&\qquad\qquad{}+\YLtratio\left[\gtr-{1\over2}\gsnj-2\g_{t_j}^{(9)}
        -{1\over2}\Ypratio\g_{t_j}^{(8)}+\left(2+{1\over2}\Ypratio\right)
        \g_{22_j}\right]\\[1ex]
      &&\qquad\qquad{}+\YLratio\left[{1\over2}\gsnj+2\left(\g_{t_j}^{(6)}
        +\g_{t_j}^{(7)}\right)+\Ypratio\g_{t_j}^{(5)}\right]\NO\\[1ex]
      &&\left.{}+\sum\limits_i\left[{1\over2}\left(\Ymratio\Ypiratio
        -\Ypratio\Ymiratio\right)\sum\limits_{k=1}^4\g_{\scr\sni\snj}^{(k)}
        +\Ymratio\Yniratio\sum\limits_{k=1}^2\g_{\scr N_i\snj}^{(k)}
        +\left(\YLratio-\YLtratio\right)\g_{\scr N_i\snj}^{(1)}\right]
        \right\}\NO\;.
    \eeqa
    The Boltzmann equations for the lepton asymmetries are given by
    \beqa
      \lefteqn{{\mbox{d}\YL\over\mbox{d}z}={-z\over sH(M_1)}\left\{
        \sum\limits_j\left[\left({1\over2}\YLratio+\ve_j\right)
        \left({1\over2}\gnj+\gsnj\right)-{1\over2}\ve_j\left(
        \Ynjratio\gnj+\Ypratio\gsnj\right)+{1\over2}\Ymratio\gsnj
        \right]\right.}\NO\\[1ex]
      &&\qquad{}+\YLratio\left(\g_{\scr A}^{\scr\D L}
        +\g_{\scr C}^{\scr\D L}\right)+\YLtratio\left(
        \g_{\scr B}^{\scr\D L}-\g_{\scr C}^{\scr\D L}\right)+\left(\YLratio
        -\YLtratio\right)\g_{\mbox{\tiny MSSM}}\\[1ex]
      &&\qquad{}\!+\sum\limits_j\left[\YLratio\left(\Ynjratio\g_{t_j}^{(3)}
        +\Ypratio\g_{t_j}^{(5)}+2\g_{t_j}^{(4)}+2\g_{t_j}^{(6)}
        +2\g_{t_j}^{(7)}\right)+2\Ymratio\left(\g_{t_j}^{(5)}
        +\g_{t_j}^{(6)}+\g_{t_j}^{(7)}\right)\right]\NO\\[1ex]
      &&\left.\qquad{}+\sum\limits_{i,j}\left(\YLratio-\YLtratio
        +\Ynjratio\Ymiratio\right)\g_{\scr N_j\sni}^{(1)}
        \right\}\;,\NO\\[2ex]
      \lefteqn{{\mbox{d}\YLt\over\mbox{d}z}={-z\over sH(M_1)}\left\{
        \sum\limits_j\left[\left({1\over2}\YLtratio+\ve_j\right)
        \left({1\over2}\gnj+\gsnj\right)-{1\over2}\ve_j\left(
        \Ynjratio\gnj+\Ypratio\gsnj\right)\right.\right.}\NO\\[1ex]
      &&\left.\left.\qquad{}-{1\over2}\Ymratio\gsnj+\left(\YLtratio+\Ymratio
        \right)\gtr+\left({1\over2}\Ypratio\YLtratio+2\YLtratio
        +3\Ymratio\right)\g_{22_j}\right]\right.\NO\\[1ex]
      &&\qquad{}+\YLtratio\left(\g_{\scr A}^{\scr\D L}
        +\g_{\scr D}^{\scr\D L}\right)+\YLratio\left(
        \g_{\scr B}^{\scr\D L}-\g_{\scr C}^{\scr\D L}\right)
        +\left(\YLtratio-\YLratio\right)\g_{\mbox{\tiny MSSM}}\\[1ex]
      &&\qquad{}+\sum\limits_j\left[\YLtratio\left(2\Ynjratio\g_{t_j}^{(0)}
        +{1\over2}\Ypratio\g_{t_j}^{(8)}+2\g_{t_j}^{(1)}+2\g_{t_j}^{(2)}
        +2\g_{t_j}^{(9)}\right)-\Ymratio\left(\g_{t_j}^{(8)}
        +2\g_{t_j}^{(9)}\right)\right]\NO\\[1ex]
      &&\left.\qquad{}+\sum\limits_{i,j}\left(\YLtratio-\YLratio
        -\Ynjratio\Ymiratio\right)\g_{\scr N_j\sni}^{(1)}\right\}\NO\;,
    \eeqa
    where we have introduced the following abbreviations for the
    lepton number violating scatterings mediated by a heavy
    (s)neutrino
    \beqa
      \g_{\scr A}^{\scr\D L}&=&2\g_{\scr N}^{(1)}+\g_{\scr N}^{(3)}
        +\g_{\scr N}^{(4)}+\g_{\scr N}^{(6)}+\g_{\scr N}^{(7)}
        +2\g_{\scr N}^{(12)}+\g_{\scr N}^{(14)}\;,\\[1ex]
      \g_{\scr B}^{\scr\D L}&=&\g_{\scr N}^{(3)}+\g_{\scr N}^{(4)}
        -\g_{\scr N}^{(6)}-\g_{\scr N}^{(7)}+\g_{\scr N}^{(14)}
        \;,\\[1ex]
      \g_{\scr C}^{\scr\D L}&=&3\g_{\scr N}^{(9)}+\g_{\scr N}^{(17)}
        +\g_{\scr N}^{(18)}+6\g_{\scr N}^{(19)}\;,\\[1ex]
      \g_{\scr D}^{\scr\D L}&=&4\g_{\scr N}^{(5)}+2\g_{\scr N}^{(8)}
        +8\g_{\scr N}^{(10)}+3\g_{\scr N}^{(9)}+4\g_{\scr N}^{(15)}
        +2\g_{\scr N}^{(16)}+\g_{\scr N}^{(17)}+\g_{\scr N}^{(18)}
        +6\g_{\scr N}^{(19)}\;.
    \eeqa
    The numerical factors in front of the reaction densities arise due
    to the change in quantum numbers in the corresponding scattering,
    e.g.\ processes transforming leptons into sleptons appear with a
    relative minus sign in the Boltzmann equations for $\YL$ and
    $\YLt$. Furthermore, any reaction density is multiplied by the
    number of different processes (cf.~section \ref{theory})
    contributing independently to the Boltzmann equations.
    
    This set of Boltzmann equations is valid for the most general case
    with arbitrary masses of the right-handed neutrinos. However, if
    the heavy neutrinos are mass degenerate, it is always possible to
    find a basis where the mass matrix $M$ and the Yukawa matrix
    $\l_{\n}$ are diagonal, i.e.\ no asymmetry is generated.
    Therefore, one has to assume a mass hierarchy for the right-handed
    neutrinos, which in turn implies that the lepton number violating
    processes induced by the lightest right-handed neutrino are in
    thermal equilibrium as long as the temperature is higher than the
    mass of this neutrino. Hence, the lepton asymmetries generated in
    the decays of the heavier right-handed neutrinos are washed out
    and the asymmetry that we observe today must have been generated
    by the lightest right-handed neutrino. We will assume that the
    first generation neutrino $N_1$ is the lightest.

    Hence, in the following we will always neglect the heavier
    right-handed neutrinos as free particles. However, they have to be
    taken into account as intermediate states since they give a
    substantial contribution to the effective lepton number
    violating processes at low energies.

    The fermionic part $\YL$ of the generated lepton asymmetry will be
    transformed into a $(B-L)$ asymmetry by the action of sphalerons.
    But since MSSM processes like the one in section \ref{MSSMsect}
    enforce the relation
    \beq
      \YL=\YLt\; ,
      \label{equal}
    \eeq
    the total lepton asymmetry $Y_L=\YL+\YLt$ will be proportional to 
    the baryon asymmetry \cite{sphal},
    \beq
      Y_B=-\left({8N_f+4N_H\over22N_f+13N_H}\right)Y_L\; ,
    \eeq
    where $N_f$ is the number of quark-lepton families and $N_H$ is
    the number of Higgs doublets. In our model with $N_f=3$ and
    $N_H=2$ we have
    \beq
      Y_B=-{8\over23}Y_L\; .
      \label{sphal}
    \eeq
    {}From eqs.~(\ref{number}) and (\ref{equal}) we can infer the 
    asymmetries that we have to generate,
    \beq
      \YL=\YLt=-(0.9-1.4)\cdot 10^{-10}\;.
      \label{requested}
    \eeq
    The additional anomalous global symmetries in supersymmetric
    theories at high temperatures have no influence on these
    considerations, since they are broken well before the electroweak
    phase transition \cite{ibanez}.
\subsection{The generated lepton asymmetry \label{gen}}
    \begin{figure}[t]
      \begin{minipage}[t]{8cm}
        \begin{center}\hspace{1cm}(a)\end{center}
        \vspace{-0.5cm}
        \epsfig{file=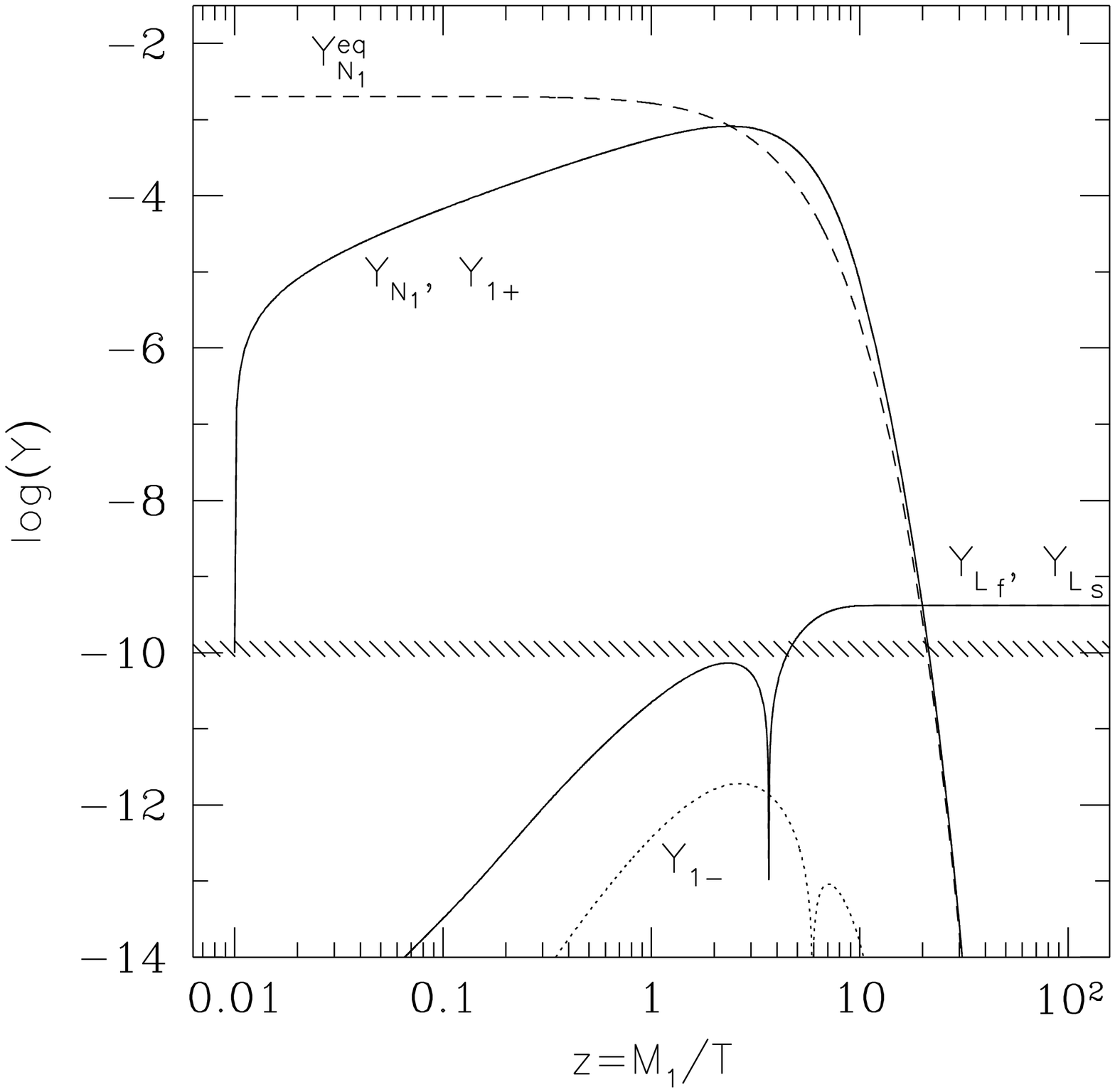,width=8cm}
      \end{minipage}
      \hspace{\fill}
      \begin{minipage}[t]{8cm}
        \begin{center}\hspace{1cm}(b)\end{center}
        \vspace{-0.5cm}
        \epsfig{file=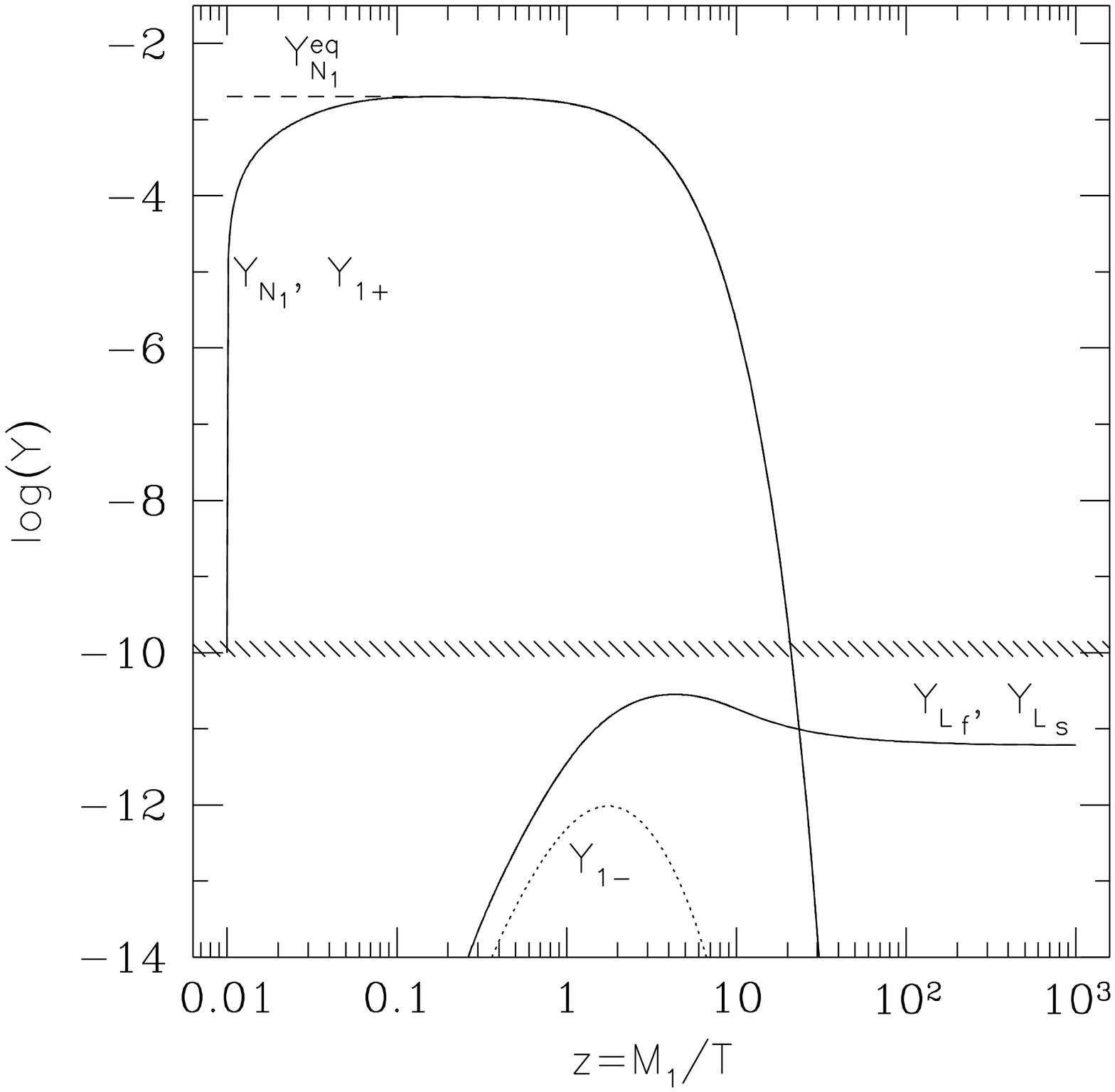,width=8cm}
     \end{minipage}  
       \caption{\it Typical solutions of the Boltzmann equations. The
        dashed line represents the equilibrium distribution for the
        neutrinos $N_1$ and the solid lines show the solutions for the
        (s)neutrino number and the absolute values of the asymmetries
        in $L_f$ and $L_s$, while the dotted line represents the
        absolute value of the scalar neutrino asymmetry $Y_{1-}$. The
        lines for $Y_{N_1}$ and $Y_{1+}$  and for the two asymmetries
        $\YL$ and $\YLt$ cannot be distinguished, since they are lying
        one upon another. The hatched area shows the measured value
        (\ref{requested}).
       \label{sol1fig}}
    \end{figure}
    Typical solutions of the Boltzmann equations are shown in
    fig.~\ref{sol1fig}, where we have assumed a neutrino mass
    $M_1=10^{10}\;$GeV and a mass hierarchy of the form 
    \beqa
      a_2=10^3\;,&\qquad&(m_{\scr D}^{\dg}m_{\scr D})_{22}
         =a_2\,\left(m_{\scr D}^{\dg}m_{\scr D}\right)_{11}\;,\\[1ex]
      a_3=10^6\;,&\qquad&(m_{\scr D}^{\dg}m_{\scr D})_{33}
         =a_3\,\left(m_{\scr D}^{\dg}m_{\scr D}\right)_{11}\;.
    \eeqa
    Furthermore we have assumed a $CP$ asymmetry $\ve_1=-10^{-6}$. The
    only difference between both figures lies in the choice of 
    $\left(m_{\scr D}^{\dg}m_{\scr D}\right)_{11}$:
    \beq
      \wt{m}_1:={(m_{\scr D}^{\dg}m_{\scr D})_{11}
        \over M_1}=\left\{
        \begin{array}{rl}
          10^{-4}\;\mbox{eV}&\mbox{for fig.~\ref{sol1fig}a,}\\[1ex]
          10^{-2}\;\mbox{eV}&\mbox{for fig.~\ref{sol1fig}b.}
        \end{array}\right.
    \eeq
    Finally, as starting condition we have assumed that all the number
    densities vanish at high temperatures $T\gg M_1$, including the
    neutrino numbers $Y_{\scr N_1}$ and $Y_{1+}$. As one can see, the
    Yukawa interactions are strong enough to create a substantial
    number of neutrinos and scalar neutrinos in fig.~\ref{sol1fig}a,
    even if $Y_{\scr N_1}$ and $Y_{1+}$ do not reach their equilibrium
    values as long as $z<1$. However, the generated asymmetries 
    \beq
      \YL=\YLt=-4\cdot10^{-10}
      \label{sol1}
    \eeq
    are of the requested magnitude. On the other hand, in
    fig.~\ref{sol1fig}b the Yukawa interactions are much stronger,
    i.e.~the neutrinos are driven into equilibrium rapidly at high
    temperatures. However, the large Yukawa couplings also increase
    the reaction rates for the lepton number violating processes which
    can wash out a generated asymmetry, i.e.\ the final asymmetries
    are much smaller than in the previous case,
    \beq
      \YL=\YLt=-6\cdot10^{-12}\;.
      \label{sol2}
    \eeq
    In both cases a small scalar neutrino asymmetry $Y_{1-}$ is
    temporarily generated. However, $Y_{1-}$ is very small and has
    virtually no influence on the generated lepton asymmetries.

    Usually it is assumed that one has a thermal population of
    right-handed neutrinos at high temperatures which decay at very
    low temperatures $T\ll M_1$ where one can neglect lepton number
    violating scatterings. Then the generated lepton asymmetry is
    proportional to the $CP$ asymmetry and the number of decaying
    neutrinos and sneutrinos \cite{kt1},
    \beq
      Y_L\;\approx\;\ve_1\,\left[\;Y_{N_1}^{\rm eq}(T\gg M_1)
        +Y_{1+}^{\rm eq}(T\gg M_1)\;\right]\approx{\ve_1\over250}\;.
    \eeq
    With $\ve_1=-10^{-6}$ this gives
    \beq
      \YL=\YLt\approx-2\cdot10^{-9}\;.
    \eeq
    By comparison with eqs.~(\ref{sol1}) and (\ref{sol2}) one sees
    that by assuming a thermal population of heavy neutrinos at high
    temperatures and neglecting the lepton number violating
    scatterings one largely overestimates the generated lepton
    asymmetries. 

\begin{figure}[t]
        \mbox{ }\hfill
        \epsfig{file=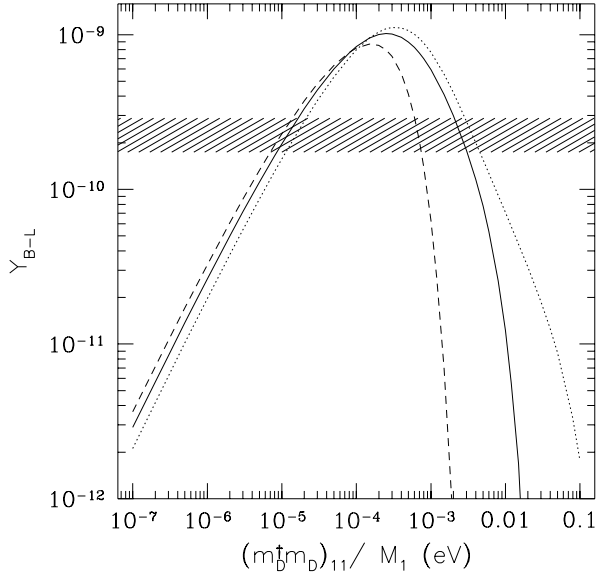,width=8cm}
        \hfill\mbox{ }
       \caption{\it Generated $(B-L)$ asymmetry as a function of 
         $\wt{m}_1$ for $M_1=10^8\;$GeV (dotted line), 
         $M_1=10^{10}\;$GeV (solid line) and $M_1=10^{12}\;$GeV
         (dashed line). The shaded area shows the measured value for
         the asymmetry.\label{fig13}}
    \end{figure}

    A characteristic feature of the non-supersymmetric version of this
    baryogenesis mechanism is that the generated asymmetry does not
    depend on the neutrino mass $M_1$ and $(m_{\scr D}^{\dg}
    m_{\scr D})_{11}$ separately but only on the ratio $\wt{m}_1$
    \cite{pluemi}.  To check if this is also the case in the
    supersymmetric scenario we have varied $\wt{m}_1$ while keeping
    all the other parameters fixed. In fig.~\ref{fig13} we have
    plotted the total lepton asymmetry $Y_L=\YL+\YLt$ as a function of
    $\wt{m}_1$ for the right-handed neutrino masses
    $M_1=10^{12}\;$GeV, $10^{10}\;$GeV and $10^8\;$GeV, and the $CP$
    asymmetry $\ve_1=-10^{-6}$.

    The main difference between the supersymmetric and
    non-supersymmetric scenarios concerns the necessary production of
    the neutrinos at high temperatures. In the non-supersymmetric
    scenario the Yukawa interactions are too weak to account for this,
    i.e.\ additional interactions of the right-handed neutrinos have
    to be introduced. This is no longer the case here. The
    supersymmetric Yukawa interactions are much more important, and
    can produce a thermal population of right-handed neutrinos, i.e.\ 
    the same vertices which are responsible for the generation of the
    asymmetry can also bring the neutrinos into thermal equilibrium at
    high temperatures. However, these lepton number violating
    processes will also erase a part of the generated asymmetry,
    hereby giving rise to the $\wt{m}_1$ dependence of the generated
    asymmetry which we shall discuss in detail.
    
    First one sees that in the whole parameter range the generated
    asymmetry is much smaller than the naively expected value
    $4\cdot10^{-9}$. For low $\wt{m}_1$ the reason being that the
    Yukawa interactions are too weak to bring the neutrinos into
    equilibrium at high temperatures, like in fig.~\ref{sol1fig}a. For
    high $\wt{m}_1$ on the other hand, the lepton number violating
    scatterings wash out a large part of the generated asymmetry at
    temperatures $T<M_1$, like in fig.~\ref{sol1fig}b. Hence, the
    requested asymmetry can only be generated if $\wt{m}_1$ is larger
    than $\sim10^{-5}\;$eV and smaller than $\sim5\cdot10^{-3}\;$eV,
    depending on the heavy neutrino mass $M_1$.
    
    The asymmetry in fig.~\ref{fig13} depends almost only on
    $\wt{m}_1$ for small $\wt{m}_1\;\ltap\;10^{-4}\;$eV, since in this
    region of parameter space the asymmetry depends mostly on the
    number of neutrinos generated at high temperatures, i.e.\ on the
    strength of the processes in which a right-handed neutrino can be
    generated or annihilated. The dominant reactions are decays,
    inverse decays and scatterings with a (s)top, which all give
    contributions proportional to $\wt{m}_1$ to the Boltzmann
    equations at high temperatures,
    \beqa
       &&{-z\over sH(M_1)}\;\gnone\propto\wt{m}_1\;,\qquad\qquad
         {-z\over sH(M_1)}\;\gsnone\propto\wt{m}_1\;,\qquad\qquad
         {-z\over sH(M_1)}\;\gtrone\propto\wt{m}_1\;,\NO\\[1ex]
       &&{-z\over sH(M_1)}\;\g_{22_1}\propto\wt{m}_1\;, \qquad\qquad
         {-z\over sH(M_1)}\;\g_{t_1}^{(i)}(T\gg M_1)\propto\wt{m}_1\;.
         \label{propor}
    \eeqa
    For large $\wt{m}_1\;\gtap\;10^{-4}\;$eV on the other hand, the
    neutrinos reach thermal equilibrium at high temperatures, i.e.\
    the generated asymmetry depends mostly on the influence of the lepton
    number violating scatterings at temperatures $T\ltap M_1$. In
    contrast to the relations of eq.~(\ref{propor}) the lepton number
    violating processes mediated by a heavy neutrino behave like
    \beq
      {-z\over sH(M_1)}\;\g_i^{\scr\D L}\propto M_1\sum\limits_j\wt{m}_j^2\;,
      \qquad i=A,\ldots,D
      \label{propor2}
    \eeq
    at low temperatures. Hence, one expects that the generated
    asymmetry becomes smaller for growing neutrino mass $M_1$ and this
    is exactly what one observes in fig.~\ref{fig13}.
    
    The lepton number violating scatterings can also explain the small
    dependence of the asymmetry on the heavy neutrino mass $M_1$ for
    $\wt{m}_1\,\ltap\,10^{-4}\;$eV. The inverse decay processes which
    take part in producing the neutrinos at high temperatures are $CP$
    violating, i.e.\ they generate a lepton asymmetry at high
    temperatures.  Due to the interplay of inverse decay processes and
    lepton number violating $2\to2$ scatterings this asymmetry has a
    different sign compared to the one generated in neutrino decays at
    low temperatures, i.e.\ the asymmetries will partially cancel each
    other, as one can see in the change of sign of the asymmetry in
    fig.~\ref{sol1fig}a.  This cancellation can only be avoided if the
    asymmetry generated at high temperatures is washed out before the
    neutrinos decay. At high temperatures the lepton number violating
    scatterings behave like
    \beq
      {-z\over sH(M_1)}\;\g_i^{\scr \D L}\propto
      M_1\sum\limits_ja_j\wt{m}_j^2\;,
      \qquad i=A,\ldots,D\;.
    \eeq 
    Hence, the wash-out processes are more efficient for
    larger neutrino masses, i.e.\ the final asymmetry should grow with
    the neutrino mass $M_1$. The finally generated asymmetry is not
    affected by the stronger wash-out processes, since for small
    $\wt{m}_1$ the neutrinos decay late, where one can neglect the
    lepton number violating scatterings.

    This change of sign in the asymmetry is not observed in
    fig.~\ref{sol1fig}b. Due to the larger $\wt{m}_1$ value the
    neutrinos are brought into equilibrium at much higher
    temperatures, where decays and inverse decays are suppressed by a
    time dilatation factor, i.e.~the (s)neutrinos are produced in CP
    invariant scatterings off a (s)top.

\section{SO(10) unification and leptogenesis \label{Yuk}}
    In ref.~\cite{bp}, it was shown that there is no direct connection
    between $CP$ violation and leptonic flavour mixing at high and low
    energies. Furthermore, the implications of non-supersymmetric
    leptogenesis on the scale of $(B-L)$ breaking and the light neutrino
    masses have been studied by assuming a
    similar pattern of masses and mixings for the leptons and the
    quarks. Here we are going to repeat this analysis for the
    supersymmetric scenario.
\subsection{Neutrino masses and mixings}
    If we choose a basis where the Majorana mass matrix $M$ and the
    Dirac mass matrix $m_l$ for the charged leptons are diagonal with
    real and positive eigenvalues,
    \beq
      m_l=\left(\begin{array}{ccc} m_e&0&0\\0&m_{\m}&0\\0&0&m_{\t}
          \end{array}\right)\;,\qquad\qquad
      M=\left(\begin{array}{ccc} M_1&0&0\\0&M_2&0\\0&0&M_3
          \end{array}\right)\;,
    \eeq
    the Dirac mass matrix of the neutrinos can be written in the form
    \beq
      m_{\scr D}=V\,\left(\begin{array}{ccc}
      m_1&0&0\\0&m_2&0\\0&0&m_3\end{array}\right)\,U^{\dag}\;,
      \label{u_def}
    \eeq
    where $V$ and $U$ are unitary matrices and the $m_i$ are real
    and positive.
    
    All the quantities relevant for baryogenesis depend only on the
    product $m_{\scr D}^{\dg}m_{\scr D}$, which is determined by the
    Dirac masses $m_i$ and the three mixing angles and six phases of
    $U$.  Five of these phases can be factored out with the Gell-Mann
    matrices $\l_i$,
    \beq
      U=\mbox{e}^{i\g}\,\mbox{e}^{i\l_3\a}\,\mbox{e}^{i\l_8\b}\,U_1\,
      \mbox{e}^{i\l_3\s}\,\mbox{e}^{i\l_8\t}\;.
    \eeq
    In analogy to the Kobayashi-Maskawa matrix for quarks the
    remaining matrix $U_1$ depends on three mixing angles and one
    phase. In unified theories based on SO(10) it is natural to
    assume a similar pattern of masses and mixings for leptons and
    quarks. This suggests the Wolfenstein parametrization 
    \cite{wolfenstein} as ansatz for $U_1$,
    \beq
      U_1=\left(\begin{array}{ccc}
      1-{\l^2\over2}  &      \l        & A\l^3(\r-i\h)\\[1ex]
        -\l         & 1-{\l^2\over2} & A\l^2 \\[1ex]
      A\l^3(1-\r-i\h) &    -A\l^2      &  1
      \end{array}\right)\;,
      \label{mm}
    \eeq
    where $A$ and $|\r+i\h|$ are of order one, while the mixing
    parameter $\l$ is assumed to be small. For the Dirac masses $m_i$
    SO(10) unification motivates a hierarchy like for the up-type
    quarks, 
    \beq
      m_1=b\l^4m_3\qquad m_2=c\l^2m_3\qquad b,c=\co(1)\;.
      \label{DMass}
    \eeq
    We have mentioned in section \ref{boltzeq} that we also need a
    hierarchy in the Majorana masses $M_i$ to get a lepton
    asymmetry. We choose a similar hierarchy as in
    eq.~(\ref{DMass}), 
    \beq
      M_1=B\l^4M_3\qquad M_2=C\l^2M_3\qquad B,C=\co(1)\;.
      \label{Majomass}
    \eeq
    Later on we will vary the parameters $B$ and $C$ to
    investigate different hierarchies for the right-handed neutrinos.
    
    The light neutrino masses read \cite{bp}
    \beqa
       m_{\n_e}&=&{b^2\over\left|C+\mbox{e}^{4i\a}\;B\right|}\;\l^4
             \;m_{\n_{\t}}+\co\left(\l^6\right)\label{mne}\\[1ex]
       m_{\n_{\m}}&=&{c^2\left|C+\mbox{e}^{4i\a}\;B\right|\over BC}
             \;\l^2\;m_{\n_{\t}}+\co\left(\l^4\right)\label{mnm}\\[1ex]
       m_{\n_{\t}}&=&{m_3^2\over M_3}+\co\left(\l^4\right)\;.\label{mnt}
    \eeqa
    We will not discuss the masses of the light scalar neutrinos here,
    since they depend on unknown soft breaking terms.

    In section \ref{gen} we have seen that the lepton asymmetry is
    largely determined by the mass parameter $\wt{m}_1$, which is
    given by
    \beq
      \wt{m}_1={c^2+A^2|\r+i\h|^2\over B}\;\l^2\;m_{\n_{\t}}\;,
      \label{mt}
    \eeq
    i.e.\ $\wt{m}_1$ is of the same order as the $\n_{\m}$ mass.
    {}From eq.~(\ref{CP}) we get the $CP$ asymmetry 
    \beq
      \ve_1={3\over8\p}\;{B\;A^2\over c^2+A^2\;|\r+i\h|^2}\;\l^4\;
      {m_3^2\over v_2^2}\;\mbox{Im}\left[(\r-i\h)^2
      \mbox{e}^{i2(\a+\sqrt{3}\b)}\right]
      \;+\;\co\left(\l^6\right)\;.
    \eeq
    In the next section we will always assume maximal phases, 
    i.e.\ we will set
    \beq
      \ve_1= \;-{3\over8\p}\;{B\;A^2\;|\r+i\h|^2\over 
       c^2+A^2\;|\r+i\h|^2}\;\l^4\;
      {m_3^2\over v_2^2}\;+\;\co\left(\l^6\right)\;.
      \label{cpa}
    \eeq
    Hence, the lepton asymmetries that we are going to calculate may
    be viewed as upper bounds on the attainable asymmetries. 

    Like in the non-supersymmetric scenario a large value of the
    Yukawa-coupling $m_3/v_2$ will be preferred by this baryogenesis
    mechanism, since $\ve_1\propto m_3^2/v_2^2$. This holds
    irrespective of our ansatz for neutrino mixings.
\subsection{Numerical results}
    The neutrino masses (\ref{mne})-(\ref{mnt}) can be used to
    constrain the free parameters of our ansatz. The strongest hint
    for a non-vanishing neutrino mass being the solar neutrino
    deficit\footnote{For a review and references, see \cite{haxt}.}, 
    we will fix the $\n_{\m}$ mass to the value preferred by the
    MSW solution \cite{msw},
    \beq
      m_{\n_{\m}}\simeq3\cdot10^{-3}\,\mbox{eV}\;.
      \label{numass}
    \eeq
    Hence the parameter $\wt{m}_1$, which is of the same order as
    $m_{\n_{\m}}$ according to eq.~(\ref{mt}), will be in the
    interval allowed by fig.~\ref{fig13}. 

\begin{figure}[t]
      \begin{minipage}[t]{8cm}
        \begin{center}\hspace{1cm}(a)\end{center}
        \vspace{-0.5cm}
        \epsfig{file=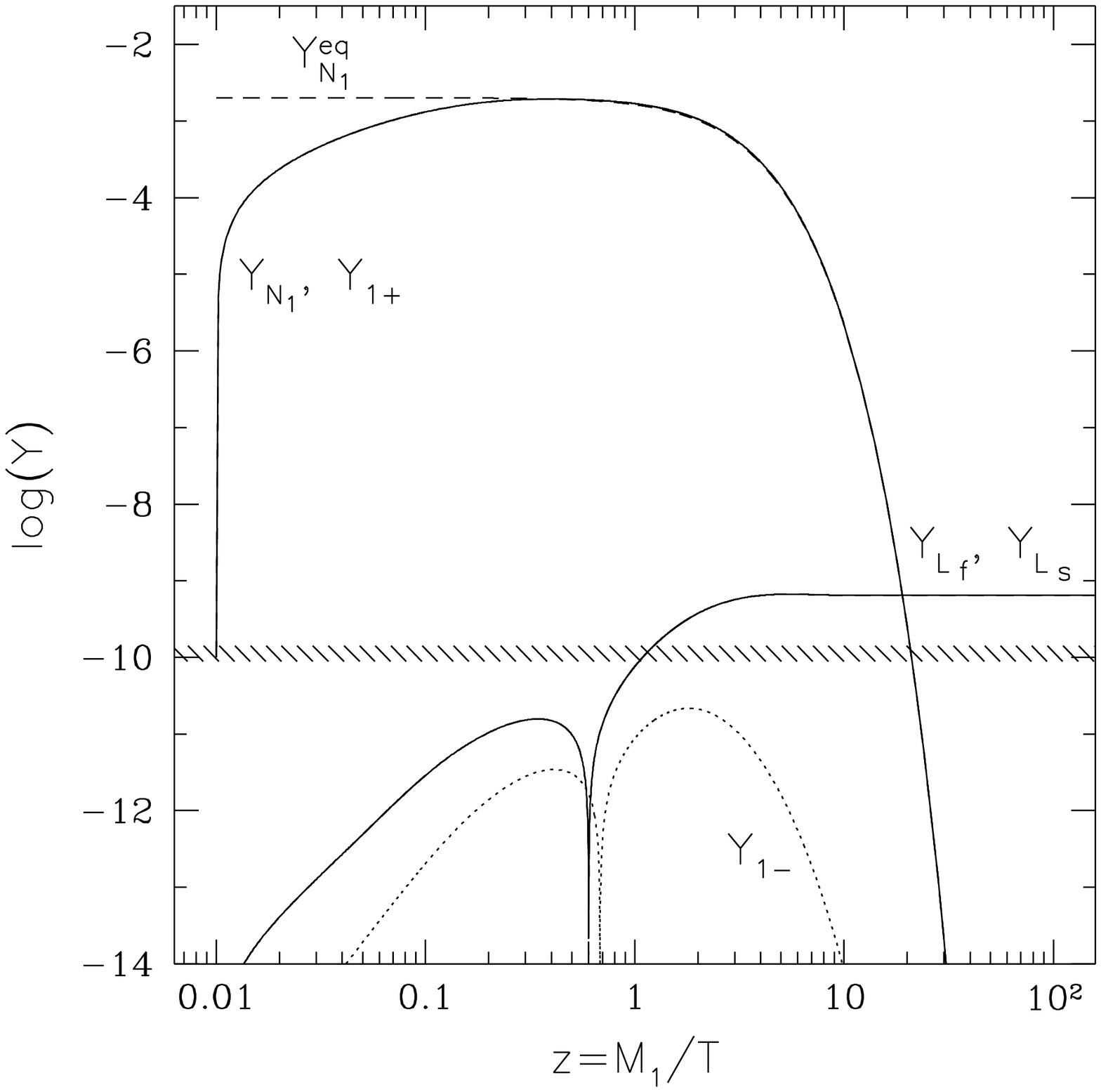,width=8cm}
      \end{minipage}
      \hspace{\fill}
      \begin{minipage}[t]{8cm}
        \begin{center}\hspace{1cm}(b)\end{center}
        \vspace{-0.5cm}
        \epsfig{file=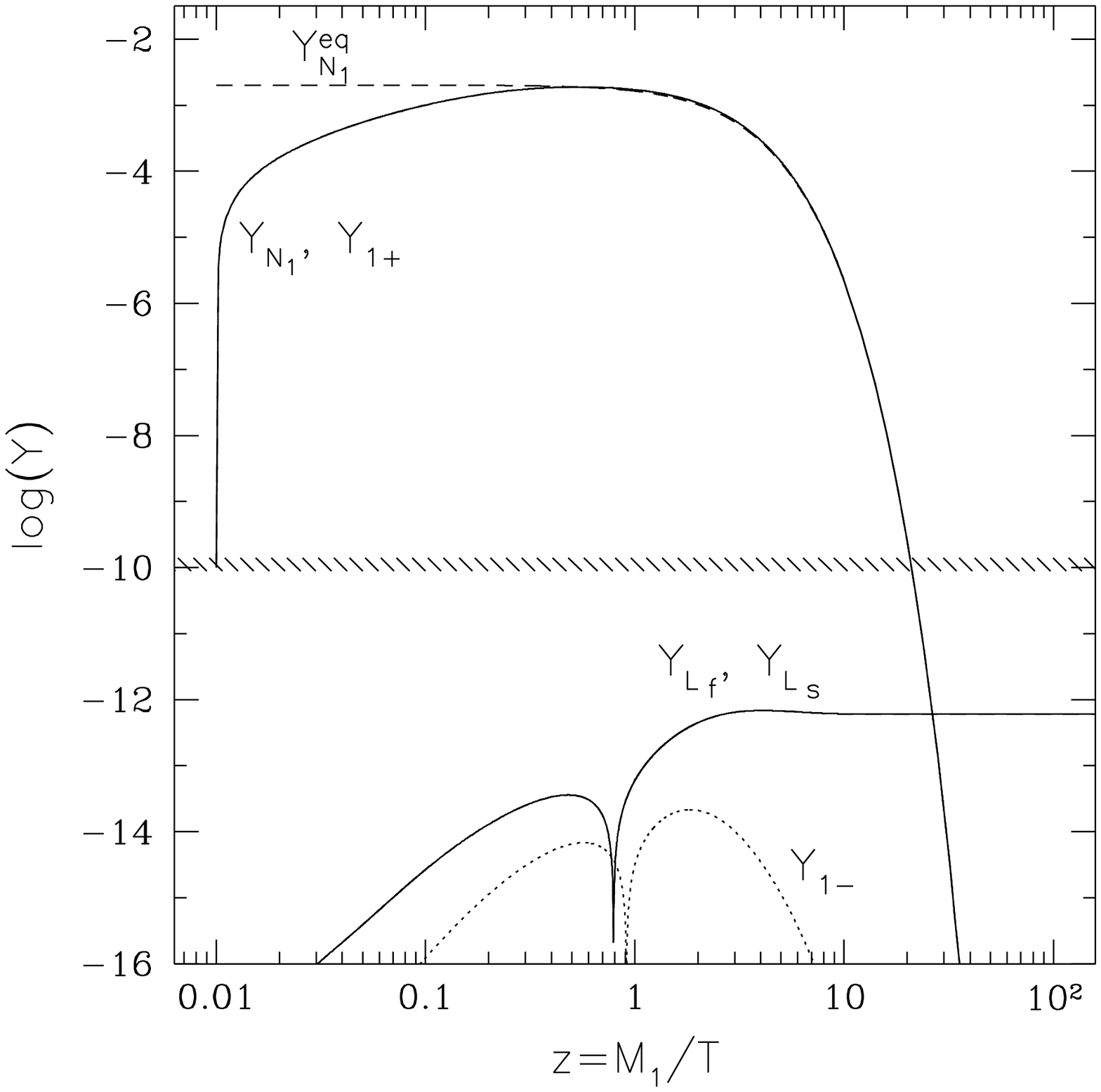,width=8cm}
     \end{minipage}  
       \caption{\it Generated asymmetry if one assumes a similar
        pattern of masses and mixings for the leptons and the
        quarks. In both figures we have $\l=0.1$ and $m_3=m_t$ (a) and
        $m_3=m_b$ (b).
       \label{sol2fig}}
    \end{figure}

    The most obvious parameter choice is to take all $\co(1)$
    parameters equal to one and to fix $\l$ to a similar
    value as the $\l$ parameter of the quark mixing matrix,
    \beq
      A=B=C=b=c=|\r+i\h|\simeq 1\; ,\qquad \l\simeq 0.1\;. 
      \label{p1}
    \eeq
    The $\n_{\m}$ mass in eqs.~(\ref{mnm}) and (\ref{numass}) then fixes
    the $\n_e$ and $\n_{\t}$ masses,
    \beq
      m_{\n_e}\simeq 8\cdot10^{-6}\;\mbox{eV}\; , \quad
      m_{\n_{\t}}\simeq 0.15\;\mbox{eV}
      \label{m1}
    \eeq
    and $\wt{m}_1$ reads
    \beq
      \wt{m}_1 \simeq 3\cdot10^{-3}\;\mbox{eV}\;.
    \eeq
    SO(10) unification suggests that the Dirac neutrino mass
    $m_3$ is equal to the top-quark mass,
    \beq
      m_3=m_t\simeq 174\;\mbox{GeV}\;.
      \label{3t}
    \eeq
    This leads to a large Majorana mass scale for the right-handed
    neutrinos,
    \beq
      M_3 \simeq 2\cdot10^{14}\;\mbox{GeV}\quad\Rightarrow\quad
      M_1\simeq 2\cdot10^{10}\;\mbox{GeV}\mbox{ and }
      M_2\simeq 2\cdot10^{12}\;\mbox{GeV}\;,
      \label{M3}
    \eeq
    and eq.~(\ref{cpa}) gives  the $CP$ asymmetry $\ve_1 \simeq 
    -6\cdot10^{-6}$. Integration of the Boltzmann equations yields 
    the $(B-L)$ asymmetry (cf.~fig.~\ref{sol2fig}a)
    \beq 
       Y_{B-L} \simeq 10^{-9}\; , 
    \eeq 
    which is of the correct order of magnitude. It is interesting
    to note that in the non-supersymmetric scenario one has $Y_{B-L} 
    \simeq 9\cdot10^{-10}$ for the same choice of parameters.

    Our assumption (\ref{3t}), $m_3 \simeq m_t$ led to a large
    Majorana mass scale $M_3$ in eq.~(\ref{M3}). To check the
    sensitivity of our result for the baryon asymmetry on this choice,
    we have envisaged a lower Dirac mass scale
    \beq 
      m_3 = m_b \simeq 4.5\; \mbox{GeV}\; , 
    \eeq 
    while keeping all other parameters in eq.~(\ref{p1}) fixed. The
    assumed $\n_{\m}$ mass (\ref{numass}) then requires a much lower
    value for the Majorana mass scale, $M_3\simeq10^{11}\;$GeV and
    the $CP$ asymmetry $\ve_1 \simeq -4\cdot10^{-9}$ becomes very small. 
    Consequently, the generated asymmetry (cf.~fig.~\ref{sol2fig}b)
    \beq 
      Y_{B-L} \simeq 10^{-12}\;, 
    \eeq 
    is too small by more than two orders of magnitude. We can conclude
    that high values for both masses $m_3$ and $M_3$ are preferred.
    This suggests that $(B-L)$ is already broken at the unification
    scale $\L_{\mbox{\tiny GUT}} \sim 10^{16}\;$GeV, without any
    intermediate scale of symmetry breaking, which is natural in
    SO(10) unification. Alternatively, a Majorana mass scale of the
    order of $10^{12}$ to $10^{14}\;$GeV can also be generated
    radiatively if SO(10) is broken into SU(5) at some high scale
    between $10^{16}\;$GeV and the Planck-scale, and SU(5) is
    subsequently broken into the MSSM gauge group at the usual GUT
    scale $\sim 10^{16}\;$GeV \cite{su5}.
    \begin{figure}[t]
        \mbox{ }\hfill
        \epsfig{file=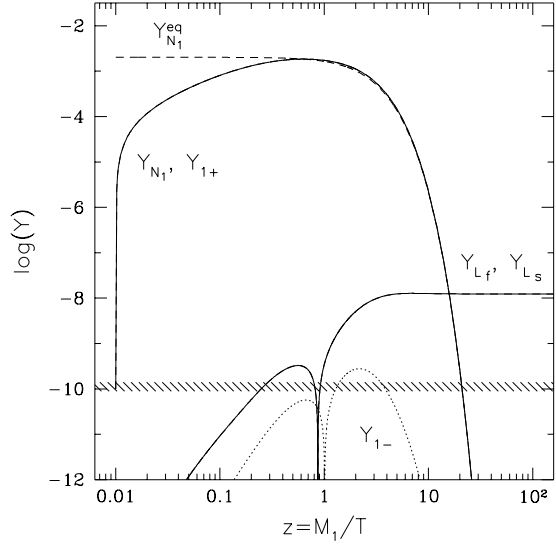,width=8cm}
        \hfill\mbox{ }
        \caption{\it Generated lepton asymmetry if one assumes a
        similar mass hierarchy for the right-handed neutrinos and the
        down-type quarks.
        \label{hierarch}}
    \end{figure}

    In eq.~(\ref{Majomass}) we had assumed a mass hierarchy for the
    heavy Majorana neutrinos like for the up-type quarks.
    Alternatively, one can assume a weaker hierarchy, like for the
    down-type quarks by choosing
    \beq
      B=10 \qquad \mbox{and} \qquad C=3\;.
    \eeq  
    Keeping the other parameters in eq.~(\ref{p1}) unchanged fixes the
    $\n_e$ and $\n_{\t}$ masses,
    \beq
      m_{\n_e} \simeq 5\cdot10^{-6}\;\mbox{eV}\;,\qquad\qquad 
        m_{\n_{\t}} \simeq 0.7\;\mbox{eV}\;,
    \eeq
    and the mass parameter $\wt{m}_1$,
    \beq
      \wt{m}_1 \simeq 10^{-3}\;\mbox{eV}\;.
    \eeq
    Choosing the Dirac mass scale (\ref{3t}) we get a large Majorana 
    mass scale 
    \beq
      M_3 \simeq 4\cdot10^{13}\;\mbox{GeV}\quad\Rightarrow\quad
      M_1 \simeq 4\cdot10^{10}\;\mbox{GeV}\mbox{ and }
        M_2 \simeq 10^{12}\;\mbox{GeV}\;.
    \eeq
    {}From eq.~(\ref{cpa}) one obtains the $CP$ asymmetry 
    $\ve_1\simeq-6\cdot10^{-5}$. The corresponding solutions of the
    Boltzmann equations are shown in fig.~\ref{hierarch}. The final
    $(B-L)$ asymmetry,
    \beq
      Y_{B-L} \simeq 2\cdot10^{-8}\;,
    \eeq
    is much larger than requested, but this value can always be
    lowered by adjusting the unknown phases. Hence, the possibility to
    generate a lepton asymmetry does not depend on the special form of
    the mass hierarchy assumed for the right-handed neutrinos.

    In the non-supersymmetric scenario one finds for the same
    parameter choice
    \beq
      Y_{B-L} \simeq 2\cdot10^{-8}\;.
    \eeq
    Hence, when comparing the supersymmetric and the
    non-supersymmetric scenario, one sees that the larger $CP$
    asymmetry in the former and the additional contributions from the
    sneutrino decays are compensated by the wash-out processes which
    are stronger than in the latter.
\section{Conclusions}
    We conclude that the cosmological baryon asymmetry can be
    generated in a supersymmetric extension of the Standard Model by
    the out-of-equilibrium decays of heavy right-handed Majorana
    neutrinos and their scalar partners. Solving the Boltzmann
    equations we have shown that, in order to be consistent, one has
    to pay attention to two phenomena which can hamper the generation
    of a lepton asymmetry.

    First, one has to take into consideration lepton number violating
    scatterings. These processes, which are usually neglected, can
    wash out a large part of the asymmetry if the Yukawa couplings of
    the right-handed neutrinos become too large.

    On the other hand the neutrinos have to be brought into thermal
    equilibrium at high temperatures. We could show that for this
    purpose it is not necessary to assume additional interactions of
    the right-handed neutrinos in our theory, since the Yukawa
    interactions can be sufficiently strong to produce a thermal
    population of heavy neutrinos at high temperatures, while still
    being weak enough to prevent the final asymmetry from being
    washed out.

    The observed baryon asymmetry can be obtained without any fine
    tuning of parameters if one assumes a similar pattern of mixings
    and Dirac masses for the neutrinos and the up-type quarks. Then
    the generated asymmetry is related to the $\n_{\m}$ mass and
    fixing this mass to the value preferred by the MSW-solution to the
    solar neutrino problem leads to a baryon asymmetry of the
    requested order, provided $(B-L)$ is broken at the unification
    scale, as suggested by supersymmetric SO(10) unification.

    In supersymmetric theories there are further possible sources of a
    $(B-L)$ asymmetry, e.g.\ it may be possible to combine inflation
    with leptogenesis (cf.~refs.~\cite{infl}). In this connection,
    possible constraints on the neutrino masses and on the reheating
    temperature from lepton number violating processes at low
    temperatures require further studies.

\pagebreak 
\noindent
{\bf\Large Acknowledgments}

    \mbox{ }\\\noindent 
    I would like to thank W.~Buchm\"uller for many fruitful
    discussions and continuous support. It is also a pleasure to thank
    J.~R.~Espinosa for useful discussions. Finally, I would like to
    thank L.~Covi, E.~Roulet and F.~Vissani for a helpful comment.\\[2ex]

\begin{appendix}
\noindent
{\bf\Large Appendix}
\section{The Superfields \label{appA}}
    \setcounter{equation}{0}
    \renewcommand{\theequation}{\mbox{\Alph{section}.\arabic{equation}}}
    In addition to the usual MSSM particles the supersymmetric SO(10)
    contains right-handed neutrinos and their scalar partners. With 
    $y=x-i\q\s\bar{\q}$ we have the following chiral 
    superfields\footnote{We are using the conventions of
    ref.~\cite{wb} with metric $\h_{\m\n}=\mbox{diag}(1,-1,-1,-1).$}
    \beqa
      H_i&=&H_i(y)+\sqrt{2}\,\q \wt{H}_i(y)+\q\q\, F_{\scr H_i}(y)\;,\\
      Q&=&\wt{q}(y)+\sqrt{2}\,\q q_{\mbox{\tiny L}}(y)
        +\q\q\, F_{\scr Q}(y)\;,\\
      L&=&\wt{l}(y)+\sqrt{2}\,\q l_{\mbox{\tiny L}}(y)
        +\q\q\, F_{l}(y)\;,\\
      U^c&=&\wt{U^c}(y)+\sqrt{2}\,\q {u_{\mbox{\tiny R}}}^c(y)
        +\q\q\, F_{\scr U^c}(y)\;,\\
      D^c&=&\wt{D^c}(y)+\sqrt{2}\,\q {d_{\mbox{\tiny R}}}^c(y)
        +\q\q\, F_{\scr D^c}(y)\;,\\
      E^c&=&\wt{E^c}(y)+\sqrt{2}\,\q {e_{\mbox{\tiny R}}}^c(y)
        +\q\q\, F_{\scr E^c}(y)\;,\\
      N^c&=&\wt{N^c}(y)+\sqrt{2}\,\q {\n_{\mbox{\tiny R}}}^c(y)
        +\q\q\, F_{\scr N^c}(y)\;.
    \eeqa
    $H_i$ denotes the two Higgs-doublets,
    \beq
      H_1=\left(\begin{array}{c}H_1^0\\-H_1^-\end{array}\right)
      \qquad\mbox{and}\qquad
      H_2=\left(\begin{array}{c}H_2^+\\H_2^0\end{array}\right)\;,
    \eeq
    $Q$ and $L$ stand for the left-handed quark and lepton doublets
    and $U^c$, $D^c$, $E^c$ and $N^c$ are the right-handed singlet
    fields.

    Besides the usual bispinors for the quarks and the charged leptons
    we can introduce Majorana-spinors for the right- and
    left-handed neutrinos
    \beq
      N=\left(\begin{array}{c}{{\n_{\mbox{\tiny R}}}^c}_{\a}\\[1ex]
      {\Bar{{\n_{\mbox{\tiny R}}}^c}}^{\;\dt{\a}}\end{array}\right)
      \qquad\mbox{and}\qquad
      \n=\left(\begin{array}{c}{\n_{\mbox{\tiny L}}}_{\a}\\[1ex]
      {\Bar{\n_{\mbox{\tiny L}}}}^{\;\dt{\a}}\end{array}\right)\;.
    \eeq
    In the symmetric phase of the MSSM no mixing occurs between the
    fermionic partners of gauge and Higgs bosons. Therefore, we have
    two Dirac higgsinos 
    \beq
      \wt{h^0}=\left(\begin{array}{c}{\wt{H_1^0}}_{\a}\\[1ex]
      {\Bar{\wt{H_2^0}}}^{\;\dt{\a}}\end{array}\right)
      \qquad\mbox{and}\qquad
      \wt{h^-}=\left(\begin{array}{c}{\wt{H_1^-}}_{\a}\\[1ex]
      {\Bar{\wt{H_2^+}}}^{\;\dt{\a}}\end{array}\right)\;,
    \eeq
    which again form an isospin doublet,
    \beq
      \wt{h}=\left(\begin{array}{c}\wt{h^0}\\[1ex]
       -\wt{h^-}\end{array}\right)\;.
    \eeq
\section{Boltzmann equations \label{appB}}
    \setcounter{equation}{0}
    \renewcommand{\theequation}{\mbox{\Alph{section}.\arabic{equation}}}
    The microscopic evolution of particle densities and asymmetries is
    governed by a network of Boltzmann equations. In the following we
    will compile some basic formulae to introduce our 
    notation\footnote{For a review and references, see \cite{kw}.}. 

    It is usually a good approximation to assume Maxwell-Boltzmann
    statistics, so that the equilibrium number density of a particle
    $i$ is given by
    \beq
     n_i^{\rm eq}(T)={g_i\over(2\p)^3}\int\dd^3p_i\,f_i^{\rm eq}
     \qquad\mbox{with}\qquad
     f^{\rm eq}_i\left(E_i,T\right)=\mbox{e}^{-E_i/T}\;,
    \eeq
    where $g_i$ is the number of internal degrees of
    freedom. This particle density can be changed by interactions and
    by the expansion of the universe. Since we are only interested in
    the effect of the interactions it is useful to scale out the
    expansion. This is done by using the number of particles per
    comoving volume element,
    \beq
      Y_i={n_i\over s}\;,
    \eeq
    where $s$ is the entropy density, as independent variable instead
    of the number density. 

    In our case elastic scatterings, which can only change the phase
    space distributions but not the particle densities, occur at a
    much higher rate than inelastic processes. Therefore, we can
    assume kinetic equilibrium, so that the phase space densities are
    given by
    \beq
     f_i(E_i,T)={n_i\over n_i^{\rm eq}}\mbox{e}^{-E_i/T}\;.
    \eeq
    In this framework the Boltzmann equation describing the evolution
    of a particle number $Y_{\j}$ in an isentropically expanding
    universe reads \cite{kw,luty}
    \beqa
    {\mbox{d}Y_{\j}\over\mbox{d}z}&=&-{z\over sH\left(m_{\j}\right)}
    \sum\limits_{a,i,j,\ldots}\left[{Y_{\j}Y_a\ldots\over
     Y_{\j}^{\rm eq}Y_a^{\rm eq}\ldots}\,\g^{\rm eq}\left(\j+a+\ldots\to 
     i+j+\ldots\right)\right.\NO\\[1ex]
     &&\qquad\qquad\qquad\left.{}-{Y_iY_j\ldots\over 
     Y_i^{\rm eq}Y_j^{\rm eq}\ldots}
     \,\g^{\rm eq}\left(i+j+\ldots\to\j+a+\ldots\right)\right]\;,
    \label{7}
    \eeqa
    where $z=m_{\j}/T$ and $H\left(m_{\j}\right)$ is the Hubble
    parameter at $T=m_{\j}$. The $\g^{\rm eq}$ are space time
    densities of scatterings for the different processes. For
    a decay one finds \cite{luty}
    \beq
     \g_D:=\g^{\rm eq}(\j\to i+j+\ldots)=
     n^{\rm eq}_{\j}{\mbox{K}_1(z)\over\mbox{K}_2(z)}\,\G\;,
     \label{decay}
    \eeq
    where K$_1$ and K$_2$ are modified Bessel functions and
    $\G$ is the usual decay width in the rest system of
    the decaying particle. Neglecting a possible $CP$ violation, one
    finds the same reaction density for the inverse decay.

    The reaction density for a two body scattering reads
    \beq
     \g^{\rm eq}({\j}+a\leftrightarrow i+j+\ldots)=
     {T\over64\p^4}\int\limits_{\left(m_{\j}+m_a\right)^2}^{\infty}
     \hspace{-0.5cm}\dd s\,\hat{\s}(s)\,\sqrt{s}\,
     \mbox{K}_1\left({\sqrt{s}\over T}\right)\;,
     \label{22scatt}
    \eeq 
    where $s$ is the squared centre of mass energy and the reduced
    cross section $\hat{\s}(s)$ for the process ${\j}+a\to i+j+\ldots$
    is related to the usual total cross section $\s(s)$ by
    \beq
      \hat{\s}(s)=
      {8\over s}
      \left[\left(p_{\j}\cdot p_a\right)^2-m_{\j}^2m_a^2\right]\,\s(s)\;.
    \eeq
    \vspace{\fill}

\section{Reduced cross sections \label{appC}}
    \setcounter{equation}{0}
    \renewcommand{\theequation}{\mbox{\Alph{section}.\arabic{equation}}}
    In this section we will collect the reduced cross sections for all
    the $2\leftrightarrow2$ and $2\leftrightarrow3$ processes that we
    had discussed in section \ref{theory}. The corresponding reaction
    densities, which can be calculated analytically in some interesting
    limiting cases, will be discussed in the next section.
\subsection{Lepton number violating processes mediated by a
    right-handed neutrino}
    We have mentioned in the main text that we have to subtract the
    contributions coming from on-shell (s)neutrinos, i.e.\ we have to
    replace the usual propagators by off-shell propagators
    \beq
      {1\over D_j(x)} := {x-a_j\over(x-a_j)^2+a_jc_j}\quad
      \mbox{and}\quad
      {1\over \wt{D}_j(x)} := {x-a_j\over(x-a_j)^2+a_j\wt{c_j}}\;.
    \eeq
    To begin with, let us specify the reduced cross sections 
    for the reactions depicted in fig.~\ref{fig03}.
    For the processes $\wt{l}+\Bar{\wt{h}}\leftrightarrow
    {\wt{l}{ }}^{\,\dg}+\wt{h}$ and $l+H_2\leftrightarrow\Bar{l}
    +H_2^{\dg}$ one has
    \beqa
      \lefteqn{\hat{\s}_{\scr N}^{(1)}(x) = \hat{\s}_{\scr N}^{(2)}(x) = 
        {1\over2\pi}\left\{\sum\limits_j\lljj^2\;{a_j\over x}\left[
        {x\over a_j}+{x\over D_j(x)}+{x^2\over 2D_j^2(x)}
        -\left(1+{x+a_j\over D_j(x)}\right)\lnaj\right]\right.}
        \NO\\[1ex]
      &&{}+\sum\limits_{n,j\atop j<n}\mbox{Re}\left[\llnj^2\right]
        {\sqrt{a_na_j}\over x}\left[{x\over D_j(x)}+{x\over D_n(x)}
        +{x^2\over D_j(x)D_n(x)}\right.\\[1ex]
      &&\left.\left.{}+\left(x+a_j\right)\left({2\over a_n-a_j}
        -{1\over D_n(x)}\right)\lnaj+\left(x+a_n\right)
        \left({2\over a_j-a_n}-{1\over D_j(x)}\right)\lnan
        \right]\right\}\;,\NO
    \eeqa
    where $n$ and $j$ are the flavour indices of the neutrinos in the 
    intermediate state. The interference terms with $n\ne j$ are
    always very small and can safely be neglected.
    \newpage
    \noindent
    The reduced cross section for the process 
    $\wt{l}+\Bar{\wt{h}}\leftrightarrow\Bar{l}+H_2^{\dg}$ reads 
    \beqa
      \lefteqn{\hat{\s}_{\scr N}^{(3)}(x) = {1\over2\pi}\left\{
        \sum\limits_j\lljj^2\;{a_j\over x}\left[{-x\over x+a_j}
        +{x\over D_j(x)}+{x^2\over 2D_j^2(x)}
        +\left(1-{a_j\over D_j(x)}\right)\lnaj\right]\right.}\NO\\[1ex]
      &&\qquad{}+\sum\limits_{n,j\atop j<n}\mbox{Re}\left[\llnj^2\right]
        {\sqrt{a_na_j}\over x}\left[{x\over D_j(x)}+{x\over D_n(x)}
        +{x^2\over D_j(x)D_n(x)}\right.\\[1ex]
      &&\left.\left.\qquad
        {}-\;a_j\left({2\over a_n-a_j}+{1\over D_n(x)}\right)\lnaj
        -a_n\left({2\over a_j-a_n}+{1\over D_j(x)}\right)\lnan
        \right]\right\}\;.\NO
    \eeqa
    The same result is valid for the $CP$ conjugated process.\\
    For the process $l+\Bar{\wt{h}}\leftrightarrow{\wt{l}{ }}^{\,\dg}
    +H_2^{\dg}$ one finds
    \beqa
      \hat{\s}_{\scr N}^{(4)}(x) &=& {1\over2\pi}\left\{
        \sum\limits_j\lljj^2\;{a_j\over x}\left[{x^2\over a_j(x+a_j)}
        +{x^2\over \wt{D_j}^2(x)}+{x\over\wt{D_j}(x)}
        \lnaj\right]\right.\NO\\[1ex]
      &&{}+\sum\limits_{n,j\atop j<n}\mbox{Re}\left[\llnj^2\right]
        {\sqrt{a_na_j}\over x}\left[{2x^2\over\wt{D_j}(x)
        \wt{D_n}(x)}+x\left({2\over a_n-a_j}+
        {1\over\wt{D_n}(x)}\right)\lnaj\right.\NO\\[1ex]
      &&\left.\left.\hspace{2cm}{}+\;x\left({2\over a_j-a_n}
        +{1\over\wt{D_j}(x)}\right)\lnan\right]\right\}\;.
    \eeqa
    For the scattering $\wt{l}+H_2\rightarrow{\wt{l}{ }}^{\,\dg}
    +\sur+\wt{q}$ and the corresponding $CP$ transformed process we
    have
    \beqa
      \lefteqn{\hat{\s}_{\scr N}^{(5)}(x) = {3\,\a_u\over8\pi^2}\left\{
        \sum\limits_j\lljj^2\;{a_j\over x}\left[{x\over a_j}
        +{x\over \wt{D_j}(x)}+{x^2\over \wt{D_j}^2(x)}
        -\left(1+{x+a_j\over\wt{D_j}(x)}\right)
        \lnaj\right]\right.}\NO\\[1ex]
      &&{}+\sum\limits_{n,j\atop j<n}\mbox{Re}\left[\llnj^2\right]
        {\sqrt{a_na_j}\over x}\left[{x\over\wt{D_j}(x)}+
        {x\over\wt{D_n}(x)}+{x^2\over\wt{D_j}(x)
        \wt{D_n}(x)}\right.\\[1ex]
      &&\left.\left.\!{}+\;\left(x+a_j\right)\left({2\over a_n-a_j}
        -{1\over\wt{D_n}(x)}\right)\lnaj
        +\left(x+a_n\right)\left({2\over a_j-a_n}
        -{1\over\wt{D_j}(x)}\right)\lnan\right]\right\}\;.\NO
    \eeqa
    Finally, we have two processes which do not violate lepton number
    but merely transform leptons into scalar leptons and vice versa.
    We have the $2\to2$ scattering $l+H_2\leftrightarrow\wt{l}
    +\Bar{\wt{h}}$,
    \beq
      \hat{\s}_{\scr N}^{(6)}(x) = {1\over4\pi}
      \sum\limits_{j,n}\left|\llnj\right|^2\;{x^2\over D_j(x)D_n(x)}\;,
    \eeq
    and the $2\to3$ process $l+\Bar{\wt{h}}\leftrightarrow\wt{l}
    +\wt{q}^{\dg}+\sur^{\dg}$,
    \beq
      \hat{\s}_{\scr N}^{(7)}(x) = {3\,\a_u\over16\pi^2}
        \sum\limits_{j,n}\left|\llnj\right|^2\;
        {x^2\over\wt{D_j}(x)\wt{D_n}(x)}\;.
    \eeq

    Let us now come to the $2\to3$ processes shown in
    fig.~\ref{fig04}.\\
    For the transition $\wt{q}+\sur\rightarrow\wt{l}+\wt{l}+H_2$ the
    reduced cross section reads\\
    \beqa
      \lefteqn{\hat{\s}_{\scr N}^{(8)}(x) = {3\,\a_u\over16\pi^2}\left\{
        \sum\limits_j\lljj^2\;{a_j\over x}\left[
        -{x\over a_j+\wt{c_j}}+{x-a_j\over\sqrt{a_j\wt{c_j}}}\atnj
        \right.\right.}\NO\\[1ex]
      &&\left.\hspace{2cm}{}-\lnpropj
        +{1\over2}\int\limits_0^x\;\mbox{d}x_1\;
        {1\over\wt{D_j}(x_1)}\ln\left({(x-x_1-a_j)^2
        +a_j\wt{c_j}\over a_j^2+a_j\wt{c_j}}\right)\right]\NO\\[1ex]
      \lefteqn{{}+2\sum\limits_{n,j\atop j<n}\mbox{Re}
        \left[\llnj^2\right]{\sqrt{a_na_j}\over x}\left[
         {1\over2}\int\limits_0^x\;\mbox{d}x_1\;
         {1\over\wt{D_n}(x_1)}\ln\left({(x-x_1-a_j)^2+a_j\wt{c_j}
         \over a_j^2+a_j\wt{c_j}}\right)
        \right.}\\[1ex]
      &&{}+2\sqrt{a_j\wt{c_j}}{x-a_n\over\left(a_j-a_n\right)^2}\atnj
        +{x-a_j\over a_j-a_n}\lnpropj\NO\\[1ex]
      &&\left.\left.
        {}+2\sqrt{a_n\wt{c_n}}{x-a_j\over\left(a_n-a_j\right)^2}\atnn
        +{x-a_n\over a_n-a_j}\lnpropn\right]\right\}\NO\;.
    \eeqa
    The remaining integral cannot be solved exactly. However,
    it can be neglected for $x>a_j,a_n$ and for $x<a_j,a_n$ it can be
    approximated by
    \beqa
      \lefteqn{{1\over2}\int\limits_0^x\;\mbox{d}x_1\;{1\over\wt{D_n}(x_1)}
        \ln\left({(x-x_1-a_j)^2+a_j\wt{c_j}
        \over a_j^2+a_j\wt{c_j}}\right)}\\[1ex]
      &&\qquad\approx\ln\left({a_j+a_n-x\over a_j}\right)
        \ln\left({a_n-x\over a_n}\right)
        +\mbox{Sp}\left({a_n\over a_n+a_j-x}\right)-
        \mbox{Sp}\left({a_n-x\over a_n+a_j-x}\right)\;,\NO
    \eeqa
    where $\mbox{Sp}(x)$ is the Spence function.\\
    For the scatterings $\wt{q}+\sur\rightarrow\wt{l}+\Bar{l}+\wt{h}$, 
    ${\wt{l}{ }}^{\,\dg}+\wt{q}\rightarrow\Bar{l}+\sur^{\dg}+\wt{h}$
    and ${\wt{l}{ }}^{\,\dg}+\sur\rightarrow\Bar{l}+\wt{q}^{\dg}+\wt{h}$
    the reduced cross sections are equal,
    \beqa
      \lefteqn{\hat{\s}_{\scr N}^{(9)}(x) = \hat{\s}_{\scr N}^{(11)}(x) = 
        {3\,\a_u\over8\pi^2x}\left\{\sum\limits_j\lljj^2
        \left[-{3\over2}x+{1\over2}(x-2a_j)\lnpropj
        \right.\right.}\NO\\[1ex]
      &&\hspace{2cm}\left.{}+{1\over2}\sqrt{a_j\over\wt{c_j}}
        \left(x-a_j+3\wt{c_j}\right)\atnj\right]\\[1ex]
      &&\hspace{-21pt}{}+2\sum\limits_{n,j\atop j<n}\left|\llnj\right|^2
        \left[-2x+a_j{x-a_j\over a_j-a_n}\lnpropj
      +a_n{x-a_n\over a_n-a_j}\lnpropn\right.\NO\\[1ex]
      &&\hspace{2cm}{}+2\sqrt{a_j\wt{c_j}}\;
        {xa_n-2a_na_j+a_j^2\over(a_j-a_n)^2}\atnj\NO\\[1ex]
      &&\left.\left.\hspace{2cm} 
        {}+2\sqrt{a_n\wt{c_n}}\;{xa_j-2a_na_j+a_n^2\over(a_n-a_j)^2}
        \atnn\right]\right\}\;.\NO
    \eeqa
    For the process ${\wt{l}{ }}^{\,\dg}+\wt{q}\rightarrow\wt{l}
    +\sur^{\dg}+H_2$ and similar reactions one gets
    \beqa
      \lefteqn{\hat{\s}_{\scr N}^{(10)}(x) = {3\,\a_u\over16\pi^2}\left\{
        \sum\limits_j\lljj^2\;{a_j\over x}\left[
        {x\over a_j}-2\ln\left({x+a_j\over a_j}\right)
        -\lnpropj\right.\right.}\NO\\[1ex]
      &&\left.\!{}+{x-a_j\over\sqrt{a_j\wt{c_j}}}\atnj
        +2\int\limits_0^x\;\mbox{d}x_1\;
        {1\over\wt{D_j}(x_1)}\left[\mbox{Sp}\left(-{x\over a_j}
        \right)-\mbox{Sp}\left(-{x_1\over a_j}\right)\right]\right]
        \NO\\[1ex]
      \lefteqn{{}+2\sum\limits_{n,j\atop j<n}\mbox{Re}
        \left[\llnj^2\right]{\sqrt{a_na_j}\over x}\left[
        2{x+a_j\over a_n-a_j}\lnaj+2{x+a_n\over a_j-a_n}\lnan
        \right.}\\[1ex]
      &&{}+{x-a_j\over a_j-a_n}\lnpropj+2\sqrt{a_j\wt{c_j}}
        {x-a_n\over(a_j-a_n)^2}\atnj\NO\\[1ex]
      &&{}+{x-a_n\over a_n-a_j}\lnpropn+2\sqrt{a_n\wt{c_n}}
        {x-a_j\over(a_n-a_j)^2}\atnn\NO\\[1ex]
      &&\left.\left.{}+\int\limits_0^x\;\mbox{d}x_1\;\left[
        {1\over\wt{D_j}(x_1)}\left(\mbox{Sp}\left(-{x\over a_n}
        \right)-\mbox{Sp}\left(-{x_1\over a_n}\right)\right)
        +{1\over\wt{D_n}(x_1)}\left(\mbox{Sp}\left(-{x\over a_j}
        \right)-\mbox{Sp}\left(-{x_1\over a_j}\right)\right)\right]
        \right]\right\}\;.\NO
    \eeqa
    The remaining integral can again not be solved exactly. However
    it can be approximated by
    \beqa
      \lefteqn{\int\limits_0^x\;\mbox{d}x_1\;
        {1\over\wt{D_j}(x_1)}\left[\mbox{Sp}\left(-{x\over a_n}
        \right)-\mbox{Sp}\left(-{x_1\over a_n}\right)\right]}\\[1ex]
      &&\approx{x\over a_n}-{\sqrt{a_j\wt{c_j}}\over a_n}\atnj
        -{x-a_j\over2a_n}\lnpropj\NO
    \eeqa
    for $x<a_n$ and for $x>a_n$ it can be neglected.

    \vspace{\fill}
    Finally, we have to compute the $t$- and $u$-channel processes
    from fig.~\ref{fig05} which give compara\-tively simple
    contributions.\\     
    For the processes $\wt{l}+\wt{l}\leftrightarrow\wt{h}+\wt{h}$ and 
    $l+l\leftrightarrow H_2^{\dg}+H_2^{\dg}$ we get
    \beqa
      \hat{\s}_{\scr N}^{(12)}(x)&=&\hat{\s}_{\scr N}^{(13)}(x)
        ={1\over2\pi}\left\{\sum\limits_j\lljj^2\;\left[{x\over x+a_j}
        +{a_j\over x+2a_j}\ln\left({x+a_j\over a_j}\right)\right]
        \right.\NO\\[1ex]
      &&{}+\sum\limits_{n,j\atop j<n}\mbox{Re}\left[\llnj^2\right]
        \sqrt{a_na_j}\left[\left({1\over x+a_n+a_j}
        +{2\over a_n-a_j}\right)\ln\left({x+a_j\over a_j}\right)
        \right.\NO\\[1ex]
      &&\left.\left.\hspace{2cm}{}+\left({1\over x+a_n+a_j}
        +{2\over a_j-a_n}\right)\ln\left({x+a_n\over a_n}\right)
        \right]\right\}\;.
    \eeqa
    At this order of perturbation theory the same result is valid for
    the $CP$ transformed processes.\\
    \newpage
    \noindent
    For the scattering $\wt{l}+l\leftrightarrow\wt{h}+H_2^{\dg}$ one
    has 
    \beqa
      \hat{\s}_{\scr N}^{(14)}(x) &=& {1\over2\pi}\left\{
        \sum\limits_j\lljj^2\;\left[{x\over x+a_j}
        -{a_j\over x+2a_j}\ln\left({x+a_j\over a_j}\right)\right]
        \right.\NO\\[1ex]
      &&{}+2\sum\limits_{n,j\atop j<n}\mbox{Re}\left[\llnj^2\right]
        \sqrt{a_na_j}\left[\left({1\over x+a_n+a_j}
        +{1\over a_n-a_j}\right)\ln\left({x+a_j\over a_j}\right)
        \right.\NO\\[1ex]
      &&\left.\left.\hspace{2cm}{}+\left({1\over x+a_n+a_j}
        +{1\over a_j-a_n}\right)\ln\left({x+a_n\over a_n}\right)
        \right]\right\}\;.
    \eeqa      
    The $2\to3$ process $H_2+\wt{q}^{\,\dg}\leftrightarrow
    {\wt{l}{}}^{\;\dg}+{\wt{l}{}}^{\;\dg}+\sur$ gives
    \beqa
      \lefteqn{\hat{\s}_{\scr N}^{(15)}(x) = {3\a_u\over8\pi^2}\left\{
        \sum\limits_j\lljj^2{a_j\over x}\;\left[{x\over a_j}
        -\left(1-{1\over2}\ln\left({x+2a_j\over a_j}\right)\right)
        \ln\left({x+a_j\over a_j}\right)\right.\right.}\NO\\[1ex]
      &&\left.\hspace{4cm}{}+{1\over2}\mbox{Sp}\left({a_j\over x+2a_j}\right)
        -{1\over2}\mbox{Sp}\left({x+a_j\over x+2a_j}\right)
        \right]\NO\\[1ex]
      &&{}+\sum\limits_{n,j\atop j<n}\mbox{Re}\left[\llnj^2\right]
        {\sqrt{a_na_j}\over x}\left[\left(2{x+a_j\over a_n-a_j}
        +\ln\left({x+a_n+a_j\over a_n}\right)\right)\ln\left(
        {x+a_j\over a_j}\right)\right.\NO\\[1ex]
      &&\hspace{3cm}{}+\left(2{x+a_n\over a_j-a_n}
        +\ln\left({x+a_n+a_j\over a_j}\right)
        \right)\ln\left({x+a_n\over a_n}\right)\\[1ex]
      &&\left.\left.{}+\mbox{Sp}\left({aj\over x+a_n+a_j}\right)
        -\mbox{Sp}\left({x+aj\over x+a_n+a_j}\right)
        +\mbox{Sp}\left({an\over x+a_n+a_j}\right)
        -\mbox{Sp}\left({x+an\over x+a_n+a_j}\right)\right]\right\}\;.\NO
    \eeqa
    For the related transition $\wt{l}+\wt{l}\leftrightarrow
    \sur+\wt{q}+H_2^{\dg}$ we have
    \beqa
      \lefteqn{\hat{\s}_{\scr N}^{(16)}(x) = {3\a_u\over16\pi^2}\left\{
        \sum\limits_j\lljj^2{a_j\over x}\;\left[{x\over a_j}
        +2\,\mbox{Sp}\left(-{x+a_j\over a_j}\right)
        +{\p^2\over6}\right.\right.}\NO\\[1ex]
      &&\left.\hspace{4cm}{}-\left(1-2\ln\left({x+2a_j\over a_j}
        \right)\right)\ln\left({x+a_j\over a_j}\right)\right]\\[1ex]
      &&\hspace{-20pt}{}+2\sum\limits_{n,j\atop j<n}
        \mbox{Re}\left[\llnj^2\right]{\sqrt{a_na_j}\over x}
        \left[\left({x+a_j\over a_n-a_j}+\ln\left({x+a_n+a_j\over 
        a_n}\right)\right)\ln\left({x+a_j\over a_j}\right)
        +\mbox{Sp}\left(-{x+a_j\over a_n}\right)\right.\NO\\[1ex]
      &&\hspace{\fill}\left.\left.{}+\left({x+a_n\over a_j-a_n}
        +\ln\left({x+a_n+a_j\over a_j}\right)\right)
        \ln\left({x+a_n\over a_n}\right)
        +\mbox{Sp}\left(-{x+a_n\over a_j}\right)
        +{\p^2\over6}+{1\over2}\ln^2\left({a_n\over a_j}\right)
        \right]\right\}\;.\NO
    \eeqa
    There are some $2\to2$ processes left which do not violate lepton
    number but simply transform leptons into scalar leptons, like in 
    the process $\wt{l}+\Bar{l}\leftrightarrow\wt{h}+H_2$
    \beqa
      &&\hspace{-2cm}\hat{\s}_{\scr N}^{(17)}(x)={1\over2\pi}\left\{
        \sum\limits_j\lljj^2\;\left[{-x\over x+a_j}
        +\ln\left({x+a_j\over a_j}\right)\right]\right.\NO\\[1ex]
      &&\left.\hspace{-1cm}{}+2\sum\limits_{n,j\atop j<n}
        \left|\llnj\right|^2
        \left[{a_j\over a_j-a_n}\ln\left({x+a_j\over a_j}\right)
        +{a_n\over a_n-a_j}\ln\left({x+a_n\over a_n}\right)
        \right]\right\}\;,
    \eeqa
    or in the similar process $\wt{l}+H_2^{\dg}\leftrightarrow\wt{h}+l$
    \beqa
      &&\hspace{-2cm}\hat{\s}_{\scr N}^{(18)}(x) ={1\over2\pi}\left\{
        \sum\limits_j\lljj^2\;\left[-2+{x+2a_j\over x}
        \ln\left({x+a_j\over a_j}\right)\right]\right.\NO\\[1ex]
      &&\left.\hspace{-1cm}{}+2\sum\limits_{n,j\atop j<n}
        \left|\llnj\right|^2\left[-1+{a_j\over x}{x+a_j\over a_j-a_n}
        \ln\left({x+a_j\over a_j}\right)
        +{a_n\over x}{x+a_n\over a_n-a_j}\ln\left({x+a_n\over a_n}\right)
        \right]\right\}\;.
    \eeqa
    Finally, the last process $l+{\wt{l}{}}^{\;\dg}\leftrightarrow
    \wt{h}+\wt{q}^{\,\dg}+\sur^{\dg}$ gives
    \beqa
      &&\hspace{-2cm}\hat{\s}_{\scr N}^{(19)}(x) ={3\a_u\over8\pi^2}
        \left\{\sum\limits_j\lljj^2\;\left[-2+{x+2a_j\over x}
        \ln\left({x+a_j\over a_j}\right)\right]\right.\NO\\[1ex]
      &&\left.\hspace{-1cm}{}+2\sum\limits_{n,j\atop j<n}
        \left|\llnj\right|^2\left[-1+{a_j\over x}{x+a_j\over a_j-a_n}
        \ln\left({x+a_j\over a_j}\right)
        +{a_n\over x}{x+a_n\over a_n-a_j}\ln\left({x+a_n\over a_n}\right)
        \right]\right\}\;.
    \eeqa
\subsection{Scattering off a top or a stop}
    For the processes specified in fig.~\ref{Ntop} the reduced cross
    sections read
    \beqa
      \hat{\s}_{t_j}^{(0)}&=&{3\,\a_u\over2}\lljj{x^2-a_j^2\over x^2}
        \;,\\[1ex]
      \hat{\s}_{t_j}^{(1)}&=&3\,\a_u\,\lljj{x-a_j\over x}\left[
        -{2x-a_j+2a_h\over x-a_j+a_h}+{x+2a_h\over x-a_j}
        \ln\left({x-a_j+a_h\over a_h}\right)\right]\;, \\[1ex]
      \hat{\s}_{t_j}^{(2)}&=&3\,\a_u\,\lljj{x-a_j\over x}
        \left[-{x-a_j\over x-a_j+2a_h}+\ln\left({x-a_j+a_h\over a_h}
        \right)\right]\;,\\[1ex]
      \hat{\s}_{t_j}^{(3)}&=&3\,\a_u\,\lljj\left({x-a_j\over x}
        \right)^2\;,\\[1ex]
      \hat{\s}_{t_j}^{(4)}&=&3\,\a_u\,\lljj{x-a_j\over x}\left[
        {x-2a_j+2a_h\over x-a_j+a_h}+{a_j-2a_h\over x-a_j}
        \ln\left({x-a_j+a_h\over a_h}\right)\right]\;.
    \eeqa
    To regularize an infrared divergence in the $t$-channel diagrams
    we had to introduce a Higgs-mass
    \beq
       a_h:=\left({\m\over M_1}\right)^2\;.
    \eeq
    In the calculations we have used the value $\m=800\;$GeV.

    The analogous processes involving a scalar neutrino
    (cf.~fig.~\ref{Nttop}) give similar contributions
    \beqa
      \hat{\s}_{t_j}^{(5)}&=&{3\,\a_u\over2}\lljj
        \left({x-a_j\over x}\right)^2\;, \\[1ex]
      \hat{\s}_{t_j}^{(6)}&=&3\,\a_u\,\lljj{x-a_j\over x}\left[
        -2+{x-a_j+2a_h\over x-a_j}\ln\left({x-a_j+a_h\over
        a_h}\right)\right]\;, \\[1ex]
      \hat{\s}_{t_j}^{(7)}&=&3\,\a_u\,\lljj\left[-{x-a_j\over 
        x-a_j+2a_h}+\ln\left({x-a_j+a_h\over a_h}\right)\right] 
        \;,\\[1ex]
      \hat{\s}_{t_j}^{(8)}&=&3\,\a_u\,\lljj{x-a_j\over x}
        \;{a_j\over x}\;,\\[1ex]
      \hat{\s}_{t_j}^{(9)}&=&3\,\a_u\,\lljj{a_j\over x}\left[-{x-a_j\over 
        x-a_j+a_h}+\ln\left({x-a_j+a_h\over a_h}\right)\right]\;.
    \eeqa
\subsection{Neutrino pair creation and annihilation}
    With the abbreviations
    \beqa
      \l_{ij}&=&\l\left(x,a_i,a_j\right)=
        \left[x-\left(\sqrt{a_i}+\sqrt{a_j}\right)^2\right]\,
        \left[x-\left(\sqrt{a_i}-\sqrt{a_j}\right)^2\right]\\[1ex]
      \mbox{L}_{ij}&=&\ln\left({x-a_i-a_j+\sqrt{\lkin}
        \over x-a_i-a_j-\sqrt{\lkin}}\right)\;,
    \eeqa
    the reduced cross sections for the right-handed neutrino pair
    creation read
    \beqa
      \hspace{-1.3cm}\hat{\s}_{N_iN_j} ^{(1)}&=&{1\over4\p}
        \left\{\lljj\llii\left[-{2\over x}\sqrt{\lkin}+\lnNN\right]
        -2\,\mbox{Re}\left[\llji^2\right]{\sqrt{a_ia_j}
        \left(a_i+a_j\right)\over x\left(x-a_i-a_j\right)}\lnNN
        \right\}\;,\\[1ex]
      \hspace{-1.3cm}\hat{\s}_{N_iN_j} ^{(2)}&=&{1\over4\p}
        \left\{\lljj\llii\left[{2\over x}\sqrt{\lkin}+
        {a_i+a_j\over x}\lnNN\right]-2\,\mbox{Re}\left[\llji^2\right]
        {\sqrt{a_ia_j}\over x-a_i-a_j}\lnNN\right\},\\[1ex]
      \hspace{-1.3cm}\hat{\s}_{N_iN_j} ^{(3)}&=&{1\over4\p}
        \left\{\left|\llji\right|^2\left[-{2\over x}\sqrt{\lkin}
        +\lnNN\right]-2\,\mbox{Re}\left[\llji^2\right]{\sqrt{a_ia_j}
        \left(a_i+a_j\right)\over x\left(x-a_i-a_j\right)}\lnNN
        \right\}\;,\\[1ex]
      \hspace{-1.3cm}\hat{\s}_{N_iN_j} ^{(4)}&=&{1\over4\p}
        \left\{\left|\llji\right|^2\left[{2\over x}\sqrt{\lkin}+
        {a_i+a_j\over x}\lnNN\right]-2\,\mbox{Re}\left[\llji^2\right]
        {\sqrt{a_ia_j}\over x-a_i-a_j}\lnNN\right\}\;.
    \eeqa
    For the scalar neutrinos one has similarly
    \beqa
      \hat{\s}_{\sni\snj} ^{(1)}&=&{1\over4\p}
        \lljj\llii\left[-{2\over x}\sqrt{\lkin}
        +{x-a_i-a_j\over x}\lnNN\right]\;,\\[1ex]
      \hat{\s}_{\sni\snj} ^{(2)}&=&{1\over4\p}
        \left\{\lljj\llii{2\over x}\sqrt{\lkin}
        -2\,\mbox{Re}\left[\llji^2\right]
        {\sqrt{a_ia_j}\over x}\lnNN\right\}\;,\\[1ex]
      \hat{\s}_{\sni\snj} ^{(3)}&=&{1\over4\p}
        \left|\llji\right|^2\left[-{2\over x}\sqrt{\lkin}
        +{x-a_i-a_j\over x}\lnNN\right]\;,\\[1ex]
      \hat{\s}_{\sni\snj} ^{(4)}&=&{1\over4\p}
        \left\{\left|\llji\right|^2{2\over x}\sqrt{\lkin}
        -2\,\mbox{Re}\left[\llji^2\right]
        {\sqrt{a_ia_j}\over x}\lnNN\right\}\;.
    \eeqa
    For the diagrams involving one neutrino and one sneutrino
    (cf.~fig.~\ref{NNt}) one finally has
    \beqa
      \hspace{-1.3cm}\hat{\s}_{N_j\sni}^{(1)}&=&{1\over4\p}
        \left\{\lljj\llii{x+a_i-a_j\over x}\lnNN
        -2\,\mbox{Re}\left[\llji^2\right]{\sqrt{a_ia_j}\over x}
        {x+a_i-a_j\over x-a_i-a_j}\lnNN\right\}\\[1ex]
      \hspace{-1.3cm}\hat{\s}_{N_j\sni} ^{(2)}&=&{1\over4\p}
        \left\{\left|\llji\right|^2{x+a_i-a_j\over x}\lnNN
        -2\,\mbox{Re}\left[\llji^2\right]{\sqrt{a_ia_j}\over x}
        {x+a_i-a_j\over x-a_i-a_j}\lnNN\right\}\;.
    \eeqa

\section{Reaction densities \label{appD}}
    \setcounter{equation}{0}
    \renewcommand{\theequation}{\mbox{\Alph{section}.\arabic{equation}}}
    In general the reaction densities corresponding to the reduced
    cross sections discussed in appendix~\ref{appC} have to be calculated
    numerically. However, there exist some interesting limiting cases
    where one can calculate them analytically.
\subsection{Lepton number violating scatterings}
    In the Boltzmann equations we do not need every reaction density
    $\g_{\scr N}^{(i)}$, $i=1,\ldots,19$ separately
    (cf.~sect.~\ref{boltzeq}). We only have to consider the combined
    reaction densities
    \beqa
      \g_{\scr A}^{\scr\D L}&=&2\g_{\scr N}^{(1)}+\g_{\scr N}^{(3)}
        +\g_{\scr N}^{(4)}+\g_{\scr N}^{(6)}+\g_{\scr N}^{(7)}
        +2\g_{\scr N}^{(12)}+\g_{\scr N}^{(14)}\;,\\[1ex]
      \g_{\scr B}^{\scr\D L}&=&\g_{\scr N}^{(3)}+\g_{\scr N}^{(4)}
        -\g_{\scr N}^{(6)}-\g_{\scr N}^{(7)}+\g_{\scr N}^{(14)}
        \;,\\[1ex]
      \g_{\scr C}^{\scr\D L}&=&3\g_{\scr N}^{(9)}+\g_{\scr N}^{(17)}
        +\g_{\scr N}^{(18)}+6\g_{\scr N}^{(19)}\;,\\[1ex]
      \g_{\scr D}^{\scr\D L}&=&4\g_{\scr N}^{(5)}+2\g_{\scr N}^{(8)}
        +8\g_{\scr N}^{(10)}+3\g_{\scr N}^{(9)}+4\g_{\scr N}^{(15)}
        +2\g_{\scr N}^{(16)}+\g_{\scr N}^{(17)}+\g_{\scr N}^{(18)}
        +6\g_{\scr N}^{(19)}\;.
    \eeqa
    For low temperatures, i.e.~$z\gg1/\sqrt{a_j}\;$, the dominant
    contribution to the integrand of the reaction densities comes from
    small centre of mass energies, i.e.~$x\ll a_j$. In this limit the
    reduced cross sections $\hat{\s}_{\scr N}^{(i)}$ for the $L+\wt{L}$ 
    violating or conserving processes behave differently. For the 
    $L+\wt{L}$ violating scatterings ($i=1,\ldots,5,8,10,12,\ldots,16$)
    one finds
    \beq
      \hat{\s}_{\scr N}^{(i)}\propto x\quad\mbox{for }x\ll a_j\;,
    \eeq
    while one has
    \beq
      \hat{\s}_{\scr N}^{(i)}\propto x^2\quad\mbox{for }x\ll a_j
    \eeq
    for the $L+\wt{L}$ conserving processes ($i=6,7,9,11,17,18,19$).
    Hence, the reaction densities can be calculated analytically in
    this limit and one finds
    \beqa
      &&\hspace{-1cm}\g_{\scr A}^{\scr\D L}={M_1^4\over\p^5}{1\over z^6}
        \left\{\sum\limits_j\lljj^2{2\over a_j}+
        \sum\limits_{n,j\atop j<n}\mbox{Re}\left[\llnj^2\right]
        {19\over4\sqrt{a_na_j}}\right\}\;,\\[1ex]
      &&\hspace{-1cm}\g_{\scr B}^{\scr\D L}={M_1^4\over\p^5}{1\over z^6}
        \left\{\sum\limits_j\lljj^2{1\over2a_j}+
        \sum\limits_{n,j\atop j<n}\mbox{Re}\left[\llnj^2\right]
        {7\over4\sqrt{a_na_j}}\right\}\;,\\[1ex]
      &&\hspace{-1cm}\g_{\scr C}^{\scr\D L}={M_1^4\over\p^5}{1\over z^8}
        \left\{\sum\limits_j\lljj^2{1\over a_j^2}\left(4+{27\a_u\over4\p}
        \right)+\sum\limits_{n,j\atop j<n}\left|\llnj\right|^2
        {1\over a_ja_n}\left(8+{18\a_u\over\p}\right)\right\}\;,\\[1ex]
      &&\hspace{-1cm}\g_{\scr D}^{\scr\D L}={M_1^4\over\p^5}{\a_u\over\p}
        {1\over z^6}\left\{\sum\limits_j\lljj^2{153\over32a_j}+
        \sum\limits_{n,j\atop j<n}\mbox{Re}\left[\llnj^2\right]
        {147\over16\sqrt{a_na_j}}\right\}\;.
    \eeqa
    For high temperatures, i.e.~$z\ll1/\sqrt{a_j}\;$, we can use
    the asymptotic expansions of the reduced cross sections to compute
    the reactions densities and we get
    \beqa
      \g_{\scr A}^{\scr\D L}&=&{M_1^4\over64\p^5}{1\over z^4}\left\{
        \left(13+{3\a_u\over4\p}\right)\sum\limits_j\lljj^2+
        \sum\limits_{n,j\atop j<n}\mbox{Re}\left[\llnj^2\right]
        {24\sqrt{a_na_j}\over a_n-a_j}\ln\left(a_n\over a_j\right)
        \right.\NO\\[1ex]
      &&\left.\hspace{3cm}{}+\left(2+{3\a_u\over2\p}\right)
        \sum\limits_{n,j\atop j<n}\left|\llnj\right|^2\right\}\;,\\[2ex]
      \g_{\scr B}^{\scr\D L}&=&{M_1^4\over64\p^5}{1\over z^4}\left\{
        \left(3-{3\a_u\over4\p}\right)\sum\limits_j\lljj^2+
        \sum\limits_{n,j\atop j<n}\mbox{Re}\left[\llnj^2\right]
        {8\sqrt{a_na_j}\over a_n-a_j}\ln\left(a_n\over a_j\right)
        \right.\NO\\[1ex]
      &&\left.\hspace{3cm}{}-\left(2+{3\a_u\over2\p}\right)
        \sum\limits_{n,j\atop j<n}\left|\llnj\right|^2\right\}\;,\\[2ex]
      \g_{\scr C}^{\scr\D L}&=&{M_1^4\over32\p^5}{1\over z^4}\left\{
        \sum\limits_j\lljj^2\left[-1-{45\a_u\over8\p}+{9\a_u\over8}
        \sqrt{a_j\over\wt{c_j}}+\left(4+{27\a_u\over2\p}\right)
        \left(\ln\left({2\over z\sqrt{a_j}}\right)
        -\g_{\mbox{\tiny E}}\right)\right]\right.\NO\\[1ex]
      &&{}+\sum\limits_{n,j\atop j<n}\left|\llnj\right|^2
        \left[2+{9\a_u\over\left(a_n-a_j\right)^2}
        \left(a_n\sqrt{a_j\wt{c_j}}+a_j\sqrt{a_n\wt{c_n}}\right)
        \right.\\[1ex]
      &&\left.\left.\hspace{2cm}{}+\left(8+{36\a_u\over\p}\right)
        \left({a_n\over a_n-a_j}\ln\left({2\over\sqrt{a_n}z}\right)
        +{a_j\over a_j-a_n}\ln\left({2\over\sqrt{a_j}z}\right)
        -\g_{\mbox{\tiny E}}\right)\right]\right\}\;,\NO\\[2ex]
      \g_{\scr D}^{\scr\D L}&=&{M_1^4\over32\p^5}{1\over z^4}\left\{
        \sum\limits_j\lljj^2\left[-1+{27\a_u\over8\p}+{39\a_u\over8}
        \sqrt{a_j\over\wt{c_j}}+\left(4+{27\a_u\over2\p}\right)
        \left(\ln\left({2\over z\sqrt{a_j}}\right)
        -\g_{\mbox{\tiny E}}\right)\right]\right.\NO\\[1ex]
      &&{}+\sum\limits_{n,j\atop j<n}\left|\llnj\right|^2
        \left[4-8\g_{\mbox{\tiny E}}
        +{8\over a_n-a_j}\left(a_n\ln\left({2\over\sqrt{a_n}z}\right)
        -a_j\ln\left({2\over\sqrt{a_j}z}\right)\right)\right]\\[1ex]
      &&\left.{}+{3\a_u\over\p}\sum\limits_{n,j\atop j<n}\mbox{Re}
        \left[\llnj^2\right]\sqrt{a_na_j}\left[
        {7\over2}{1\over a_n-a_j}\ln\left({a_n\over a_j}\right)
        +{5\p\over\left(a_n-a_j\right)^2}\left(\sqrt{a_j\wt{c_j}}
        +\sqrt{a_n\wt{c_n}}\right)\right]\right\}\NO\;,
    \eeqa
    where $\g_{\mbox{\tiny E}}=0.577216$ is Euler's constant. These
    reaction densities are therefore proportional to $T^4$ at high
    temperatures, as expected on purely dimensional grounds.

    For intermediate temperatures $z\sim 1/\sqrt{a_j}$ the reaction
    densities have to be computed numerically. This becomes
    increasingly difficult in the narrow width limit, where $1/D_j(x)$
    has two very sharp peaks. However, in the limit $c_j\to0$
    the two peaks in $1/D_j(x)$ cancel each other, since they
    have a different sign, while the peaks in $1/D^2_j(x)$ add up.
    Therefore, the terms proportional to $1/D_j(x)$ or
    $1/D_j(x)D_n(x)$ with $n\ne j$ can be neglected in the narrow width
    limit, while $1/D^2_j(x)$ can be approximated by a $\d$-function
    \beq 
      {1\over D^2_j(x)}\approx{\p\over2\sqrt{a_jc_j}}\,
        \d\left(x-a_j\right)\;.
    \eeq
    An analogous relation holds for $1/\wt{D_j}^2(x)$.

    These relations allow to calculate the contributions from the
    $s$-channel diagrams to the reaction densities analytically in the
    limit $c_j\to0$, while the contributions from the $t$-channel
    diagrams can easily be calculated numerically. 
\subsection{Interactions with quarks and squarks}
    The reaction densities $\g_{t_j}^{(i)}$ for the interaction of a
    (s)neutrino with a top or a stop can also be calculated
    analytically in the limit of high temperatures
    $z\ll1/\sqrt{a_j}\;$. For the $s$-channel processes one finds
    \beqa
      \g_{t_j}^{(0)}&=&{3\a_uM_1^4\over64\p^4}
        \lljj a_j{\mbox{K}_2\left(z\sqrt{a_j}\right)\over z^2}\;,\\[1ex]
      \g_{t_j}^{(3)}&=&2\g_{t_j}^{(0)}\;,\qquad\qquad
        \g_{t_j}^{(5)}=\g_{t_j}^{(0)}\;.
    \eeqa
    For the $t$-channel reaction densities one has analogously
    \beqa
      &&\hspace{-1.8cm}\g_{t_j}^{(1)}={3\a_uM_1^4\over8\p^4}
        \lljj{1\over z^4}\left[\left(1-{z^2a_j\over4}\right)
        \mbox{K}_0\left(z\sqrt{a_j}\right)
        +{z^2a_j\over4}\left(\ln\left({a_j\over a_h}\right)-1
        \right)\mbox{K}_2\left(z\sqrt{a_j}\right)\right]\;,\\[1ex]
      &&\hspace{-1.8cm}\g_{t_j}^{(2)}={3\a_uM_1^4\over8\p^4}
        \lljj{1\over z^4}\left[
        \left(1-{z^2a_j\over4}\right)\mbox{K}_0\left(z\sqrt{a_j}\right)
        +{z^2a_j\over4}\ln\left({a_j\over a_h}\right)
        \mbox{K}_2\left(z\sqrt{a_j}\right)\right]\;,\\[1ex]
      &&\g_{t_j}^{(4)}=2\g_{t_j}^{(0)}\;,\qquad\qquad
        \g_{t_j}^{(6)}=\g_{t_j}^{(1)}\;,\qquad\qquad
        \g_{t_j}^{(7)}=\g_{t_j}^{(2)}\;.
    \eeqa
    $\g_{t_j}^{(8)}$ and $\g_{t_j}^{(9)}$ are several orders of
    magnitude smaller than the other $\g_{t_j}^{(i)}$ for small $z$
    and can therefore be neglected at high temperatures. 
    
    By using the series expansions of the Bessel functions, one sees
    that the processes with a higgsino in the $t$-channel, i.e.\ 
    $\g_{t_j}^{(1)}$, $\g_{t_j}^{(2)}$, $\g_{t_j}^{(6)}$ and
    $\g_{t_j}^{(7)}$, behave like $T^4\ln(T/M_j)$ at high
    temperatures, whereas the other reaction densities are
    proportional to $T^4$.

\subsection{Pair creation and annihilation of neutrinos}
    In the Boltzmann equations we only need certain combinations of
    reaction densities which can easily be evaluated for high
    temperatures:
    \beqa
      \sum\limits_{k=1}^2\g_{\scr N_iN_j}^{(k)}&=&
        \sum\limits_{k=1}^2\g_{\scr \sni\snj}^{(k)}=
        \g_{\scr N_j\sni}^{(1)}=\\[1ex]
       &=&{M_1^4\over16\p^5}{1\over z^4}\lljj\llii\left\{
        \left[1-{z^2\over4}\left(\sqrt{a_i}+\sqrt{a_j}\right)^2\right]
        \mbox{K}_0\left(z\left(\sqrt{a_i}+\sqrt{a_j}\right)\right)
        \right.\NO\\[1ex]
     &&\left.{}+{z^2\over4}\left(\sqrt{a_i}+\sqrt{a_j}\right)^2\left[
        1+\ln\left(2+{a_i+a_j\over\sqrt{a_ia_j}}\right)\right]
        \mbox{K}_2\left(z\left(\sqrt{a_i}+\sqrt{a_j}\right)\right)
        \right\}\;,\NO\\[1ex]
     \sum\limits_{k=3}^4\g_{\scr N_iN_j}^{(k)}&=&
        \sum\limits_{k=3}^4\g_{\scr \sni\snj}^{(k)}=
        \g_{\scr N_j\sni}^{(2)}=\\[1ex]
     &=&{M_1^4\over16\p^5}{1\over z^4}
        \left|\llji\right|^2
        \left\{\left[1-{z^2\over4}\left(\sqrt{a_i}+\sqrt{a_j}\right)^2\right]
        \mbox{K}_0\left(z\left(\sqrt{a_i}+\sqrt{a_j}\right)\right)
        \right.\NO\\[1ex]
     &&\left.{}+{z^2\over4}\left(\sqrt{a_i}+\sqrt{a_j}\right)^2\left[
        1+\ln\left(2+{a_i+a_j\over\sqrt{a_ia_j}}\right)\right]
        \mbox{K}_2\left(z\left(\sqrt{a_i}+\sqrt{a_j}\right)\right)
        \right\}\;,\NO
    \eeqa
    i.e.\ these reaction densities are proportional to
    $T^4\ln(T/(M_i+M_j))$ at high temperatures.
\end{appendix}

\end{document}

%% file: Fig01.tex
\begin{center}
\parbox[c]{12.5cm}{
\pspicture(0,0)(3.7,2.6)
\psline[linewidth=1pt](0.6,1.3)(1.6,1.3)
\psline[linewidth=1pt](1.6,1.3)(2.3,0.6)
\psline[linewidth=1pt,linestyle=dashed](1.6,1.3)(2.3,2.0)
\psline[linewidth=1pt]{<-}(1.9,1.0)(2.0,0.9)
\psline[linewidth=1pt]{->}(2.03,1.73)(2.13,1.83)
\rput[cc]{0}(0.3,1.3){$N_j$}
\rput[cc]{0}(2.6,0.6){$\tilde{h}$}
\rput[cc]{0}(2.6,2.0){$\tilde{l}$}
\rput[cc]{0}(3.5,1.3){$+$}
\endpspicture
\pspicture(-0.5,0)(4.2,2.6)
\psline[linewidth=1pt](0.6,1.3)(1.3,1.3)
\psline[linewidth=1pt,linestyle=dashed](1.3,1.3)(2.0,1.3)
\psline[linewidth=1pt](2,1.3)(2.5,1.8)
\psline[linewidth=1pt,linestyle=dashed](2.5,1.8)(3,2.3)
\psline[linewidth=1pt](2,1.3)(3,0.3)
\psarc[linewidth=1pt](2,1.3){0.7}{45}{180}
\psline[linewidth=1pt]{<-}(1.53,1.3)(1.63,1.3)
\psline[linewidth=1pt]{->}(1.7,1.93)(1.8,1.99)
\psline[linewidth=1pt]{->}(2.75,2.05)(2.85,2.15)
\psline[linewidth=1pt]{<-}(2.4,0.9)(2.5,0.8)
\rput[cc]{0}(0.3,1.3){$N_j$}
\rput[cc]{0}(1.65,0.9){$\tilde{l}$}
\rput[cc]{0}(2,2.3){$\tilde{h}$}
\rput[cc]{0}(2.6,1.45){$N$}
\rput[cc]{0}(3.3,2.3){$\tilde{l}$}
\rput[cc]{0}(3.3,0.3){$\tilde{h}$}
\rput[cc]{0}(4.0,1.3){$+$}
\endpspicture
\pspicture(-0.5,0)(3.5,2.6)
\psline[linewidth=1pt](0.6,1.3)(1.3,1.3)
\psline[linewidth=1pt,linestyle=dashed](1.3,1.3)(2.0,1.3)
\psline[linewidth=1pt,linestyle=dashed](2,1.3)(2.5,0.8)
\psline[linewidth=1pt](2.5,0.8)(3,0.3)
\psline[linewidth=1pt,linestyle=dashed](2,1.3)(3,2.3)
\psarc[linewidth=1pt](2,1.3){0.7}{-180}{-45}
\psline[linewidth=1pt]{<-}(1.53,1.3)(1.63,1.3)
\psline[linewidth=1pt]{<-}(1.7,0.67)(1.8,0.64)
\psline[linewidth=1pt]{<-}(2.16,1.14)(2.26,1.04)
\psline[linewidth=1pt]{<-}(2.7,0.6)(2.8,0.5)
\psline[linewidth=1pt]{->}(2.5,1.8)(2.6,1.9)
\rput[cc]{0}(0.3,1.3){$N_j$}
\rput[cc]{0}(1.65,1.7){$H_2$}
\rput[cc]{0}(2,0.3){$l$}
\rput[cc]{0}(2.7,1.15){$\widetilde{N^c}$}
\rput[cc]{0}(3.3,0.3){$\tilde{h}$}
\rput[cc]{0}(3.3,2.3){$\tilde{l}$}
\endpspicture\\
\pspicture(0,0)(3.7,2.6)
\rput[cc]{0}(3.5,1.3){$+$}
\endpspicture
\pspicture(-0.5,0)(4.2,2.6)
\psline[linewidth=1pt](0.5,1.3)(0.9,1.3)
\psline[linewidth=1pt](1.7,1.3)(2.3,1.3)
\psarc[linewidth=1pt,linestyle=dashed](1.3,1.3){0.4}{-180}{0}
\psarc[linewidth=1pt](1.3,1.3){0.4}{0}{180}
\psline[linewidth=1pt]{->}(1.32,1.69)(1.42,1.69)
\psline[linewidth=1pt]{<-}(1.18,0.91)(1.28,0.91)
\psline[linewidth=1pt](2.3,1.3)(3.0,0.6)
\psline[linewidth=1pt,linestyle=dashed](2.3,1.3)(3.0,2.0)
\psline[linewidth=1pt]{<-}(2.6,1.0)(2.7,0.9)
\psline[linewidth=1pt]{->}(2.73,1.73)(2.83,1.83)
\rput[cc]{0}(1.3,0.5){$\tilde{l}$}
\rput[cc]{0}(1.3,2.1){$\tilde{h}$}
\rput[cc]{0}(2.05,1.55){$N$}
\rput[cc]{0}(0.2,1.3){$N_j$}
\rput[cc]{0}(3.3,0.6){$\tilde{h}$}
\rput[cc]{0}(3.3,2.0){$\tilde{l}$}
\rput[cc]{0}(4.0,1.3){$+$}
\endpspicture
\pspicture(-0.5,0)(3.5,2.6)
\psline[linewidth=1pt](0.5,1.3)(0.9,1.3)
\psline[linewidth=1pt](1.7,1.3)(2.3,1.3)
\psarc[linewidth=1pt](1.3,1.3){0.4}{-180}{0}
\psarc[linewidth=1pt,linestyle=dashed](1.3,1.3){0.4}{0}{180}
\psline[linewidth=1pt]{<-}(1.18,1.69)(1.28,1.69)
\psline[linewidth=1pt]{<-}(1.18,0.91)(1.28,0.91)
\psline[linewidth=1pt](2.3,1.3)(3.0,0.6)
\psline[linewidth=1pt,linestyle=dashed](2.3,1.3)(3.0,2.0)
\psline[linewidth=1pt]{<-}(2.6,1.0)(2.7,0.9)
\psline[linewidth=1pt]{->}(2.73,1.73)(2.83,1.83)
\rput[cc]{0}(1.3,0.5){$l$}
\rput[cc]{0}(1.3,2){$H_2$}
\rput[cc]{0}(2.05,1.55){$N$}
\rput[cc]{0}(0.2,1.3){$N_j$}
\rput[cc]{0}(3.3,0.6){$\tilde{h}$}
\rput[cc]{0}(3.3,2.0){$\tilde{l}$}
\endpspicture\\[3ex]
\pspicture(0,0)(3.7,2.6)
\psline[linewidth=1pt](0.6,1.3)(1.6,1.3)
\psline[linewidth=1pt](1.6,1.3)(2.3,0.6)
\psline[linewidth=1pt,linestyle=dashed](1.6,1.3)(2.3,2.0)
\psline[linewidth=1pt]{->}(2.03,0.87)(2.13,0.77)
\psline[linewidth=1pt]{->}(2.03,1.73)(2.13,1.83)
\rput[cc]{0}(0.3,1.3){$N_j$}
\rput[cc]{0}(2.6,0.6){$l$}
\rput[cc]{0}(2.6,2.0){$H_2$}
\rput[cc]{0}(3.5,1.3){$+$}
\endpspicture
\pspicture(-0.5,0)(4.2,2.6)
\psline[linewidth=1pt](0.6,1.3)(1.3,1.3)
\psline[linewidth=1pt,linestyle=dashed](1.3,1.3)(2.0,1.3)
\psline[linewidth=1pt](2,1.3)(2.5,1.8)
\psline[linewidth=1pt,linestyle=dashed](2.5,1.8)(3,2.3)
\psline[linewidth=1pt](2,1.3)(3,0.3)
\psarc[linewidth=1pt](2,1.3){0.7}{45}{180}
\psline[linewidth=1pt]{<-}(1.53,1.3)(1.63,1.3)
\psline[linewidth=1pt]{<-}(1.7,1.93)(1.8,1.96)
\psline[linewidth=1pt]{->}(2.75,2.05)(2.85,2.15)
\psline[linewidth=1pt]{->}(2.5,0.8)(2.6,0.7)
\rput[cc]{0}(0.3,1.3){$N_j$}
\rput[cc]{0}(1.65,0.9){$H_2$}
\rput[cc]{0}(2,2.3){$l$}
\rput[cc]{0}(2.6,1.45){$N$}
\rput[cc]{0}(3.3,2.3){$H_2$}
\rput[cc]{0}(3.3,0.3){$l$}
\rput[cc]{0}(4.0,1.3){$+$}
\endpspicture
\pspicture(-0.5,0)(3.5,2.6)
\psline[linewidth=1pt](0.6,1.3)(1.3,1.3)
\psline[linewidth=1pt,linestyle=dashed](1.3,1.3)(2.0,1.3)
\psline[linewidth=1pt,linestyle=dashed](2,1.3)(2.5,0.8)
\psline[linewidth=1pt](2.5,0.8)(3,0.3)
\psline[linewidth=1pt,linestyle=dashed](2,1.3)(3,2.3)
\psarc[linewidth=1pt](2,1.3){0.7}{-180}{-45}
\psline[linewidth=1pt]{<-}(1.53,1.3)(1.63,1.3)
\psline[linewidth=1pt]{->}(1.7,0.69)(1.8,0.64)
\psline[linewidth=1pt]{<-}(2.16,1.14)(2.26,1.04)
\psline[linewidth=1pt]{->}(2.7,0.6)(2.8,0.5)
\psline[linewidth=1pt]{->}(2.5,1.8)(2.6,1.9)
\rput[cc]{0}(0.3,1.3){$N_j$}
\rput[cc]{0}(1.65,1.7){$\tilde{l}$}
\rput[cc]{0}(2,0.3){$\tilde{h}$}
\rput[cc]{0}(2.7,1.15){$\widetilde{N^c}$}
\rput[cc]{0}(3.3,0.3){$l$}
\rput[cc]{0}(3.3,2.3){$H_2$}
\endpspicture\\
\pspicture(0,0)(3.7,2.6)
\rput[cc]{0}(3.5,1.3){$+$}
\endpspicture
\pspicture(-0.5,0)(4.2,2.6)
\psline[linewidth=1pt](0.5,1.3)(0.9,1.3)
\psline[linewidth=1pt](1.7,1.3)(2.3,1.3)
\psarc[linewidth=1pt,linestyle=dashed](1.3,1.3){0.4}{-180}{0}
\psarc[linewidth=1pt](1.3,1.3){0.4}{0}{180}
\psline[linewidth=1pt]{->}(1.32,1.69)(1.42,1.69)
\psline[linewidth=1pt]{<-}(1.18,0.91)(1.28,0.91)
\psline[linewidth=1pt](2.3,1.3)(3.0,0.6)
\psline[linewidth=1pt,linestyle=dashed](2.3,1.3)(3.0,2.0)
\psline[linewidth=1pt]{->}(2.73,0.87)(2.83,0.77)
\psline[linewidth=1pt]{->}(2.73,1.73)(2.83,1.83)
\rput[cc]{0}(1.3,0.5){$\tilde{l}$}
\rput[cc]{0}(1.3,2.1){$\tilde{h}$}
\rput[cc]{0}(2.05,1.55){$N$}
\rput[cc]{0}(0.2,1.3){$N_j$}
\rput[cc]{0}(3.3,0.6){$l$}
\rput[cc]{0}(3.3,2.0){$H_2$}
\rput[cc]{0}(4.0,1.3){$+$}
\endpspicture
\pspicture(-0.5,0)(3.5,2.6)
\psline[linewidth=1pt](0.5,1.3)(0.9,1.3)
\psline[linewidth=1pt](1.7,1.3)(2.3,1.3)
\psarc[linewidth=1pt](1.3,1.3){0.4}{-180}{0}
\psarc[linewidth=1pt,linestyle=dashed](1.3,1.3){0.4}{0}{180}
\psline[linewidth=1pt]{<-}(1.18,1.69)(1.28,1.69)
\psline[linewidth=1pt]{<-}(1.18,0.91)(1.28,0.91)
\psline[linewidth=1pt](2.3,1.3)(3.0,0.6)
\psline[linewidth=1pt,linestyle=dashed](2.3,1.3)(3.0,2.0)
\psline[linewidth=1pt]{->}(2.73,0.87)(2.83,0.77)
\psline[linewidth=1pt]{->}(2.73,1.73)(2.83,1.83)
\rput[cc]{0}(1.3,0.5){$l$}
\rput[cc]{0}(1.3,2){$H_2$}
\rput[cc]{0}(2.05,1.55){$N$}
\rput[cc]{0}(0.2,1.3){$N_j$}
\rput[cc]{0}(3.3,0.6){$l$}
\rput[cc]{0}(3.3,2.0){$H_2$}
\endpspicture\\[3ex]
\pspicture(0,0)(3.7,2.6)
\psline[linewidth=1pt,linestyle=dashed](0.6,1.3)(1.6,1.3)
\psline[linewidth=1pt](1.6,1.3)(2.3,2.0)
\psline[linewidth=1pt](1.6,1.3)(2.3,0.6)
\psline[linewidth=1pt]{<-}(0.84,1.3)(0.94,1.3)
\psline[linewidth=1pt]{<-}(1.9,1.0)(2.0,0.9)
\psline[linewidth=1pt]{->}(2.0,1.7)(2.1,1.8)
\rput[cc]{0}(0.3,1.3){$\sn$}
\rput[cc]{0}(2.6,0.6){$\tilde{h}$}
\rput[cc]{0}(2.6,2.0){$l$}
\rput[cc]{0}(3.5,1.3){$+$}
\endpspicture
\pspicture(-0.5,0)(4.2,2.6)
\psline[linewidth=1pt,linestyle=dashed](0.6,1.3)(1.3,1.3)
\psline[linewidth=1pt,linestyle=dashed](1.3,1.3)(2.0,1.3)
\psline[linewidth=1pt](2,1.3)(2.5,1.8)
\psline[linewidth=1pt](2.5,1.8)(3,2.3)
\psline[linewidth=1pt](2,1.3)(3,0.3)
\psarc[linewidth=1pt,linestyle=dashed](2,1.3){0.7}{45}{180}
\psline[linewidth=1pt]{<-}(0.82,1.3)(0.92,1.3)
\psline[linewidth=1pt]{<-}(1.53,1.3)(1.63,1.3)
\psline[linewidth=1pt]{<-}(1.52,1.81)(1.62,1.88)
\psline[linewidth=1pt]{->}(2.75,2.05)(2.85,2.15)
\psline[linewidth=1pt]{<-}(2.5,0.8)(2.6,0.7)
\rput[cc]{0}(0.3,1.3){$\sn$}
\rput[cc]{0}(1.65,0.9){$\tilde{l}$}
\rput[cc]{0}(2,2.3){$H_2$}
\rput[cc]{0}(2.6,1.45){$N$}
\rput[cc]{0}(3.3,2.3){$l$}
\rput[cc]{0}(3.3,0.3){$\tilde{h}$}
\rput[cc]{0}(4.0,1.3){$+$}
\endpspicture
\pspicture(-0.5,0)(3.5,2.6)
\psline[linewidth=1pt,linestyle=dashed](0.3,1.3)(0.9,1.3)
\psline[linewidth=1pt,linestyle=dashed](1.7,1.3)(2.3,1.3)
\psline[linewidth=1pt]{<-}(1.9,1.3)(2.0,1.3)
\psline[linewidth=1pt]{<-}(0.5,1.3)(0.6,1.3)
\psarc[linewidth=1pt,linestyle=dashed](1.3,1.3){0.4}{-180}{0}
\psarc[linewidth=1pt,linestyle=dashed](1.3,1.3){0.4}{0}{180}
\psline[linewidth=1pt]{<-}(1.18,1.69)(1.28,1.69)
\psline[linewidth=1pt]{<-}(1.18,0.91)(1.28,0.91)
\psline[linewidth=1pt](2.3,1.3)(3.0,0.6)
\psline[linewidth=1pt](2.3,1.3)(3.0,2.0)
\psline[linewidth=1pt]{<-}(2.6,1.0)(2.7,0.9)
\psline[linewidth=1pt]{->}(2.73,1.73)(2.83,1.83)
\rput[cc]{0}(1.3,0.5){$\tilde{l}$}
\rput[cc]{0}(1.3,2){$H_2$}
\rput[cc]{0}(2.05,1.6){$\widetilde{N^c}$}
\rput[cc]{0}(0,1.3){$\snj$}
\rput[cc]{0}(3.3,0.6){$\tilde{h}$}
\rput[cc]{0}(3.3,2.0){$l$}
\endpspicture\\[3ex]
\pspicture(0,0)(3.7,2.6)
\psline[linewidth=1pt,linestyle=dashed](0.6,1.3)(1.6,1.3)
\psline[linewidth=1pt,linestyle=dashed](1.6,1.3)(2.3,2.0)
\psline[linewidth=1pt,linestyle=dashed](1.6,1.3)(2.3,0.6)
\psline[linewidth=1pt]{->}(1.0,1.3)(1.1,1.3)
\psline[linewidth=1pt]{->}(2.04,1.74)(2.14,1.84)
\psline[linewidth=1pt]{->}(2.04,0.86)(2.14,0.76)
\rput[cc]{0}(0.3,1.3){$\sn$}
\rput[cc]{0}(2.6,2.0){$\tilde{l}$}
\rput[cc]{0}(2.6,0.6){$H_2$}
\rput[cc]{0}(3.5,1.3){$+$}
\endpspicture
\pspicture(-0.5,0)(4.2,2.6)
\psline[linewidth=1pt,linestyle=dashed](0.6,1.3)(1.3,1.3)
\psline[linewidth=1pt](1.3,1.3)(2.0,1.3)
\psline[linewidth=1pt](2,1.3)(2.5,1.8)
\psline[linewidth=1pt,linestyle=dashed](2.5,1.8)(3,2.3)
\psline[linewidth=1pt,linestyle=dashed](2,1.3)(3,0.3)
\psarc[linewidth=1pt](2,1.3){0.7}{45}{180}
\psline[linewidth=1pt]{->}(0.97,1.3)(1.07,1.3)
\psline[linewidth=1pt]{<-}(1.53,1.3)(1.63,1.3)
\psline[linewidth=1pt]{->}(1.7,1.93)(1.8,1.99)
\psline[linewidth=1pt]{->}(2.75,2.05)(2.85,2.15)
\psline[linewidth=1pt]{->}(2.5,0.8)(2.6,0.7)
\rput[cc]{0}(0.3,1.3){$\sn$}
\rput[cc]{0}(1.65,0.9){$l$}
\rput[cc]{0}(2,2.3){$\tilde{h}$}
\rput[cc]{0}(2.6,1.45){$N$}
\rput[cc]{0}(3.3,2.3){$\tilde{l}$}
\rput[cc]{0}(3.3,0.3){$H_2$}
\rput[cc]{0}(4.0,1.3){$+$}
\endpspicture
\pspicture(-0.5,0)(3.5,2.6)
\psline[linewidth=1pt,linestyle=dashed](0.3,1.3)(0.9,1.3)
\psline[linewidth=1pt,linestyle=dashed](1.7,1.3)(2.3,1.3)
\psline[linewidth=1pt]{->}(2.0,1.3)(2.1,1.3)
\psline[linewidth=1pt]{->}(0.6,1.3)(0.7,1.3)
\psarc[linewidth=1pt](1.3,1.3){0.4}{-180}{0}
\psarc[linewidth=1pt](1.3,1.3){0.4}{0}{180}
\psline[linewidth=1pt]{->}(1.32,0.91)(1.42,0.91)
\psline[linewidth=1pt]{<-}(1.18,1.69)(1.28,1.69)
\psline[linewidth=1pt,linestyle=dashed](2.3,1.3)(3.0,0.6)
\psline[linewidth=1pt,linestyle=dashed](2.3,1.3)(3.0,2.0)
\psline[linewidth=1pt]{->}(2.74,0.86)(2.84,0.76)
\psline[linewidth=1pt]{->}(2.73,1.73)(2.83,1.83)
\rput[cc]{0}(1.3,0.5){$\tilde{h}$}
\rput[cc]{0}(1.3,2){$l$}
\rput[cc]{0}(2.05,1.65){$\widetilde{N^c}$}
\rput[cc]{0}(0,1.3){$\snj$}
\rput[cc]{0}(3.3,0.6){$H_2$}
\rput[cc]{0}(3.3,2.0){$\tilde{l}$}
\endpspicture
}
\end{center}

%% file: Fig02.tex
\begin{center}
\parbox[c]{12cm}{
\pspicture(-0.6,0)(3,2)
\psline[linewidth=1pt,linestyle=dashed](0.6,1)(2.6,1)
\psline[linewidth=1pt,linestyle=dashed](1.6,1)(2.3,1.7)
\psline[linewidth=1pt,linestyle=dashed](1.6,1)(2.3,0.3)
\psline[linewidth=1pt]{->}(1.05,1)(1.15,1)
\psline[linewidth=1pt]{->}(2.25,1)(2.35,1)
\psline[linewidth=1pt]{<-}(1.95,1.35)(2.05,1.45)
\psline[linewidth=1pt]{->}(2.04,0.56)(2.14,0.46)
\rput[cc]{0}(-0.6,1){(a)}
\rput[cc]{0}(0.3,1){$\sn$}
\rput[cc]{0}(2.6,1.7){$\tilde{l}$}
\rput[cc]{0}(3.0,1){$\sur$}
\rput[cc]{0}(2.6,0.3){$\tilde{q}$}
\endpspicture\hspace{\fill}
\pspicture(-0.6,0)(2.5,2)
\psline[linewidth=1pt,linestyle=dashed](0.6,0.3)(1.3,1)
\psline[linewidth=1pt,linestyle=dashed](0.6,1.7)(1.3,1)
\psline[linewidth=1pt,linestyle=dashed](1.3,1)(2,1.7)
\psline[linewidth=1pt,linestyle=dashed](1.3,1)(2,0.3)
\psline[linewidth=1pt]{->}(0.83,0.53)(0.93,0.63)
\psline[linewidth=1pt]{->}(0.83,1.47)(0.93,1.37)
\psline[linewidth=1pt]{->}(1.72,1.42)(1.82,1.52)
\psline[linewidth=1pt]{->}(1.72,0.58)(1.82,0.48)
\rput[cc]{0}(-0.6,1){(b)}
\rput[cc]{0}(0.3,1.7){$\sn$}
\rput[cc]{0}(0.3,0.3){$\tilde{l}$}
\rput[cc]{0}(2.3,0.3){$\sur$}
\rput[cc]{0}(2.3,1.7){$\tilde{q}$}
\endpspicture
}
\end{center}

%% file: Fig03.tex
\begin{center}
\parbox[c]{11.5cm}{
\pspicture(-2,0)(3.5,2)
\rput[rc]{0}(-1,1){$\g_{\scr N}^{(1)}$:}
\psline[linewidth=1pt](0.6,0.3)(1.3,1)
\psline[linewidth=1pt,linestyle=dashed](0.6,1.7)(1.3,1)
\psline[linewidth=1pt](1.3,1)(2.3,1)
\psline[linewidth=1pt,linestyle=dashed](2.3,1)(3,1.7)
\psline[linewidth=1pt](2.3,1)(3,0.3)
\psline[linewidth=1pt]{<-}(0.8,0.5)(0.9,0.6)
\psline[linewidth=1pt]{->}(0.85,1.45)(0.95,1.35)
\psline[linewidth=1pt]{<-}(2.65,1.35)(2.75,1.45)
\psline[linewidth=1pt]{->}(2.7,0.6)(2.8,0.5)
\rput[cc]{0}(0.3,1.7){$\widetilde{l}$}
\rput[cc]{0}(0.3,0.3){$\widetilde{h}$}
\rput[cc]{0}(1.8,1.3){$N$}
\rput[cc]{0}(3.3,0.3){$\widetilde{h}$}
\rput[cc]{0}(3.3,1.7){$\widetilde{l}$}
\rput[cc]{0}(4.0,1.0){$+$}
%\rput[cc]{0}(6.5,0.9){$N.1.1$}
\endpspicture\hspace{5ex}
\pspicture(0,-0.1)(3.5,1.8)
\psline[linewidth=1pt,linestyle=dashed](0.6,1.5)(1.8,1.5)
\psline[linewidth=1pt](1.8,1.5)(3,1.5)
\psline[linewidth=1pt](1.8,1.5)(1.8,0.3)
\psline[linewidth=1pt](0.6,0.3)(1.8,0.3)
\psline[linewidth=1pt,linestyle=dashed](1.8,0.3)(3,0.3)
\psline[linewidth=1pt]{->}(1.2,1.5)(1.3,1.5)
\psline[linewidth=1pt]{->}(2.42,1.5)(2.52,1.5)
\psline[linewidth=1pt]{<-}(1.05,0.3)(1.15,0.3)
\psline[linewidth=1pt]{<-}(2.3,0.3)(2.4,0.3)
\rput[cc]{0}(0.3,1.5){$\widetilde{l}$}
\rput[cc]{0}(0.3,0.3){$\widetilde{h}$}
\rput[cc]{0}(2.1,0.9){$N$}
\rput[cc]{0}(3.3,1.5){$\widetilde{h}$}
\rput[cc]{0}(3.3,0.3){$\widetilde{l}$}
%\rput[cc]{0}(6.5,0.9){$N.1.2$}
\endpspicture\\[2ex]
\pspicture(-2,0)(3.5,2)
\rput[rc]{0}(-1,1){$\g_{\scr N}^{(2)}$:}
\psline[linewidth=1pt,linestyle=dashed](0.6,0.3)(1.3,1)
\psline[linewidth=1pt](0.6,1.7)(1.3,1)
\psline[linewidth=1pt](1.3,1)(2.3,1)
\psline[linewidth=1pt](2.3,1)(3,1.7)
\psline[linewidth=1pt,linestyle=dashed](2.3,1)(3,0.3)
\psline[linewidth=1pt]{->}(0.85,0.55)(0.95,0.65)
\psline[linewidth=1pt]{->}(0.85,1.45)(0.95,1.35)
\psline[linewidth=1pt]{<-}(2.65,1.35)(2.75,1.45)
\psline[linewidth=1pt]{<-}(2.65,0.65)(2.75,0.55)
\rput[cc]{0}(0.3,1.7){$l$}
\rput[cc]{0}(0.3,0.3){$H_2$}
\rput[cc]{0}(1.8,1.3){$N$}
\rput[cc]{0}(3.3,0.3){$H_2$}
\rput[cc]{0}(3.3,1.7){$l$}
\rput[cc]{0}(4.0,1.0){$+$}
\endpspicture
\hspace{5ex}
\pspicture(0,-0.1)(3.5,1.8)
\psline[linewidth=1pt](0.6,1.5)(1.8,1.5)
\psline[linewidth=1pt,linestyle=dashed](1.8,1.5)(3,1.5)
\psline[linewidth=1pt](1.8,1.5)(1.8,0.3)
\psline[linewidth=1pt,linestyle=dashed](0.6,0.3)(1.8,0.3)
\psline[linewidth=1pt](1.8,0.3)(3,0.3)
\psline[linewidth=1pt]{->}(1.2,1.5)(1.3,1.5)
\psline[linewidth=1pt]{<-}(2.3,1.5)(2.4,1.5)
\psline[linewidth=1pt]{->}(1.2,0.3)(1.3,0.3)
\psline[linewidth=1pt]{<-}(2.3,0.3)(2.4,0.3)
\rput[cc]{0}(0.3,1.5){$l$}
\rput[cc]{0}(0.3,0.3){$H_2$}
\rput[cc]{0}(2.1,0.9){$N$}
\rput[cc]{0}(3.3,1.5){$H_2$}
\rput[cc]{0}(3.3,0.3){$l$}
\endpspicture\\[2ex]
\pspicture(-2,0)(3.5,2)
\rput[rc]{0}(-1,1){$\g_{\scr N}^{(3)}$:}
\psline[linewidth=1pt](0.6,0.3)(1.3,1)
\psline[linewidth=1pt,linestyle=dashed](0.6,1.7)(1.3,1)
\psline[linewidth=1pt](1.3,1)(2.3,1)
\psline[linewidth=1pt](2.3,1)(3,1.7)
\psline[linewidth=1pt,linestyle=dashed](2.3,1)(3,0.3)
\psline[linewidth=1pt]{<-}(0.8,0.5)(0.9,0.6)
\psline[linewidth=1pt]{->}(0.85,1.45)(0.95,1.35)
\psline[linewidth=1pt]{<-}(2.65,1.35)(2.75,1.45)
\psline[linewidth=1pt]{<-}(2.65,0.65)(2.75,0.55)
\rput[cc]{0}(0.3,1.7){$\widetilde{l}$}
\rput[cc]{0}(0.3,0.3){$\widetilde{h}$}
\rput[cc]{0}(1.8,1.3){$N$}
\rput[cc]{0}(3.3,0.3){$H_2$}
\rput[cc]{0}(3.3,1.7){$l$}
\rput[cc]{0}(4.0,1.0){$+$}
%\rput[cc]{0}(6.5,0.9){$N.3.1$}
\endpspicture\hspace{5ex}
\pspicture(0,-0.1)(3.5,1.8)
\psline[linewidth=1pt,linestyle=dashed](0.6,1.5)(1.8,1.5)
\psline[linewidth=1pt,linestyle=dashed](1.8,1.5)(3,1.5)
\psline[linewidth=1pt,linestyle=dashed](1.8,1.5)(1.8,0.3)
\psline[linewidth=1pt](0.6,0.3)(1.8,0.3)
\psline[linewidth=1pt](1.8,0.3)(3,0.3)
\psline[linewidth=1pt]{->}(1.2,1.5)(1.3,1.5)
\psline[linewidth=1pt]{<-}(2.3,1.5)(2.4,1.5)
\psline[linewidth=1pt]{->}(1.8,0.86)(1.8,0.76)
\psline[linewidth=1pt]{<-}(1.05,0.3)(1.15,0.3)
\psline[linewidth=1pt]{<-}(2.3,0.3)(2.4,0.3)
\rput[cc]{0}(0.3,1.5){$\widetilde{l}$}
\rput[cc]{0}(0.3,0.3){$\widetilde{h}$}
\rput[cc]{0}(2.15,0.9){$\widetilde{N^c}$}
\rput[cc]{0}(3.3,1.5){$H_2$}
\rput[cc]{0}(3.3,0.3){$l$}
%\rput[cc]{0}(6.5,0.9){$N.3.2$}
\endpspicture\\[2ex]
\pspicture(-2,0)(3.5,2)
\rput[rc]{0}(-1,1){$\g_{\scr N}^{(4)}$:}
\psline[linewidth=1pt](0.6,0.3)(1.3,1)
\psline[linewidth=1pt](0.6,1.7)(1.3,1)
\psline[linewidth=1pt,linestyle=dashed](1.3,1)(2.3,1)
\psline[linewidth=1pt,linestyle=dashed](2.3,1)(3,1.7)
\psline[linewidth=1pt,linestyle=dashed](2.3,1)(3,0.3)
\psline[linewidth=1pt]{<-}(0.8,0.5)(0.9,0.6)
\psline[linewidth=1pt]{->}(0.85,1.45)(0.95,1.35)
\psline[linewidth=1pt]{<-}(1.53,1)(1.63,1)
\psline[linewidth=1pt]{<-}(2.65,1.35)(2.75,1.45)
\psline[linewidth=1pt]{<-}(2.65,0.65)(2.75,0.55)
\rput[cc]{0}(0.3,1.7){$l$}
\rput[cc]{0}(0.3,0.3){$\widetilde{h}$}
\rput[cc]{0}(1.8,1.35){$\widetilde{N^c}$}
\rput[cc]{0}(3.3,0.3){$H_2$}
\rput[cc]{0}(3.3,1.7){$\widetilde{l}$}
\rput[cc]{0}(4.0,1.0){$+$}
%\rput[cc]{0}(6.5,0.9){$N.4.1$}
\endpspicture\hspace{5ex}
\pspicture(0,-0.1)(3.5,1.8)
\psline[linewidth=1pt](0.6,1.5)(1.8,1.5)
\psline[linewidth=1pt,linestyle=dashed](1.8,1.5)(3,1.5)
\psline[linewidth=1pt](1.8,1.5)(1.8,0.3)
\psline[linewidth=1pt](0.6,0.3)(1.8,0.3)
\psline[linewidth=1pt,linestyle=dashed](1.8,0.3)(3,0.3)
\psline[linewidth=1pt]{->}(1.2,1.5)(1.3,1.5)
\psline[linewidth=1pt]{<-}(2.3,1.5)(2.4,1.5)
\psline[linewidth=1pt]{<-}(1.05,0.3)(1.15,0.3)
\psline[linewidth=1pt]{<-}(2.3,0.3)(2.4,0.3)
\rput[cc]{0}(0.3,1.5){$l$}
\rput[cc]{0}(0.3,0.3){$\widetilde{h}$}
\rput[cc]{0}(2.1,0.9){$N$}
\rput[cc]{0}(3.3,1.5){$H_2$}
\rput[cc]{0}(3.3,0.3){$\widetilde{l}$}
%\rput[cc]{0}(6.5,0.9){$N.4.2$}
\endpspicture\\[2ex]
\pspicture(-2,0)(4.0,2)
\rput[rc]{0}(-1,1){$\g_{\scr N}^{(5)}$:}
\psline[linewidth=1pt,linestyle=dashed](0.6,0.3)(1.3,1)
\psline[linewidth=1pt,linestyle=dashed](0.6,1.7)(1.3,1)
\psline[linewidth=1pt,linestyle=dashed](1.3,1)(2.3,1)
\psline[linewidth=1pt,linestyle=dashed](2.3,1)(3.3,1)
\psline[linewidth=1pt,linestyle=dashed](2.3,1)(3,1.7)
\psline[linewidth=1pt,linestyle=dashed](2.3,1)(3,0.3)
\psline[linewidth=1pt]{->}(0.85,0.55)(0.95,0.65)
\psline[linewidth=1pt]{->}(0.85,1.45)(0.95,1.35)
\psline[linewidth=1pt]{->}(1.7,1)(1.8,1)
\psline[linewidth=1pt]{<-}(2.65,1.35)(2.75,1.45)
\psline[linewidth=1pt]{->}(2.95,1)(3.05,1)
\psline[linewidth=1pt]{->}(2.73,0.57)(2.83,0.47)
\rput[cc]{0}(0.3,1.7){$\widetilde{l}$}
\rput[cc]{0}(0.3,0.3){$H_2$}
\rput[cc]{0}(1.8,1.4){$\widetilde{N^c}$}
\rput[cc]{0}(3.3,0.3){$\widetilde{q}$}
\rput[cc]{0}(3.3,1.7){$\widetilde{l}$}
\rput[cc]{0}(3.6,1){$\widetilde{U^c}$}
\rput[cc]{0}(4.5,1.0){$+$}
%\rput[cc]{0}(6.5,1){$N.5.1$}
\endpspicture\hspace{5ex}
\pspicture(0,0)(3.5,2.1)
\psline[linewidth=1pt,linestyle=dashed](0.6,1.8)(1.8,1.8)
\psline[linewidth=1pt,linestyle=dashed](1.8,1.8)(3,1.8)
\psline[linewidth=1pt,linestyle=dashed](1.8,1.8)(1.8,0.6)
\psline[linewidth=1pt,linestyle=dashed](0.6,0.6)(1.8,0.6)
\psline[linewidth=1pt,linestyle=dashed](1.8,0.6)(3,0.9)
\psline[linewidth=1pt,linestyle=dashed](1.8,0.6)(3,0.3)
\psline[linewidth=1pt]{->}(1.22,1.8)(1.32,1.8)
%\psline[linewidth=1pt]{<-}(1.06,1.8)(1.16,1.8)
%\psline[linewidth=1pt]{->}(2.44,1.8)(2.54,1.8)
\psline[linewidth=1pt]{<-}(2.28,1.8)(2.38,1.8)
%\psline[linewidth=1pt]{->}(1.8,1.2)(1.8,1.3)
\psline[linewidth=1pt]{<-}(1.8,1.08)(1.8,1.18)
\psline[linewidth=1pt]{->}(1.22,0.6)(1.32,0.6)
%\psline[linewidth=1pt]{<-}(1.06,0.6)(1.16,0.6)
\psline[linewidth=1pt]{->}(2.42,0.45)(2.52,0.42)
%\psline[linewidth=1pt]{<-}(2.3,0.48)(2.4,0.45)
\psline[linewidth=1pt]{->}(2.42,0.75)(2.52,0.78)
%\psline[linewidth=1pt]{<-}(2.3,0.72)(2.4,0.75)
\rput[cc]{0}(0.3,1.8){$H_2$}
\rput[cc]{0}(0.3,0.6){$\widetilde{l}$}
\rput[cc]{0}(1.4,1.2){$\widetilde{N^c}$}
\rput[cc]{0}(3.3,1.8){$\widetilde{l}$}
\rput[cc]{0}(3.3,0.9){$\widetilde{U^c}$}
\rput[cc]{0}(3.3,0.3){$\widetilde{q}$}
%\rput[cc]{0}(6.5,1.2){$N.5.2$}
\endpspicture
}
\parbox[c]{14cm}{
\pspicture(-2,0)(3.5,2)
\rput[rc]{0}(-1,1){$\g_{\scr N}^{(6)}$:}
\psline[linewidth=1pt,linestyle=dashed](0.6,0.3)(1.3,1)
\psline[linewidth=1pt](0.6,1.7)(1.3,1)
\psline[linewidth=1pt](1.3,1)(2.3,1)
\psline[linewidth=1pt,linestyle=dashed](2.3,1)(3,1.7)
\psline[linewidth=1pt](2.3,1)(3,0.3)
\psline[linewidth=1pt]{->}(0.85,0.55)(0.95,0.65)
\psline[linewidth=1pt]{->}(0.85,1.45)(0.95,1.35)
\psline[linewidth=1pt]{->}(2.75,1.45)(2.85,1.55)
\psline[linewidth=1pt]{<-}(2.65,0.65)(2.75,0.55)
\rput[cc]{0}(0.3,1.7){$l$}
\rput[cc]{0}(0.3,0.3){$H_2$}
\rput[cc]{0}(1.8,1.3){$N$}
\rput[cc]{0}(3.3,0.3){$\widetilde{h}$}
\rput[cc]{0}(3.3,1.7){$\widetilde{l}$}
%\rput[cc]{0}(6.5,0.9){$N.6$}
\endpspicture\hspace{10ex}
\pspicture(-2,0)(4.0,2)
\rput[rc]{0}(-1,1){$\g_{\scr N}^{(7)}$:}
\psline[linewidth=1pt](0.6,0.3)(1.3,1)
\psline[linewidth=1pt](0.6,1.7)(1.3,1)
\psline[linewidth=1pt,linestyle=dashed](1.3,1)(2.3,1)
\psline[linewidth=1pt,linestyle=dashed](2.3,1)(3.3,1)
\psline[linewidth=1pt,linestyle=dashed](2.3,1)(3,1.7)
\psline[linewidth=1pt,linestyle=dashed](2.3,1)(3,0.3)
\psline[linewidth=1pt]{<-}(0.8,0.5)(0.9,0.6)
\psline[linewidth=1pt]{->}(0.85,1.45)(0.95,1.35)
\psline[linewidth=1pt]{<-}(1.55,1)(1.65,1)
\psline[linewidth=1pt]{<-}(2.8,1)(2.9,1)
\psline[linewidth=1pt]{->}(2.73,1.43)(2.83,1.53)
\psline[linewidth=1pt]{<-}(2.65,0.65)(2.75,0.55)
\rput[cc]{0}(0.3,1.7){$l$}
\rput[cc]{0}(0.3,0.3){$\widetilde{h}$}
\rput[cc]{0}(1.8,1.4){$\widetilde{N^c}$}
\rput[cc]{0}(3.3,0.3){$\widetilde{U^c}$}
\rput[cc]{0}(3.3,1.7){$\widetilde{l}$}
\rput[cc]{0}(3.6,1){$\widetilde{q}$}
%\rput[cc]{0}(6.5,1){$N.7$}
\endpspicture
}
\end{center}

%% file: Fig04.tex
\begin{center}
\parbox[c]{11.5cm}{
\pspicture(-2,0)(4.0,2.3)
\rput[rc]{0}(-1,1.3){$\g_{\scr N}^{(8)}$:}
\psline[linewidth=1pt,linestyle=dashed](0.6,0.6)(1.8,1.3)
\psline[linewidth=1pt,linestyle=dashed](0.6,2)(1.8,1.3)
\psline[linewidth=1pt,linestyle=dashed](1.8,1.3)(3,2)
\psline[linewidth=1pt,linestyle=dashed](1.8,1.3)(2.4,0.95)
\psline[linewidth=1pt,linestyle=dashed](2.4,0.95)(3.17,0.9)
\psline[linewidth=1pt,linestyle=dashed](2.4,0.95)(2.83,0.3)
\psline[linewidth=1pt]{->}(1.2,0.95)(1.3,1.01)
\psline[linewidth=1pt]{->}(1.2,1.65)(1.3,1.59)
\psline[linewidth=1pt]{->}(2.4,1.65)(2.5,1.71)
\psline[linewidth=1pt]{->}(2.1,1.13)(2.2,1.07)
\psline[linewidth=1pt]{->}(2.8,0.92)(2.9,0.91)
\psline[linewidth=1pt]{->}(2.65,0.57)(2.68,0.52)
\rput[cc]{0}(0.3,2){$\widetilde{q}$}
\rput[cc]{0}(0.3,0.6){$\widetilde{U^c}$}
\rput[cc]{0}(3.3,2){$\widetilde{l}$}
\rput[cc]{0}(1.9,0.8){$\widetilde{N^c}$}
\rput[cc]{0}(3.47,0.9){$\widetilde{l}$}
\rput[cc]{0}(3.13,0.3){$H_2$}
\rput[cc]{0}(4.5,1.3){$+$}
%\rput[cc]{0}(6.5,1.3){$N.8.1$}
\endpspicture\hspace{5ex}
\pspicture(0,0)(4.0,2.3)
\psline[linewidth=1pt,linestyle=dashed](0.6,0.6)(1.8,1.3)
\psline[linewidth=1pt,linestyle=dashed](0.6,2)(1.8,1.3)
\psline[linewidth=1pt,linestyle=dashed](1.8,1.3)(2.4,0.8)
\psline[linewidth=1pt,linestyle=dashed](2.4,0.8)(3.0,0.1)
\psline[linewidth=1pt,linestyle=dashed](2.4,0.8)(3.0,2.0)
\psline[linewidth=1pt,linestyle=dashed](1.8,1.3)(3.2,1.3)
\psline[linewidth=1pt]{->}(1.2,0.95)(1.3,1.01)
\psline[linewidth=1pt]{->}(1.2,1.65)(1.3,1.59)
\psline[linewidth=1pt]{->}(2.22,1.3)(2.32,1.3)
\psline[linewidth=1pt]{->}(2.83,1.66)(2.88,1.76)
\psline[linewidth=1pt]{->}(2.1,1.05)(2.2,0.97)
\psline[linewidth=1pt]{->}(2.76,0.38)(2.86,0.28)
\rput[cc]{0}(0.3,2){$\widetilde{q}$}
\rput[cc]{0}(0.3,0.6){$\widetilde{U^c}$}
\rput[cc]{0}(3.2,2){$\widetilde{l}$}
\rput[cc]{0}(1.9,0.7){$\widetilde{N^c}$}
\rput[cc]{0}(3.47,1.3){$\widetilde{l}$}
\rput[cc]{0}(3.3,0.1){$H_2$}
%\rput[cc]{0}(6.5,1.3){$N.8.1$}
\endpspicture\\[2ex]
\pspicture(-2,0)(4.0,2.3)
\rput[rc]{0}(-1,1.3){$\g_{\scr N}^{(10)}$:}
\psline[linewidth=1pt,linestyle=dashed](0.6,0.6)(1.8,1.3)
\psline[linewidth=1pt,linestyle=dashed](0.6,2)(1.8,1.3)
\psline[linewidth=1pt,linestyle=dashed](1.8,1.3)(3,2)
\psline[linewidth=1pt,linestyle=dashed](1.8,1.3)(2.4,0.95)
\psline[linewidth=1pt,linestyle=dashed](2.4,0.95)(3.17,0.9)
\psline[linewidth=1pt,linestyle=dashed](2.4,0.95)(2.83,0.3)
\psline[linewidth=1pt]{->}(1.2,0.95)(1.3,1.01)
\psline[linewidth=1pt]{<-}(1.1,1.71)(1.2,1.65)
\psline[linewidth=1pt]{<-}(2.31,1.6)(2.41,1.66)
\psline[linewidth=1pt]{->}(2.1,1.13)(2.2,1.07)
\psline[linewidth=1pt]{->}(2.8,0.92)(2.9,0.91)
\psline[linewidth=1pt]{->}(2.65,0.57)(2.68,0.52)
\rput[cc]{0}(0.3,2){$\widetilde{l}$}
\rput[cc]{0}(0.3,0.6){$\widetilde{q}$}
\rput[cc]{0}(3.3,2){$\widetilde{U^c}$}
\rput[cc]{0}(1.9,0.8){$\widetilde{N^c}$}
\rput[cc]{0}(3.47,0.9){$\widetilde{l}$}
\rput[cc]{0}(3.13,0.3){$H_2$}
\rput[cc]{0}(4.5,1.3){$+$}
%\rput[cc]{0}(6.5,1.3){$N.10.1$}
\endpspicture\hspace{5ex}
\pspicture(0,0)(3.5,2.1)
\psline[linewidth=1pt,linestyle=dashed](0.6,1.8)(1.8,1.8)
\psline[linewidth=1pt,linestyle=dashed](1.8,1.8)(3,1.8)
\psline[linewidth=1pt,linestyle=dashed](1.8,1.8)(1.8,0.6)
\psline[linewidth=1pt,linestyle=dashed](0.6,0.6)(1.8,0.6)
\psline[linewidth=1pt,linestyle=dashed](1.8,0.6)(3,0.9)
\psline[linewidth=1pt,linestyle=dashed](1.8,0.6)(3,0.3)
%\psline[linewidth=1pt]{->}(1.22,1.8)(1.32,1.8)
\psline[linewidth=1pt]{<-}(1.06,1.8)(1.16,1.8)
\psline[linewidth=1pt]{->}(2.44,1.8)(2.54,1.8)
%\psline[linewidth=1pt]{<-}(2.28,1.8)(2.38,1.8)
\psline[linewidth=1pt]{->}(1.8,1.2)(1.8,1.3)
%\psline[linewidth=1pt]{<-}(1.8,1.08)(1.8,1.18)
\psline[linewidth=1pt]{->}(1.22,0.6)(1.32,0.6)
%\psline[linewidth=1pt]{<-}(1.06,0.6)(1.16,0.6)
\psline[linewidth=1pt]{->}(2.42,0.45)(2.52,0.42)
%\psline[linewidth=1pt]{<-}(2.3,0.48)(2.4,0.45)
%\psline[linewidth=1pt]{->}(2.42,0.75)(2.52,0.78)
\psline[linewidth=1pt]{<-}(2.3,0.72)(2.4,0.75)
\rput[cc]{0}(0.3,1.8){$\widetilde{l}$}
\rput[cc]{0}(0.3,0.6){$\widetilde{q}$}
\rput[cc]{0}(1.4,1.2){$\widetilde{N^c}$}
\rput[cc]{0}(3.3,1.8){$H_2$}
\rput[cc]{0}(3.3,0.9){$\widetilde{U^c}$}
\rput[cc]{0}(3.3,0.3){$\widetilde{l}$}
%\rput[cc]{0}(6.5,1.2){$N.10.2$}
\endpspicture
}
\parbox[c]{14cm}{
\pspicture(-2,0)(3.5,2.3)
\rput[rc]{0}(-1,1){$\g_{\scr N}^{(9)}$:}
\psline[linewidth=1pt,linestyle=dashed](0.6,0.6)(1.8,1.3)
\psline[linewidth=1pt,linestyle=dashed](0.6,2)(1.8,1.3)
\psline[linewidth=1pt,linestyle=dashed](1.8,1.3)(3,2)
\psline[linewidth=1pt,linestyle=dashed](1.8,1.3)(2.4,0.95)
\psline[linewidth=1pt](2.4,0.95)(3.17,0.9)
\psline[linewidth=1pt](2.4,0.95)(2.83,0.3)
\psline[linewidth=1pt]{->}(1.2,0.95)(1.3,1.01)
\psline[linewidth=1pt]{->}(1.2,1.65)(1.3,1.59)
\psline[linewidth=1pt]{->}(2.4,1.65)(2.5,1.71)
\psline[linewidth=1pt]{->}(2.1,1.13)(2.2,1.07)
\psline[linewidth=1pt]{->}(2.8,0.92)(2.9,0.91)
\psline[linewidth=1pt]{<-}(2.55,0.7)(2.58,0.66)
\rput[cc]{0}(0.3,2){$\widetilde{q}$}
\rput[cc]{0}(0.3,0.6){$\widetilde{U^c}$}
\rput[cc]{0}(3.3,2){$\widetilde{l}$}
\rput[cc]{0}(1.9,0.8){$\widetilde{N^c}$}
\rput[cc]{0}(3.47,0.9){$\widetilde{h}$}
\rput[cc]{0}(3.13,0.3){$l$}
%\rput[cc]{0}(6.5,1.3){$N.9$}
\endpspicture\hspace{\fill}
\pspicture(-2,0)(4.0,2.3)
\rput[rc]{0}(-1,1){$\g_{\scr N}^{(11)}$:}
\psline[linewidth=1pt,linestyle=dashed](0.6,0.6)(1.8,1.3)
\psline[linewidth=1pt,linestyle=dashed](0.6,2)(1.8,1.3)
\psline[linewidth=1pt,linestyle=dashed](1.8,1.3)(3,2)
\psline[linewidth=1pt,linestyle=dashed](1.8,1.3)(2.4,0.95)
\psline[linewidth=1pt](2.4,0.95)(3.17,0.9)
\psline[linewidth=1pt](2.4,0.95)(2.83,0.3)
\psline[linewidth=1pt]{->}(1.2,0.95)(1.3,1.01)
\psline[linewidth=1pt]{<-}(1.1,1.71)(1.2,1.65)
\psline[linewidth=1pt]{<-}(2.31,1.6)(2.41,1.66)
\psline[linewidth=1pt]{->}(2.1,1.13)(2.2,1.07)
\psline[linewidth=1pt]{->}(2.8,0.92)(2.9,0.91)
\psline[linewidth=1pt]{<-}(2.55,0.7)(2.58,0.66)
\rput[cc]{0}(0.3,2){$\widetilde{l}$}
\rput[cc]{0}(0.3,0.6){$\widetilde{q}$}
\rput[cc]{0}(3.3,2){$\widetilde{U^c}$}
\rput[cc]{0}(1.9,0.8){$\widetilde{N^c}$}
\rput[cc]{0}(3.47,0.9){$\widetilde{h}$}
\rput[cc]{0}(3.13,0.3){$l$}
%\rput[cc]{0}(6.5,1.3){$N.11$}
\endpspicture
}
\end{center}

%% file: Fig05.tex
\begin{center}
\parbox[c]{11.5cm}{
\pspicture(-2,0)(3.5,1.8)
\rput[rc]{0}(-1,1){$\g_{\scr N}^{(12)}$:}
\psline[linewidth=1pt](0.6,1.5)(1.8,1.5)
\psline[linewidth=1pt,linestyle=dashed](1.8,1.5)(3,1.5)
\psline[linewidth=1pt](1.8,1.5)(1.8,0.3)
\psline[linewidth=1pt](0.6,0.3)(1.8,0.3)
\psline[linewidth=1pt,linestyle=dashed](1.8,0.3)(3,0.3)
\psline[linewidth=1pt]{<-}(1.05,1.5)(1.15,1.5)
\psline[linewidth=1pt]{<-}(2.3,1.5)(2.4,1.5)
\psline[linewidth=1pt]{<-}(1.05,0.3)(1.15,0.3)
\psline[linewidth=1pt]{<-}(2.3,0.3)(2.4,0.3)
\rput[cc]{0}(0.3,1.5){$\widetilde{h}$}
\rput[cc]{0}(0.3,0.3){$\widetilde{h}$}
\rput[cc]{0}(2.1,0.9){$N$}
\rput[cc]{0}(3.3,1.5){$\widetilde{l}$}
\rput[cc]{0}(3.3,0.3){$\widetilde{l}$}
\rput[cc]{0}(4.0,1.0){$+$}
%\rput[cc]{0}(6.5,0.9){$N.12.1$}
\endpspicture\hspace{5ex}
\pspicture(0,0)(3.5,1.8)
\psline[linewidth=1pt](0.6,1.5)(1.8,1.5)
\psline[linewidth=1pt,linestyle=dashed](1.8,1.5)(3,0.3)
\psline[linewidth=1pt](1.8,1.5)(1.8,0.3)
\psline[linewidth=1pt](0.6,0.3)(1.8,0.3)
\psline[linewidth=1pt,linestyle=dashed](1.8,0.3)(3,1.5)
\psline[linewidth=1pt]{<-}(1.05,1.5)(1.15,1.5)
\psline[linewidth=1pt]{<-}(2.62,1.12)(2.72,1.22)
\psline[linewidth=1pt]{<-}(1.05,0.3)(1.15,0.3)
\psline[linewidth=1pt]{<-}(2.62,0.68)(2.72,0.58)
\rput[cc]{0}(0.3,1.5){$\widetilde{h}$}
\rput[cc]{0}(0.3,0.3){$\widetilde{h}$}
\rput[cc]{0}(1.5,0.9){$N$}
\rput[cc]{0}(3.3,1.5){$\widetilde{l}$}
\rput[cc]{0}(3.3,0.3){$\widetilde{l}$}
%\rput[cc]{0}(6.5,0.9){$N.12.1$}
\endpspicture\\[2ex]
\pspicture(-2,0)(3.5,1.8)
\rput[rc]{0}(-1,1){$\g_{\scr N}^{(13)}$:}
\psline[linewidth=1pt](0.6,1.5)(1.8,1.5)
\psline[linewidth=1pt](1.8,1.5)(1.8,0.3)
\psline[linewidth=1pt](0.6,0.3)(1.8,0.3)
\psline[linewidth=1pt,linestyle=dashed](1.8,0.3)(3,0.3)
\psline[linewidth=1pt,linestyle=dashed](1.8,1.5)(3,1.5)
\psline[linewidth=1pt]{->}(1.2,1.5)(1.3,1.5)
\psline[linewidth=1pt]{<-}(2.3,1.5)(2.4,1.5)
\psline[linewidth=1pt]{->}(1.2,0.3)(1.3,0.3)
\psline[linewidth=1pt]{<-}(2.3,0.3)(2.4,0.3)
\rput[cc]{0}(0.3,1.5){$l$}
\rput[cc]{0}(0.3,0.3){$l$}
\rput[cc]{0}(2.1,0.9){$N$}
\rput[cc]{0}(3.3,1.5){$H_2$}
\rput[cc]{0}(3.3,0.3){$H_2$}
\rput[cc]{0}(4.0,1.0){$+$}
\endpspicture\hspace{5ex}
\pspicture(0,0)(3.5,1.8)
\psline[linewidth=1pt](0.6,1.5)(1.8,1.5)
\psline[linewidth=1pt](1.8,1.5)(1.8,0.3)
\psline[linewidth=1pt](0.6,0.3)(1.8,0.3)
\psline[linewidth=1pt,linestyle=dashed](1.8,0.3)(3,1.5)
\psline[linewidth=1pt,linestyle=dashed](1.8,1.5)(3,0.3)
\psline[linewidth=1pt]{->}(1.2,1.5)(1.3,1.5)
\psline[linewidth=1pt]{->}(1.2,0.3)(1.3,0.3)
\psline[linewidth=1pt]{<-}(2.62,1.12)(2.72,1.22)
\psline[linewidth=1pt]{<-}(2.62,0.68)(2.72,0.58)
\rput[cc]{0}(0.3,1.5){$l$}
\rput[cc]{0}(0.3,0.3){$l$}
\rput[cc]{0}(1.5,0.9){$N$}
\rput[cc]{0}(3.3,1.5){$H_2$}
\rput[cc]{0}(3.3,0.3){$H_2$}
\endpspicture\\[2ex]
\pspicture(-2,0)(3.5,1.8)
\rput[rc]{0}(-1,1){$\g_{\scr N}^{(14)}$:}
\psline[linewidth=1pt,linestyle=dashed](0.6,1.5)(1.8,1.5)
\psline[linewidth=1pt](0.6,0.3)(1.8,0.3)
\psline[linewidth=1pt](1.8,1.5)(1.8,0.3)
\psline[linewidth=1pt](1.8,1.5)(3,1.5)
\psline[linewidth=1pt,linestyle=dashed](1.8,0.3)(3,0.3)
\psline[linewidth=1pt]{->}(1.2,1.5)(1.3,1.5)
\psline[linewidth=1pt]{->}(1.2,0.3)(1.3,0.3)
\psline[linewidth=1pt]{->}(2.4,1.5)(2.5,1.5)
\psline[linewidth=1pt]{<-}(2.3,0.3)(2.4,0.3)
\rput[cc]{0}(0.3,1.5){$\widetilde{l}$}
\rput[cc]{0}(0.3,0.3){$l$}
\rput[cc]{0}(2.1,0.9){$N$}
\rput[cc]{0}(3.3,1.5){$\widetilde{h}$}
\rput[cc]{0}(3.3,0.3){$H_2$}
\rput[cc]{0}(4.0,0.9){$+$}
%\rput[cc]{0}(6.5,0.9){$N.14.1$}
\endpspicture\hspace{5ex}
\pspicture(0,0)(3.5,1.8)
\psline[linewidth=1pt,linestyle=dashed](0.6,1.5)(1.8,1.5)
\psline[linewidth=1pt,linestyle=dashed](1.8,1.5)(3,1.5)
\psline[linewidth=1pt,linestyle=dashed](1.8,1.5)(1.8,0.3)
\psline[linewidth=1pt](0.6,0.3)(1.8,0.3)
\psline[linewidth=1pt](1.8,0.3)(3,0.3)
\psline[linewidth=1pt]{->}(1.2,1.5)(1.3,1.5)
\psline[linewidth=1pt]{<-}(2.3,1.5)(2.4,1.5)
\psline[linewidth=1pt]{->}(1.8,0.86)(1.8,0.76)
\psline[linewidth=1pt]{->}(1.2,0.3)(1.3,0.3)
\psline[linewidth=1pt]{->}(2.45,0.3)(2.55,0.3)
\rput[cc]{0}(0.3,1.5){$\widetilde{l}$}
\rput[cc]{0}(0.3,0.3){$l$}
\rput[cc]{0}(2.15,0.9){$\widetilde{N^c}$}
\rput[cc]{0}(3.3,1.5){$H_2$}
\rput[cc]{0}(3.3,0.3){$\widetilde{h}$}
%\rput[cc]{0}(6.5,0.9){$N.14.2$}
\endpspicture\\[2ex]
\pspicture(-2,0)(3.5,2.1)
\rput[rc]{0}(-1,1.3){$\g_{\scr N}^{(15)}$:}
\psline[linewidth=1pt,linestyle=dashed](0.6,1.8)(1.8,1.8)
\psline[linewidth=1pt,linestyle=dashed](1.8,1.8)(3,1.8)
\psline[linewidth=1pt,linestyle=dashed](1.8,1.8)(1.8,0.6)
\psline[linewidth=1pt,linestyle=dashed](0.6,0.6)(1.8,0.6)
\psline[linewidth=1pt,linestyle=dashed](1.8,0.6)(3,0.9)
\psline[linewidth=1pt,linestyle=dashed](1.8,0.6)(3,0.3)
\psline[linewidth=1pt]{->}(1.22,1.8)(1.32,1.8)
\psline[linewidth=1pt]{<-}(2.28,1.8)(2.38,1.8)
\psline[linewidth=1pt]{<-}(1.8,1.08)(1.8,1.18)
\psline[linewidth=1pt]{<-}(1.06,0.6)(1.16,0.6)
\psline[linewidth=1pt]{->}(2.42,0.45)(2.52,0.42)
\psline[linewidth=1pt]{<-}(2.3,0.72)(2.4,0.75)
\rput[cc]{0}(0.3,1.8){$H_2$}
\rput[cc]{0}(0.3,0.6){$\widetilde{q}$}
\rput[cc]{0}(1.4,1.2){$\widetilde{N^c}$}
\rput[cc]{0}(3.3,1.8){$\widetilde{l}$}
\rput[cc]{0}(3.3,0.3){$\widetilde{U^c}$}
\rput[cc]{0}(3.3,0.9){$\widetilde{l}$}
\rput[cc]{0}(4.0,1.3){$+$}
%\rput[cc]{0}(6.5,1.2){$N.15.1$}
\endpspicture\hspace{5ex}
\pspicture(0,0)(3.5,2.1)
\psline[linewidth=1pt,linestyle=dashed](0.6,1.8)(1.8,1.8)
\psline[linewidth=1pt,linestyle=dashed](1.8,1.8)(3,0.9)
\psline[linewidth=1pt,linestyle=dashed](1.8,1.8)(1.8,0.6)
\psline[linewidth=1pt,linestyle=dashed](0.6,0.6)(1.8,0.6)
\psline[linewidth=1pt,linestyle=dashed](1.8,0.6)(3,1.8)
\psline[linewidth=1pt,linestyle=dashed](1.8,0.6)(3,0.3)
\psline[linewidth=1pt]{->}(1.22,1.8)(1.32,1.8)
\psline[linewidth=1pt]{<-}(1.8,1.08)(1.8,1.18)
\psline[linewidth=1pt]{<-}(1.08,0.6)(1.18,0.6)
\psline[linewidth=1pt]{->}(2.42,0.45)(2.52,0.42)
\psline[linewidth=1pt]{<-}(2.63,1.43)(2.73,1.53)
\psline[linewidth=1pt]{<-}(2.64,1.17)(2.74,1.1)
\rput[cc]{0}(0.3,1.8){$H_2$}
\rput[cc]{0}(0.3,0.6){$\widetilde{q}$}
\rput[cc]{0}(1.4,1.2){$\widetilde{N^c}$}
\rput[cc]{0}(3.3,1.8){$\widetilde{l}$}
\rput[cc]{0}(3.3,0.3){$\widetilde{U^c}$}
\rput[cc]{0}(3.3,0.9){$\widetilde{l}$}
%\rput[cc]{0}(6.5,1.2){$N.15.2$}
\endpspicture\\[2ex]
\pspicture(-2,0)(3.5,2.1)
\rput[rc]{0}(-1,1.3){$\g_{\scr N}^{(16)}$:}
\psline[linewidth=1pt,linestyle=dashed](0.6,1.8)(1.8,1.8)
\psline[linewidth=1pt,linestyle=dashed](1.8,1.8)(3,1.8)
\psline[linewidth=1pt,linestyle=dashed](1.8,1.8)(1.8,0.6)
\psline[linewidth=1pt,linestyle=dashed](0.6,0.6)(1.8,0.6)
\psline[linewidth=1pt,linestyle=dashed](1.8,0.6)(3,0.9)
\psline[linewidth=1pt,linestyle=dashed](1.8,0.6)(3,0.3)
\psline[linewidth=1pt]{->}(1.22,1.8)(1.32,1.8)
\psline[linewidth=1pt]{<-}(2.28,1.8)(2.38,1.8)
\psline[linewidth=1pt]{<-}(1.8,1.08)(1.8,1.18)
\psline[linewidth=1pt]{->}(1.22,0.6)(1.32,0.6)
\psline[linewidth=1pt]{->}(2.42,0.45)(2.52,0.42)
\psline[linewidth=1pt]{->}(2.42,0.75)(2.52,0.78)
\rput[cc]{0}(0.3,1.8){$\widetilde{l}$}
\rput[cc]{0}(0.3,0.6){$\widetilde{l}$}
\rput[cc]{0}(1.4,1.2){$\widetilde{N^c}$}
\rput[cc]{0}(3.3,1.8){$H_2$}
\rput[cc]{0}(3.3,0.9){$\widetilde{U^c}$}
\rput[cc]{0}(3.3,0.3){$\widetilde{q}$}
\rput[cc]{0}(4.0,1.3){$+$}
%\rput[cc]{0}(6.5,1.2){$N.16.1$}
\endpspicture\hspace{5ex}
\pspicture(0,0)(3.5,2.1)
\psline[linewidth=1pt,linestyle=dashed](0.6,0.6)(1.8,1.8)
\psline[linewidth=1pt,linestyle=dashed](0.6,1.8)(1.8,0.6)
\psline[linewidth=1pt,linestyle=dashed](1.8,1.8)(3,1.8)
\psline[linewidth=1pt,linestyle=dashed](1.8,1.8)(1.8,0.6)
\psline[linewidth=1pt,linestyle=dashed](1.8,0.6)(3,0.9)
\psline[linewidth=1pt,linestyle=dashed](1.8,0.6)(3,0.3)
\psline[linewidth=1pt]{->}(0.87,0.87)(0.97,0.97)
\psline[linewidth=1pt]{->}(0.87,1.53)(0.97,1.43)
%\psline[linewidth=1pt]{->}(2.44,1.8)(2.54,1.8)
\psline[linewidth=1pt]{<-}(2.28,1.8)(2.38,1.8)
%\psline[linewidth=1pt]{->}(1.8,1.2)(1.8,1.3)
\psline[linewidth=1pt]{<-}(1.8,1.08)(1.8,1.18)
\psline[linewidth=1pt]{->}(2.42,0.45)(2.52,0.42)
%\psline[linewidth=1pt]{<-}(2.3,0.48)(2.4,0.45)
\psline[linewidth=1pt]{->}(2.42,0.75)(2.52,0.78)
%\psline[linewidth=1pt]{<-}(2.3,0.72)(2.4,0.75)
\rput[cc]{0}(0.3,1.8){$\widetilde{l}$}
\rput[cc]{0}(0.3,0.6){$\widetilde{l}$}
\rput[cc]{0}(2.2,1.3){$\widetilde{N^c}$}
\rput[cc]{0}(3.3,1.8){$H_2$}
\rput[cc]{0}(3.3,0.9){$\widetilde{U^c}$}
\rput[cc]{0}(3.3,0.3){$\widetilde{q}$}
%\rput[cc]{0}(6.5,1.2){$N.16.2$}
\endpspicture\\
}
\parbox[c]{16cm}{
\pspicture(-1,0)(4.0,1.8)
\rput[rc]{0}(-0.3,1){$\g_{\scr N}^{(17)}$:}
\psline[linewidth=1pt,linestyle=dashed](0.6,1.5)(1.8,1.5)
\psline[linewidth=1pt](1.8,1.5)(3,1.5)
\psline[linewidth=1pt](1.8,1.5)(1.8,0.3)
\psline[linewidth=1pt](0.6,0.3)(1.8,0.3)
\psline[linewidth=1pt,linestyle=dashed](1.8,0.3)(3,0.3)
\psline[linewidth=1pt]{->}(1.2,1.5)(1.3,1.5)
\psline[linewidth=1pt]{->}(2.42,1.5)(2.52,1.5)
\psline[linewidth=1pt]{<-}(1.05,0.3)(1.15,0.3)
\psline[linewidth=1pt]{->}(2.42,0.3)(2.52,0.3)
\rput[cc]{0}(0.3,1.5){$\widetilde{l}$}
\rput[cc]{0}(0.3,0.3){$l$}
\rput[cc]{0}(2.1,0.9){$N$}
\rput[cc]{0}(3.3,1.5){$\widetilde{h}$}
\rput[cc]{0}(3.3,0.3){$H_2$}
%\rput[cc]{0}(6.5,0.9){$N.17$}
\endpspicture\hspace{\fill}
\pspicture(-1,0)(3.5,1.8)
\rput[rc]{0}(-0.3,1){$\g_{\scr N}^{(18)}$:}
\psline[linewidth=1pt,linestyle=dashed](0.6,1.5)(1.8,1.5)
\psline[linewidth=1pt](1.8,1.5)(3,1.5)
\psline[linewidth=1pt](1.8,1.5)(1.8,0.3)
\psline[linewidth=1pt,linestyle=dashed](0.6,0.3)(1.8,0.3)
\psline[linewidth=1pt](1.8,0.3)(3,0.3)
\psline[linewidth=1pt]{->}(1.2,1.5)(1.3,1.5)
\psline[linewidth=1pt]{->}(2.42,1.5)(2.52,1.5)
\psline[linewidth=1pt]{<-}(1.05,0.3)(1.15,0.3)
\psline[linewidth=1pt]{->}(2.42,0.3)(2.52,0.3)
\rput[cc]{0}(0.3,1.5){$\widetilde{l}$}
\rput[cc]{0}(0.3,0.3){$H_2$}
\rput[cc]{0}(2.1,0.9){$N$}
\rput[cc]{0}(3.3,1.5){$\widetilde{h}$}
\rput[cc]{0}(3.3,0.3){$l$}
%\rput[cc]{0}(6.5,0.9){$N.18$}
\endpspicture\hspace{\fill}
\pspicture(-1,0.3)(3.5,1.8)
\rput[rc]{0}(-0.3,1.3){$\g_{\scr N}^{(19)}$:}
\psline[linewidth=1pt](0.6,1.8)(1.8,1.8)
\psline[linewidth=1pt](1.8,1.8)(3,1.8)
\psline[linewidth=1pt,linestyle=dashed](1.8,1.8)(1.8,0.6)
\psline[linewidth=1pt,linestyle=dashed](0.6,0.6)(1.8,0.6)
\psline[linewidth=1pt,linestyle=dashed](1.8,0.6)(3,0.9)
\psline[linewidth=1pt,linestyle=dashed](1.8,0.6)(3,0.3)
\psline[linewidth=1pt]{->}(1.22,1.8)(1.32,1.8)
%\psline[linewidth=1pt]{<-}(1.06,1.8)(1.16,1.8)
\psline[linewidth=1pt]{->}(2.44,1.8)(2.54,1.8)
%\psline[linewidth=1pt]{<-}(2.28,1.8)(2.38,1.8)
\psline[linewidth=1pt]{->}(1.8,1.2)(1.8,1.3)
%\psline[linewidth=1pt]{<-}(1.8,1.08)(1.8,1.18)
%\psline[linewidth=1pt]{->}(1.22,0.6)(1.32,0.6)
\psline[linewidth=1pt]{<-}(1.06,0.6)(1.16,0.6)
%\psline[linewidth=1pt]{->}(2.42,0.45)(2.52,0.42)
\psline[linewidth=1pt]{<-}(2.3,0.48)(2.4,0.45)
%\psline[linewidth=1pt]{->}(2.42,0.75)(2.52,0.78)
\psline[linewidth=1pt]{<-}(2.3,0.72)(2.4,0.75)
\rput[cc]{0}(0.3,1.8){$l$}
\rput[cc]{0}(0.3,0.6){$\widetilde{l}$}
\rput[cc]{0}(1.4,1.2){$\widetilde{N^c}$}
\rput[cc]{0}(3.3,1.8){$\widetilde{h}$}
\rput[cc]{0}(3.3,0.9){$\widetilde{U^c}$}
\rput[cc]{0}(3.3,0.3){$\widetilde{q}$}
%\rput[cc]{0}(6.5,1.2){$N.19$}
\endpspicture
}
\end{center}

%% file: Fig06.tex
\begin{center}
\parbox[c]{16cm}{
\pspicture(-1,0)(3.5,2)
\rput[rc]{0}(-0.3,1){$\g_{t_j}^{(0)}$:}
\psline[linewidth=1pt,linestyle=dashed](0.6,0.3)(1.3,1)
\psline[linewidth=1pt](0.6,1.7)(1.3,1)
\psline[linewidth=1pt](1.3,1)(2.3,1)
\psline[linewidth=1pt](2.3,1)(3,1.7)
\psline[linewidth=1pt,linestyle=dashed](2.3,1)(3,0.3)
\psline[linewidth=1pt]{->}(0.83,0.53)(0.93,0.63)
\psline[linewidth=1pt]{->}(1.78,1)(1.88,1)
\psline[linewidth=1pt]{->}(2.73,1.43)(2.83,1.53)
\psline[linewidth=1pt]{->}(2.73,0.57)(2.83,0.47)
\rput[cc]{0}(0.3,1.7){$N_j$}
\rput[cc]{0}(0.3,0.3){$\widetilde{l}$}
\rput[cc]{0}(1.8,1.3){$\widetilde{h}$}
\rput[cc]{0}(3.3,0.3){$\widetilde{U^c}$}
\rput[cc]{0}(3.3,1.7){$q$}
\endpspicture\hspace{\fill}
\pspicture(-1,0)(3.5,1.8)
\rput[rc]{0}(-0.3,1){$\g_{t_j}^{(1)}$:}
\psline[linewidth=1pt](0.6,1.5)(1.8,1.5)
\psline[linewidth=1pt,linestyle=dashed](1.8,1.5)(3,1.5)
\psline[linewidth=1pt](1.8,1.5)(1.8,0.3)
\psline[linewidth=1pt](0.6,0.3)(1.8,0.3)
\psline[linewidth=1pt,linestyle=dashed](1.8,0.3)(3,0.3)
\psline[linewidth=1pt]{<-}(2.3,1.5)(2.4,1.5)
\psline[linewidth=1pt]{->}(1.8,0.9)(1.8,0.8)
\psline[linewidth=1pt]{->}(1.2,0.3)(1.1,0.3)
\psline[linewidth=1pt]{<-}(2.5,0.3)(2.4,0.3)
\rput[cc]{0}(0.3,1.5){$N_j$}
\rput[cc]{0}(0.3,0.3){$q$}
\rput[cc]{0}(2.1,0.9){$\widetilde{h}$}
\rput[cc]{0}(3.3,1.5){$\widetilde{l}$}
\rput[cc]{0}(3.3,0.3){$\widetilde{U^c}$}
\endpspicture\hspace{\fill}
\pspicture(-1,0)(3.5,1.8)
\rput[rc]{0}(-0.3,1){$\g_{t_j}^{(2)}$:}
\psline[linewidth=1pt](0.6,1.5)(1.8,1.5)
\psline[linewidth=1pt,linestyle=dashed](1.8,1.5)(3,1.5)
\psline[linewidth=1pt](1.8,1.5)(1.8,0.3)
\psline[linewidth=1pt,linestyle=dashed](0.6,0.3)(1.8,0.3)
\psline[linewidth=1pt](1.8,0.3)(3,0.3)
\psline[linewidth=1pt]{<-}(2.3,1.5)(2.4,1.5)
\psline[linewidth=1pt]{->}(1.8,0.9)(1.8,0.8)
\psline[linewidth=1pt]{->}(1.2,0.3)(1.1,0.3)
\psline[linewidth=1pt]{<-}(2.5,0.3)(2.4,0.3)
\rput[cc]{0}(0.3,1.5){$N_j$}
\rput[cc]{0}(0.3,0.3){$\widetilde{U^c}$}
\rput[cc]{0}(2.1,0.9){$\widetilde{h}$}
\rput[cc]{0}(3.3,1.5){$\widetilde{l}$}
\rput[cc]{0}(3.3,0.3){$q$}
\endpspicture\\[2ex]
\mbox{ }\hspace{\fill}
\pspicture(-1,0)(3.5,2)
\rput[rc]{0}(-0.3,1){$\g_{t_j}^{(3)}$:}
\psline[linewidth=1pt](0.6,0.3)(1.3,1)
\psline[linewidth=1pt](0.6,1.7)(1.3,1)
\psline[linewidth=1pt,linestyle=dashed](1.3,1)(2.3,1)
\psline[linewidth=1pt](2.3,1)(3,1.7)
\psline[linewidth=1pt](2.3,1)(3,0.3)
\psline[linewidth=1pt]{->}(0.83,0.53)(0.93,0.63)
\psline[linewidth=1pt]{->}(1.7,1)(1.6,1)
\psline[linewidth=1pt]{->}(2.67,1.37)(2.77,1.47)
\psline[linewidth=1pt]{->}(2.77,0.53)(2.67,0.63)
\rput[cc]{0}(0.3,1.7){$N_j$}
\rput[cc]{0}(0.3,0.3){$l$}
\rput[cc]{0}(1.8,1.3){$H_2$}
\rput[cc]{0}(3.3,0.3){$u$}
\rput[cc]{0}(3.3,1.7){$q$}
\endpspicture\hspace{\fill}
\pspicture(-1,0)(3.5,1.8)
\rput[rc]{0}(-0.3,1){$\g_{t_j}^{(4)}$:}
\psline[linewidth=1pt](0.6,1.5)(1.8,1.5)
\psline[linewidth=1pt](1.8,1.5)(3,1.5)
\psline[linewidth=1pt,linestyle=dashed](1.8,1.5)(1.8,0.3)
\psline[linewidth=1pt](0.6,0.3)(1.8,0.3)
\psline[linewidth=1pt](1.8,0.3)(3,0.3)
\psline[linewidth=1pt]{<-}(2.3,1.5)(2.4,1.5)
\psline[linewidth=1pt]{->}(1.8,0.9)(1.8,1)
\psline[linewidth=1pt]{->}(1.1,0.3)(1.2,0.3)
\psline[linewidth=1pt]{<-}(2.5,0.3)(2.4,0.3)
\rput[cc]{0}(0.3,1.5){$N_j$}
\rput[cc]{0}(0.3,0.3){$u$}
\rput[cc]{0}(2.2,0.9){$H_2$}
\rput[cc]{0}(3.3,1.5){$l$}
\rput[cc]{0}(3.3,0.3){$q$}
\endpspicture\hspace{\fill}
}
\end{center}

%% file: Fig07.tex
\begin{center}
\parbox[c]{16cm}{
\pspicture(-1,0)(3.5,2)
\rput[rc]{0}(-0.3,1){$\g_{t_j}^{(5)}$:}
\psline[linewidth=1pt](0.6,0.3)(1.3,1)
\psline[linewidth=1pt,linestyle=dashed](0.6,1.7)(1.3,1)
\psline[linewidth=1pt](1.3,1)(2.3,1)
\psline[linewidth=1pt,linestyle=dashed](2.3,1)(3,1.7)
\psline[linewidth=1pt](2.3,1)(3,0.3)
\psline[linewidth=1pt]{->}(0.83,0.53)(0.93,0.63)
\psline[linewidth=1pt]{->}(0.83,1.47)(0.93,1.37)
\psline[linewidth=1pt]{->}(1.78,1)(1.88,1)
\psline[linewidth=1pt]{->}(2.73,1.43)(2.83,1.53)
\psline[linewidth=1pt]{->}(2.73,0.57)(2.83,0.47)
\rput[cc]{0}(0.3,1.7){$\snj$}
\rput[cc]{0}(0.3,0.3){$l$}
\rput[cc]{0}(1.8,1.3){$\widetilde{h}$}
\rput[cc]{0}(3.3,0.3){$q$}
\rput[cc]{0}(3.3,1.7){$\widetilde{U^c}$}
%\rput[cc]{0}(6.5,1){$T.5$}
\endpspicture\hspace{\fill}
\pspicture(-1,0)(3.5,1.8)
\rput[rc]{0}(-0.3,1){$\g_{t_j}^{(6)}$:}
\psline[linewidth=1pt,linestyle=dashed](0.6,1.5)(1.8,1.5)
\psline[linewidth=1pt](1.8,1.5)(3,1.5)
\psline[linewidth=1pt](1.8,1.5)(1.8,0.3)
\psline[linewidth=1pt,linestyle=dashed](0.6,0.3)(1.8,0.3)
\psline[linewidth=1pt](1.8,0.3)(3,0.3)
\psline[linewidth=1pt]{->}(1.2,1.5)(1.3,1.5)
\psline[linewidth=1pt]{<-}(2.3,1.5)(2.4,1.5)
\psline[linewidth=1pt]{->}(1.8,0.9)(1.8,0.8)
\psline[linewidth=1pt]{->}(1.2,0.3)(1.1,0.3)
\psline[linewidth=1pt]{<-}(2.5,0.3)(2.4,0.3)
\rput[cc]{0}(0.3,1.5){$\snj$}
\rput[cc]{0}(0.3,0.3){$\widetilde{U^c}$}
\rput[cc]{0}(2.1,0.9){$\widetilde{h}$}
\rput[cc]{0}(3.3,1.5){$l$}
\rput[cc]{0}(3.3,0.3){$q$}
%\rput[cc]{0}(6.5,0.9){$T.6$}
\endpspicture\hspace{\fill}
\pspicture(-1,0)(3.5,1.8)
\rput[rc]{0}(-0.3,1){$\g_{t_j}^{(7)}$:}
\psline[linewidth=1pt,linestyle=dashed](0.6,1.5)(1.8,1.5)
\psline[linewidth=1pt](1.8,1.5)(3,1.5)
\psline[linewidth=1pt](1.8,1.5)(1.8,0.3)
\psline[linewidth=1pt](0.6,0.3)(1.8,0.3)
\psline[linewidth=1pt,linestyle=dashed](1.8,0.3)(3,0.3)
\psline[linewidth=1pt]{->}(1.2,1.5)(1.3,1.5)
\psline[linewidth=1pt]{<-}(2.3,1.5)(2.4,1.5)
\psline[linewidth=1pt]{->}(1.8,0.9)(1.8,0.8)
\psline[linewidth=1pt]{->}(1.2,0.3)(1.1,0.3)
\psline[linewidth=1pt]{<-}(2.5,0.3)(2.4,0.3)
\rput[cc]{0}(0.3,1.5){$\snj$}
\rput[cc]{0}(0.3,0.3){$q$}
\rput[cc]{0}(2.1,0.9){$\widetilde{h}$}
\rput[cc]{0}(3.3,1.5){$l$}
\rput[cc]{0}(3.3,0.3){$\widetilde{U^c}$}
%\rput[cc]{0}(6.5,0.9){$T.7$}
\endpspicture\\[2ex]
\mbox{ }\hspace{\fill}
\pspicture(-1,0)(3.5,2)
\rput[rc]{0}(-0.3,1){$\g_{t_j}^{(8)}$:}
\psline[linewidth=1pt,linestyle=dashed](0.6,0.3)(1.3,1)
\psline[linewidth=1pt,linestyle=dashed](0.6,1.7)(1.3,1)
\psline[linewidth=1pt,linestyle=dashed](1.3,1)(2.3,1)
\psline[linewidth=1pt](2.3,1)(3,1.7)
\psline[linewidth=1pt](2.3,1)(3,0.3)
\psline[linewidth=1pt]{->}(0.86,0.56)(0.76,0.46)
\psline[linewidth=1pt]{->}(0.83,1.47)(0.93,1.37)
\psline[linewidth=1pt]{->}(1.7,1)(1.8,1)
\psline[linewidth=1pt]{->}(2.67,1.37)(2.77,1.47)
\psline[linewidth=1pt]{->}(2.77,0.53)(2.67,0.63)
\rput[cc]{0}(0.3,1.7){$\snj$}
\rput[cc]{0}(0.3,0.3){$\widetilde{l}$}
\rput[cc]{0}(1.8,1.3){$H_2$}
\rput[cc]{0}(3.3,0.3){$q$}
\rput[cc]{0}(3.3,1.7){$u$}
%\rput[cc]{0}(6.5,1){$T.8$}
\endpspicture\hspace{\fill}
\pspicture(-2,0)(3.5,1.8)
\rput[rc]{0}(-1,1){$\g_{t_j}^{(9)}$:}
\psline[linewidth=1pt,linestyle=dashed](0.6,1.5)(1.8,1.5)
\psline[linewidth=1pt,linestyle=dashed](1.8,1.5)(3,1.5)
\psline[linewidth=1pt,linestyle=dashed](1.8,1.5)(1.8,0.3)
\psline[linewidth=1pt](0.6,0.3)(1.8,0.3)
\psline[linewidth=1pt](1.8,0.3)(3,0.3)
\psline[linewidth=1pt]{->}(1.2,1.5)(1.3,1.5)
\psline[linewidth=1pt]{<-}(2.5,1.5)(2.4,1.5)
\psline[linewidth=1pt]{->}(1.8,0.9)(1.8,0.8)
\psline[linewidth=1pt]{->}(1.1,0.3)(1.2,0.3)
\psline[linewidth=1pt]{<-}(2.5,0.3)(2.4,0.3)
\rput[cc]{0}(0.3,1.5){$\snj$}
\rput[cc]{0}(0.3,0.3){$q$}
\rput[cc]{0}(2.2,0.9){$H_2$}
\rput[cc]{0}(3.3,1.5){$\widetilde{l}$}
\rput[cc]{0}(3.3,0.3){$u$}
%\rput[cc]{0}(6.5,0.9){$T.9$}
\endpspicture\hspace{\fill}
}
\end{center}

%% file: Fig08.tex
\begin{center}
\parbox[c]{10.7cm}{
\pspicture(-2,0)(3.5,1.8)
\rput[rc]{0}(-1,1){$\g_{\scr N_iN_j}^{(1)}$:}
\psline[linewidth=1pt](0.6,1.5)(1.8,1.5)
\psline[linewidth=1pt,linestyle=dashed](1.8,1.5)(3,1.5)
\psline[linewidth=1pt](1.8,1.5)(1.8,0.3)
\psline[linewidth=1pt](0.6,0.3)(1.8,0.3)
\psline[linewidth=1pt,linestyle=dashed](1.8,0.3)(3,0.3)
\psline[linewidth=1pt]{<-}(2.3,1.5)(2.4,1.5)
\psline[linewidth=1pt]{->}(1.8,0.9)(1.8,0.8)
\psline[linewidth=1pt]{<-}(2.5,0.3)(2.4,0.3)
\rput[cc]{0}(0.3,1.5){$N_i$}
\rput[cc]{0}(0.3,0.3){$N_j$}
\rput[cc]{0}(2.1,0.9){$\widetilde{h}$}
\rput[cc]{0}(3.3,1.5){$\widetilde{l}$}
\rput[cc]{0}(3.3,0.3){$\widetilde{l}$}
\rput[cc]{0}(4.0,0.9){$+$}
\endpspicture\hspace{5ex}
\pspicture(0,0)(3.5,1.8)
\psline[linewidth=1pt](0.6,1.5)(1.8,0.3)
\psline[linewidth=1pt](0.6,0.3)(1.8,1.5)
\psline[linewidth=1pt](1.8,1.5)(1.8,0.3)
\psline[linewidth=1pt,linestyle=dashed](1.8,0.3)(3,0.3)
\psline[linewidth=1pt,linestyle=dashed](1.8,1.5)(3,1.5)
\psline[linewidth=1pt]{<-}(2.3,1.5)(2.4,1.5)
\psline[linewidth=1pt]{->}(1.8,0.9)(1.8,0.8)
\psline[linewidth=1pt]{<-}(2.5,0.3)(2.4,0.3)
\rput[cc]{0}(0.3,1.5){$N_i$}
\rput[cc]{0}(0.3,0.3){$N_j$}
\rput[cc]{0}(2.1,0.9){$\widetilde{h}$}
\rput[cc]{0}(3.3,1.5){$\widetilde{l}$}
\rput[cc]{0}(3.3,0.3){$\widetilde{l}$}
\endpspicture\\[2ex]
\pspicture(-2,0)(3.5,1.8)
\rput[rc]{0}(-1,1){$\g_{\scr N_iN_j}^{(2)}$:}
\psline[linewidth=1pt](0.6,1.5)(1.8,1.5)
\psline[linewidth=1pt](1.8,1.5)(3,1.5)
\psline[linewidth=1pt,linestyle=dashed](1.8,1.5)(1.8,0.3)
\psline[linewidth=1pt](0.6,0.3)(1.8,0.3)
\psline[linewidth=1pt](1.8,0.3)(3,0.3)
\psline[linewidth=1pt]{<-}(2.3,1.5)(2.4,1.5)
\psline[linewidth=1pt]{->}(1.8,0.9)(1.8,1)
\psline[linewidth=1pt]{<-}(2.5,0.3)(2.4,0.3)
\rput[cc]{0}(0.3,1.5){$N_i$}
\rput[cc]{0}(0.3,0.3){$N_j$}
\rput[cc]{0}(2.2,0.9){$H_2$}
\rput[cc]{0}(3.3,1.5){$l$}
\rput[cc]{0}(3.3,0.3){$l$}
\rput[cc]{0}(4.0,0.9){$+$}
\endpspicture\hspace{5ex}
\pspicture(0,0)(3.5,1.8)
\psline[linewidth=1pt](0.6,1.5)(1.8,0.3)
\psline[linewidth=1pt](1.8,1.5)(3,1.5)
\psline[linewidth=1pt,linestyle=dashed](1.8,1.5)(1.8,0.3)
\psline[linewidth=1pt](0.6,0.3)(1.8,1.5)
\psline[linewidth=1pt](1.8,0.3)(3,0.3)
\psline[linewidth=1pt]{<-}(2.3,1.5)(2.4,1.5)
\psline[linewidth=1pt]{->}(1.8,0.9)(1.8,1)
\psline[linewidth=1pt]{<-}(2.5,0.3)(2.4,0.3)
\rput[cc]{0}(0.3,1.5){$N_i$}
\rput[cc]{0}(0.3,0.3){$N_j$}
\rput[cc]{0}(2.2,0.9){$H_2$}
\rput[cc]{0}(3.3,1.5){$l$}
\rput[cc]{0}(3.3,0.3){$l$}
\endpspicture\\[2ex]
\pspicture(-2,0)(3.5,1.8)
\rput[rc]{0}(-1,1){$\g_{\scr N_iN_j}^{(3)}$:}
\psline[linewidth=1pt](0.6,1.5)(1.8,1.5)
\psline[linewidth=1pt,linestyle=dashed](1.8,1.5)(3,1.5)
\psline[linewidth=1pt](1.8,1.5)(1.8,0.3)
\psline[linewidth=1pt](0.6,0.3)(1.8,0.3)
\psline[linewidth=1pt,linestyle=dashed](1.8,0.3)(3,0.3)
\psline[linewidth=1pt]{<-}(2.3,1.5)(2.4,1.5)
\psline[linewidth=1pt]{<-}(1.8,1.0)(1.8,0.9)
\psline[linewidth=1pt]{<-}(2.5,0.3)(2.4,0.3)
\rput[cc]{0}(0.3,1.5){$N_i$}
\rput[cc]{0}(0.3,0.3){$N_j$}
\rput[cc]{0}(2.1,0.9){$l$}
\rput[cc]{0}(3.3,1.5){$H_2$}
\rput[cc]{0}(3.3,0.3){$H_2$}
\rput[cc]{0}(4.0,0.9){$+$}
\endpspicture\hspace{5ex}
\pspicture(0,0)(3.5,1.8)
\psline[linewidth=1pt](0.6,1.5)(1.8,0.3)
\psline[linewidth=1pt,linestyle=dashed](1.8,1.5)(3,1.5)
\psline[linewidth=1pt](1.8,1.5)(1.8,0.3)
\psline[linewidth=1pt](0.6,0.3)(1.8,1.5)
\psline[linewidth=1pt,linestyle=dashed](1.8,0.3)(3,0.3)
\psline[linewidth=1pt]{<-}(2.3,1.5)(2.4,1.5)
\psline[linewidth=1pt]{<-}(1.8,1.0)(1.8,0.9)
\psline[linewidth=1pt]{<-}(2.5,0.3)(2.4,0.3)
\rput[cc]{0}(0.3,1.5){$N_i$}
\rput[cc]{0}(0.3,0.3){$N_j$}
\rput[cc]{0}(2.1,0.9){$l$}
\rput[cc]{0}(3.3,1.5){$H_2$}
\rput[cc]{0}(3.3,0.3){$H_2$}
\endpspicture\\[2ex]
\pspicture(-2,0)(3.5,1.8)
\rput[rc]{0}(-1,1){$\g_{\scr N_iN_j}^{(4)}$:}
\psline[linewidth=1pt](0.6,1.5)(1.8,1.5)
\psline[linewidth=1pt](1.8,1.5)(3,1.5)
\psline[linewidth=1pt,linestyle=dashed](1.8,1.5)(1.8,0.3)
\psline[linewidth=1pt](0.6,0.3)(1.8,0.3)
\psline[linewidth=1pt](1.8,0.3)(3,0.3)
\psline[linewidth=1pt]{<-}(2.3,1.5)(2.4,1.5)
\psline[linewidth=1pt]{<-}(1.8,0.8)(1.8,0.9)
\psline[linewidth=1pt]{<-}(2.5,0.3)(2.4,0.3)
\rput[cc]{0}(0.3,1.5){$N_i$}
\rput[cc]{0}(0.3,0.3){$N_j$}
\rput[cc]{0}(2.2,0.9){$\widetilde{l}$}
\rput[cc]{0}(3.3,1.5){$\widetilde{h}$}
\rput[cc]{0}(3.3,0.3){$\widetilde{h}$}
\rput[cc]{0}(4.0,0.9){$+$}
\endpspicture\hspace{5ex}
\pspicture(0,0)(3.5,1.8)
\psline[linewidth=1pt](0.6,1.5)(1.8,0.3)
\psline[linewidth=1pt](1.8,1.5)(3,1.5)
\psline[linewidth=1pt,linestyle=dashed](1.8,1.5)(1.8,0.3)
\psline[linewidth=1pt](0.6,0.3)(1.8,1.5)
\psline[linewidth=1pt](1.8,0.3)(3,0.3)
\psline[linewidth=1pt]{<-}(2.3,1.5)(2.4,1.5)
\psline[linewidth=1pt]{<-}(1.8,0.8)(1.8,0.9)
\psline[linewidth=1pt]{<-}(2.5,0.3)(2.4,0.3)
\rput[cc]{0}(0.3,1.5){$N_i$}
\rput[cc]{0}(0.3,0.3){$N_j$}
\rput[cc]{0}(2.2,0.9){$\widetilde{l}$}
\rput[cc]{0}(3.3,1.5){$\widetilde{h}$}
\rput[cc]{0}(3.3,0.3){$\widetilde{h}$}
\endpspicture
}
\end{center}

%% file: Fig09.tex
\begin{center}
\parbox[c]{13cm}{
\pspicture(-2,0)(3.5,1.8)
\rput[rc]{0}(-1,1){$\g_{\scr\sni\snj}^{(1)}$:}
\psline[linewidth=1pt,linestyle=dashed](0.6,1.5)(1.8,1.5)
\psline[linewidth=1pt](1.8,1.5)(3,1.5)
\psline[linewidth=1pt](1.8,1.5)(1.8,0.3)
\psline[linewidth=1pt,linestyle=dashed](0.6,0.3)(1.8,0.3)
\psline[linewidth=1pt](1.8,0.3)(3,0.3)
\psline[linewidth=1pt]{->}(1.2,1.5)(1.3,1.5)
\psline[linewidth=1pt]{<-}(2.3,1.5)(2.4,1.5)
\psline[linewidth=1pt]{->}(1.8,0.9)(1.8,0.8)
\psline[linewidth=1pt]{->}(1.2,0.3)(1.1,0.3)
\psline[linewidth=1pt]{<-}(2.5,0.3)(2.4,0.3)
\rput[cc]{0}(0.3,1.5){$\widetilde{N^c_i}$}
\rput[cc]{0}(0.3,0.3){$\widetilde{N^c_j}$}
\rput[cc]{0}(2.1,0.9){$\widetilde{h}$}
\rput[cc]{0}(3.3,1.5){$l$}
\rput[cc]{0}(3.3,0.3){$l$}
\endpspicture\hspace{\fill}
\pspicture(-2,0)(3.5,1.8)
\rput[rc]{0}(-1,1){$\g_{\scr\sni\snj}^{(3)}$:}
\psline[linewidth=1pt,linestyle=dashed](0.6,0.3)(1.8,0.3)
\psline[linewidth=1pt](1.8,0.3)(3,0.3)
\psline[linewidth=1pt](1.8,0.3)(1.8,1.5)
\psline[linewidth=1pt,linestyle=dashed](0.6,1.5)(1.8,1.5)
\psline[linewidth=1pt](1.8,1.5)(3,1.5)
\psline[linewidth=1pt]{->}(1.2,1.5)(1.3,1.5)
\psline[linewidth=1pt]{<-}(2.3,0.3)(2.4,0.3)
\psline[linewidth=1pt]{->}(1.8,0.9)(1.8,1.0)
\psline[linewidth=1pt]{->}(1.2,0.3)(1.1,0.3)
\psline[linewidth=1pt]{<-}(2.5,1.5)(2.4,1.5)
\rput[cc]{0}(0.3,1.5){$\widetilde{N^c_i}$}
\rput[cc]{0}(0.3,0.3){$\widetilde{N^c_j}$}
\rput[cc]{0}(2.1,0.9){$l$}
\rput[cc]{0}(3.3,0.3){$\widetilde{h}$}
\rput[cc]{0}(3.3,1.5){$\widetilde{h}$}
\endpspicture\\[2ex]
\mbox{ }\hspace{\fill}
\pspicture(-2,0)(3.5,1.8)
\rput[rc]{0}(-1,1){$\g_{\scr\sni\snj}^{(2)}$:}
\psline[linewidth=1pt,linestyle=dashed](0.6,1.5)(1.8,1.5)
\psline[linewidth=1pt,linestyle=dashed](1.8,1.5)(3,1.5)
\psline[linewidth=1pt,linestyle=dashed](1.8,1.5)(1.8,0.3)
\psline[linewidth=1pt,linestyle=dashed](0.6,0.3)(1.8,0.3)
\psline[linewidth=1pt,linestyle=dashed](1.8,0.3)(3,0.3)
\psline[linewidth=1pt]{->}(1.2,1.5)(1.3,1.5)
\psline[linewidth=1pt]{<-}(2.5,1.5)(2.4,1.5)
\psline[linewidth=1pt]{->}(1.8,0.9)(1.8,0.8)
\psline[linewidth=1pt]{->}(1.2,0.3)(1.1,0.3)
\psline[linewidth=1pt]{<-}(2.3,0.3)(2.4,0.3)
\rput[cc]{0}(0.3,1.5){$\widetilde{N^c_i}$}
\rput[cc]{0}(0.3,0.3){$\widetilde{N^c_j}$}
\rput[cc]{0}(2.2,0.9){$H_2$}
\rput[cc]{0}(3.3,1.5){$\widetilde{l}$}
\rput[cc]{0}(3.3,0.3){$\widetilde{l}$}
\rput[cc]{0}(4.0,0.9){$+$}
\endpspicture\hspace{5ex}
\pspicture(-0.6,0)(2.5,2)
\psline[linewidth=1pt,linestyle=dashed](0.6,0.3)(1.3,1)
\psline[linewidth=1pt,linestyle=dashed](0.6,1.7)(1.3,1)
\psline[linewidth=1pt,linestyle=dashed](1.3,1)(2,1.7)
\psline[linewidth=1pt,linestyle=dashed](1.3,1)(2,0.3)
\psline[linewidth=1pt]{->}(0.85,1.45)(0.95,1.35)
\psline[linewidth=1pt]{->}(1.72,1.42)(1.82,1.52)
\psline[linewidth=1pt]{->}(0.88,0.58)(0.78,0.48)
\psline[linewidth=1pt]{->}(1.78,0.52)(1.68,0.62)
\rput[cc]{0}(0.3,1.7){$\sni$}
\rput[cc]{0}(0.3,0.3){$\snj$}
\rput[cc]{0}(2.3,0.3){$\wt{l}$}
\rput[cc]{0}(2.3,1.7){$\wt{l}$}
\endpspicture\hspace{\fill}\mbox{ }\\[2ex]
\mbox{ }\hspace{\fill}
\pspicture(-2,0)(3.5,1.8)
\rput[rc]{0}(-1,1){$\g_{\scr\sni\snj}^{(4)}$:}
\psline[linewidth=1pt,linestyle=dashed](0.6,1.5)(1.8,1.5)
\psline[linewidth=1pt,linestyle=dashed](1.8,1.5)(3,1.5)
\psline[linewidth=1pt,linestyle=dashed](1.8,1.5)(1.8,0.3)
\psline[linewidth=1pt,linestyle=dashed](0.6,0.3)(1.8,0.3)
\psline[linewidth=1pt,linestyle=dashed](1.8,0.3)(3,0.3)
\psline[linewidth=1pt]{->}(1.2,1.5)(1.3,1.5)
\psline[linewidth=1pt]{<-}(2.5,1.5)(2.4,1.5)
\psline[linewidth=1pt]{->}(1.8,0.9)(1.8,0.8)
\psline[linewidth=1pt]{->}(1.2,0.3)(1.1,0.3)
\psline[linewidth=1pt]{<-}(2.3,0.3)(2.4,0.3)
\rput[cc]{0}(0.3,1.5){$\widetilde{N^c_i}$}
\rput[cc]{0}(0.3,0.3){$\widetilde{N^c_j}$}
\rput[cc]{0}(2.2,0.9){$\widetilde{l}$}
\rput[cc]{0}(3.3,1.5){$H_2$}
\rput[cc]{0}(3.3,0.3){$H_2$}
\rput[cc]{0}(4.0,0.9){$+$}
\endpspicture\hspace{5ex}
\pspicture(-0.6,0)(2.5,2)
\psline[linewidth=1pt,linestyle=dashed](0.6,0.3)(1.3,1)
\psline[linewidth=1pt,linestyle=dashed](0.6,1.7)(1.3,1)
\psline[linewidth=1pt,linestyle=dashed](1.3,1)(2,1.7)
\psline[linewidth=1pt,linestyle=dashed](1.3,1)(2,0.3)
\psline[linewidth=1pt]{->}(0.85,1.45)(0.95,1.35)
\psline[linewidth=1pt]{->}(1.72,1.42)(1.82,1.52)
\psline[linewidth=1pt]{->}(0.88,0.58)(0.78,0.48)
\psline[linewidth=1pt]{->}(1.78,0.52)(1.68,0.62)
\rput[cc]{0}(0.3,1.7){$\sni$}
\rput[cc]{0}(0.3,0.3){$\snj$}
\rput[cc]{0}(2.3,0.3){$H_2$}
\rput[cc]{0}(2.3,1.7){$H_2$}
\endpspicture\hspace{\fill}\mbox{ }
}
\end{center}

%% file: Fig10.tex
\begin{center}
\parbox[c]{10.7cm}{
\pspicture(-2,0)(3.5,1.8)
\rput[rc]{0}(-1,1){$\g_{\scr N_j\sni}^{(1)}$:}
\psline[linewidth=1pt,linestyle=dashed](0.6,1.5)(1.8,1.5)
\psline[linewidth=1pt](1.8,1.5)(3,1.5)
\psline[linewidth=1pt](1.8,1.5)(1.8,0.3)
\psline[linewidth=1pt](0.6,0.3)(1.8,0.3)
\psline[linewidth=1pt,linestyle=dashed](1.8,0.3)(3,0.3)
\psline[linewidth=1pt]{->}(1.2,1.5)(1.3,1.5)
\psline[linewidth=1pt]{<-}(2.3,1.5)(2.4,1.5)
\psline[linewidth=1pt]{->}(1.8,0.9)(1.8,0.8)
\psline[linewidth=1pt]{<-}(2.5,0.3)(2.4,0.3)
\rput[cc]{0}(0.3,1.5){$\widetilde{N^c_i}$}
\rput[cc]{0}(0.3,0.3){$N_j$}
\rput[cc]{0}(2.1,0.9){$\widetilde{h}$}
\rput[cc]{0}(3.3,1.5){$l$}
\rput[cc]{0}(3.3,0.3){$\widetilde{l}$}
\rput[cc]{0}(4.0,0.9){$+$}
\endpspicture\hspace{5ex}
\pspicture(0,0)(3.5,1.8)
\psline[linewidth=1pt,linestyle=dashed](0.6,1.5)(1.8,1.5)
\psline[linewidth=1pt,linestyle=dashed](1.8,1.5)(3,1.5)
\psline[linewidth=1pt,linestyle=dashed](1.8,1.5)(1.8,0.3)
\psline[linewidth=1pt](0.6,0.3)(1.8,0.3)
\psline[linewidth=1pt](1.8,0.3)(3,0.3)
\psline[linewidth=1pt]{->}(1.2,1.5)(1.3,1.5)
\psline[linewidth=1pt]{<-}(2.5,1.5)(2.4,1.5)
\psline[linewidth=1pt]{->}(1.8,0.9)(1.8,0.8)
\psline[linewidth=1pt]{<-}(2.4,0.3)(2.5,0.3)
\rput[cc]{0}(0.3,1.5){$\widetilde{N^c_i}$}
\rput[cc]{0}(0.3,0.3){$N_j$}
\rput[cc]{0}(2.2,0.9){$H_2$}
\rput[cc]{0}(3.3,1.5){$\widetilde{l}$}
\rput[cc]{0}(3.3,0.3){$l$}
\endpspicture\\[2ex]
\pspicture(-2,0)(3.5,1.8)
\rput[rc]{0}(-1,1){$\g_{\scr N_j\sni}^{(2)}$:}
\psline[linewidth=1pt,linestyle=dashed](0.6,1.5)(1.8,1.5)
\psline[linewidth=1pt](1.8,1.5)(3,1.5)
\psline[linewidth=1pt](1.8,1.5)(1.8,0.3)
\psline[linewidth=1pt](0.6,0.3)(1.8,0.3)
\psline[linewidth=1pt,linestyle=dashed](1.8,0.3)(3,0.3)
\psline[linewidth=1pt]{->}(1.2,1.5)(1.3,1.5)
\psline[linewidth=1pt]{<-}(2.5,1.5)(2.4,1.5)
\psline[linewidth=1pt]{->}(1.8,0.9)(1.8,1.0)
\psline[linewidth=1pt]{<-}(2.5,0.3)(2.4,0.3)
\rput[cc]{0}(0.3,1.5){$\widetilde{N^c_i}$}
\rput[cc]{0}(0.3,0.3){$N_j$}
\rput[cc]{0}(2.1,0.9){$l$}
\rput[cc]{0}(3.3,1.5){$\widetilde{h}$}
\rput[cc]{0}(3.3,0.3){$H_2$}
\rput[cc]{0}(4.0,0.9){$+$}
\endpspicture\hspace{5ex}
\pspicture(0,0)(3.5,1.8)
\psline[linewidth=1pt,linestyle=dashed](0.6,1.5)(1.8,1.5)
\psline[linewidth=1pt,linestyle=dashed](1.8,1.5)(3,1.5)
\psline[linewidth=1pt,linestyle=dashed](1.8,1.5)(1.8,0.3)
\psline[linewidth=1pt](0.6,0.3)(1.8,0.3)
\psline[linewidth=1pt](1.8,0.3)(3,0.3)
\psline[linewidth=1pt]{->}(1.2,1.5)(1.3,1.5)
\psline[linewidth=1pt]{<-}(2.5,1.5)(2.4,1.5)
\psline[linewidth=1pt]{->}(1.8,0.9)(1.8,0.8)
\psline[linewidth=1pt]{<-}(2.5,0.3)(2.4,0.3)
\rput[cc]{0}(0.3,1.5){$\widetilde{N^c_i}$}
\rput[cc]{0}(0.3,0.3){$N_j$}
\rput[cc]{0}(2.2,0.9){$\widetilde{l}$}
\rput[cc]{0}(3.3,1.5){$H_2$}
\rput[cc]{0}(3.3,0.3){$\widetilde{h}$}
\endpspicture
}
\end{center}

%% file: Fig11.tex
\begin{center}
\parbox[c]{9.5cm}{
\pspicture(0,0)(4.0,1.8)
\psline[linewidth=1pt](0.6,1.5)(1.8,1.5)
\psline[linewidth=1pt](1.8,1.5)(1.8,0.3)
\psline[linewidth=1pt](0.6,0.3)(1.8,0.3)
\psline[linewidth=1pt,linestyle=dashed](1.8,1.5)(3,1.5)
\psline[linewidth=1pt,linestyle=dashed](1.8,0.3)(3,0.3)
\psline[linewidth=1pt]{->}(1.2,1.5)(1.3,1.5)
\psline[linewidth=1pt]{->}(1.2,0.3)(1.3,0.3)
\psline[linewidth=1pt]{<-}(2.5,1.5)(2.4,1.5)
\psline[linewidth=1pt]{<-}(2.5,0.3)(2.4,0.3)
\rput[cc]{0}(0.3,1.5){$e$}
\rput[cc]{0}(0.3,0.3){$e$}
\rput[cc]{0}(2.2,0.9){$\widetilde{\g}$}
\rput[lc]{0}(3.3,1.5){$\widetilde{e}$}
\rput[lc]{0}(3.3,0.3){$\widetilde{e}$}
\rput[cc]{0}(4.7,1.0){$+$}
\endpspicture\hspace{8ex}
\pspicture(0,0)(3.5,1.8)
\psline[linewidth=1pt](0.6,1.5)(1.8,1.5)
\psline[linewidth=1pt](1.8,1.5)(1.8,0.3)
\psline[linewidth=1pt](0.6,0.3)(1.8,0.3)
\psline[linewidth=1pt,linestyle=dashed](1.8,1.5)(3,0.3)
\psline[linewidth=1pt,linestyle=dashed](1.8,0.3)(3,1.5)
\psline[linewidth=1pt]{->}(1.2,1.5)(1.3,1.5)
\psline[linewidth=1pt]{->}(1.2,0.3)(1.3,0.3)
\psline[linewidth=1pt]{->}(2.7,1.2)(2.8,1.3)
\psline[linewidth=1pt]{->}(2.7,0.6)(2.8,0.5)
\rput[cc]{0}(0.3,1.5){$e$}
\rput[cc]{0}(0.3,0.3){$e$}
\rput[cc]{0}(1.4,0.9){$\widetilde{\g}$}
\rput[lc]{0}(3.3,1.5){$\widetilde{e}$}
\rput[lc]{0}(3.3,0.3){$\widetilde{e}$}
\endpspicture
}
\end{center}